\documentclass[
reprint,
superscriptaddress,
 aps,
]{revtex4-2}
\usepackage{amsmath,amssymb,mathrsfs,multirow}
\usepackage{graphicx}
\usepackage{dcolumn}
\usepackage{bm,slashed}
\usepackage[colorlinks=true,linkcolor=blue,urlcolor=blue,citecolor=blue]{hyperref}
\usepackage[compat=1.1.0]{tikz-feynman}
\usepackage{amsmath}
\usepackage[utf8]{inputenc}
\usepackage{slashed}
\usepackage{subfig}
\usepackage{graphicx}
\usepackage{float}
\usepackage{multirow}
\usepackage{amssymb}
\usepackage{mathrsfs}  
\usepackage{dcolumn}
\usepackage{bm}
\begin{document}

\title{\boldmath Lepton flavor violating top quark FCNC processes at the $\mu$TRISTAN}
\author{Abhik Sarkar}
\email{sarkar.abhik@iitg.ac.in}
\affiliation{Department of Physics, Indian Institute of Technology Guwahati, North Guwahati, 781039, India}

\begin{abstract}
{We investigate charged lepton flavor violating top quark flavor-changing neutral current interactions at the proposed asymmetric muon-electron collision stage of the $\mu$TRISTAN collider, operating at a center-of-mass energy of 346 GeV. Specifically, we study the process $\mu^{+} e^{-} \rightarrow t q$ ($t \overline{q} + \overline{t}q$) with $q = u, c$, within the framework of three classes of four-fermion contact interactions: scalar, vector, and tensor operators. A cut-based analysis is performed using boost-invariant kinematic observables, followed by a likelihood-based statistical treatment to derive projected sensitivities for each operator. For an integrated luminosity of $100~\text{fb}^{-1}$, the projected constraints improve upon current LHC bounds on the corresponding effective couplings by approximately an order of magnitude. Projections for $1~\text{ab}^{-1}$ indicate even stronger sensitivity, indicating improved reach at higher luminosity. Additionally, we explore the impact of initial-state beam polarization on these projections, showing how it can further enhance sensitivity to specific operator structures.}
\end{abstract}
\maketitle
\section{Introduction}
\label{sec:introduction}
The Standard Model (SM) of particle physics has been remarkably successful in explaining a wide range of experimental observations. One of its intriguing features is the flavor structure, with each type of fermion appearing in three generations. Flavor mixing arises through the Yukawa interactions, which couple the flavor and mass eigenstates. In the quark sector, such mixing appears in the charged current interactions and is suppressed by the Glashow–Iliopoulos–Maiani (GIM) mechanism~\cite{Glashow:1970gm}. In contrast, the SM originally contained no analogous mixing in the lepton sector.

In the quark sector, flavor-changing neutral current (FCNC) processes are forbidden at tree level and highly suppressed at loop level within the SM. These include radiative decays such as $b \to s \gamma$ ($B \to X_{s} \gamma$~\cite{Belle:2014nmp,BaBar:2012eja,BaBar:2012fqh}, $B \to K^{*} \gamma$~\cite{Belle:2017hum,BaBar:2009byi}), semileptonic decays $b \to s\ell^{+} \ell^{-}$ ($B \to K \ell^{+} \ell^{-}$~\cite{BELLE:2019xld,BaBar:2008jdv}, $B \to K^{*} \ell^{+} \ell^{-}$~\cite{BaBar:2008jdv,Belle:2009zue}, $B_{s} \to \phi \ell^{+} \ell^{-}$~\cite{LHCb:2021zwz}), neutral meson mixing ($B^{0}-\overline{B}^{0}$~\cite{Belle-II:2023bps}, $B^{0}_{s}-\overline{B}^{0}_{s}$~\cite{LHCb:2021moh}, $D^{0}-\overline{D}^{0}$~\cite{LHCb:2013zpr}), and rare kaon decays ($K^{+} \to \pi^{+} \nu \overline{\nu}$~\cite{NA62:2021zjw}), among others. {Although these precision observables provide sensitive probes of potential new physics, in most cases, the current measurements are largely consistent with the Standard Model within experimental and theoretical uncertainties, and the flavor sector continues to offer strong constraints complementary to direct collider searches.} Similar FCNC searches have also been conducted in the context of top quark production and decay~\cite{CMS:2024ubt,ATLAS:2021amo,ATLAS:2022per,ATLAS:2023qzr,ATLAS:2024njy,CMS:2022ztx}. However, no significant deviations from the SM predictions have been observed so far, primarily due to limited precision in current top quark measurements. As the heaviest particle in the SM, the top quark offers a promising window into potential new physics. Consequently, FCNC studies involving the top quark are crucial for probing physics beyond the SM at current and future experimental frontiers.

In the lepton sector, lepton flavor violation (LFV) is strictly forbidden in the SM with massless neutrinos. However, the discovery of neutrino oscillations~\cite{Super-Kamiokande:1998kpq,SNO:2002tuh}, implying non-zero neutrino masses and lepton flavor mixing, confirms LFV in the neutral lepton sector. In contrast, charged LFV remains unobserved despite extensive experimental efforts. Searches have targeted processes such as radiative decays ($\mu \to e \gamma$~\cite{MEGII:2023ltw}, $\tau \to \ell \gamma$, with $\ell = e, \mu$~\cite{BaBar:2009hkt,Belle:2021ysv}), three-body decays ($\mu \to 3e$~\cite{SINDRUM:1987nra}, $\tau \to 3\ell$~\cite{Hayasaka:2010np}), and coherent $\mu \to e$ conversion in nuclei~\cite{Badertscher:1980bt,SINDRUMII:1993gxf,SINDRUMII:1996fti,SINDRUMII:2006dvw}. In the minimally extended SM with neutrino masses and mixing, contributions to charged LFV processes are extremely suppressed, with radiative decay branching ratios as low as $\sim 10^{-54}$~\cite{Davidson:2022jai}. Therefore, any experimental observation of charged LFV would constitute a clear signal of new physics in the charged lepton sector.

Top quark FCNC processes have been investigated by phenomenological studies~\cite{Hou:2020chc,Fuyuto:2015gmk,Bhattacharya:2025mlg,Bhattacharya:2023beo,Sun:2023cuf,GhasemiBostanabad:2025xua,Atkinson:2024hqp,Chen:2022dzc,Gaitan:2017tka,Shi:2019epw,Arroyo-Urena:2019qhl,Liu:2020kxt,Liu:2021crr,Gutierrez:2020eby,Bie:2020sro,Balaji:2020qjg,Frank:2006ku,dEnterria:2023wjq,Kala:2025srq} in various beyond the Standard Model (BSM) scenarios, and within the effective field theory (EFT) framework, in the context of both current and future collider experiments. Similarly, charged LFV has also been extensively studied in the literature~\cite{CMS:2016cvq,CMS:2017con,OPAL:1995grn,DELPHI:1996iox,ATLAS:2014vur,ATLAS:2020zlz,ATLAS:2021bdj,OPAL:2001qhh,Han:2010sa,Davidson:2012wn,Cai:2015poa,Cai:2018cog,Angelescu:2020uug,Murakami:2014tna,Cho:2018mro,Etesami:2021hex,Altmannshofer:2023tsa,Altmannshofer:2025nbp,Moreno-Sanchez:2025bzz,Palavric:2024gvu,Calibbi:2025fzi,Kriewald:2024cnt,Batell:2024cdl,Calibbi:2024rcm,Ding:2024zaj,Santiago:2024zpc,Heeck:2024uiz,Goudelis:2023yni,Lichtenstein:2023iut,Jahedi:2024kvi}. In this work, we aim to simultaneously address both phenomena by exploring charged lepton flavor violating top quark FCNC interactions through $tq$ ($t \overline{q} + \overline{t}q$) production, with $q = u, c$, at the $\mu^{+} e^{-}$ asymmetric collision stage of the proposed $\mu$TRISTAN collider~\cite{Hamada:2022mua}. {Such charged lepton flavor violating top quark FCNC interactions can arise naturally in a variety of BSM scenarios. For instance, models with extended Higgs sectors, such as two Higgs doublet models (2HDMs) with flavor-violating Yukawa couplings~\cite{Chen:2023eof,Crivellin:2013wna,Goncalves:2023ydf}, can generate tree-level top FCNCs together with charged LFV. Leptoquark models, which introduce colored scalar or vector particles coupling simultaneously to quarks and leptons, can induce both top quark FCNC transitions and LFV processes, with their flavor structure determined by underlying symmetries or coupling hierarchies~\cite{Dorsner:2016wpm,Davidson:1993qk}. Similarly, models with an additional $Z'$ gauge boson that has flavor-dependent couplings to SM fermions can enhance top FCNC interactions while simultaneously generating charged LFV~\cite{Langacker:2008yv,Langacker:2000ju}, without conflicting with stringent low-energy constraints. While experimental measurements strongly constrain either charged LFV processes or quark FCNCs individually, scenarios involving simultaneous violations in both the lepton and top quark sectors are relatively less restricted, allowing more freedom for phenomenological exploration. Additionally, such scenarios can also arise from composite Higgs models~\cite{Agashe:2004rs,Feruglio:2015gka}. In all these frameworks, interactions involving up-type quarks, particularly the top quark, are more natural, as down-type FCNCs are strongly constrained by precision $B$-physics measurements. This provides a theoretically well-motivated rationale for focusing on lepton flavor violating processes involving the top quark.} {For detailed BSM scenarios, refer Appendix~\ref{app:uv}.} Experimental searches at the LHC~\cite{CMS:2022ztx,ATLAS:2024njy} and some phenomenological studies~\cite{Altmannshofer:2025lun} have already explored charged lepton flavor violating top quark FCNC interactions in the context of low-energy experiments, hadron colliders, and same-flavor lepton colliders. However, different-flavor lepton colliders~\cite{Lu:2020dkx,Bouzas:2021sif,Bouzas:2023vba} offer a cleaner environment for probing charged LFV due to reduced background contamination and the possibility of introducing flavor violation directly at the initial collision vertex.

The proposed $\mu$TRISTAN collider is an asymmetric muon–electron collider designed to operate within the existing TRISTAN~\cite{Iwata:1993qk} storage ring. It aims to collide ultra-cold $\mu^{+}$ beams with energies of 1 TeV against high-intensity $e^{-}$ beams of 30 GeV, achieving a center-of-mass energy of $\sqrt{s} = 346$ GeV. The baseline design targets an instantaneous luminosity of approximately $4.6 \times 10^{33}\,\text{cm}^{-2}\,\text{s}^{-1}$, enabling the accumulation of around $100\,\text{fb}^{-1}$ of data over a few operational years. With potential upgrades to beam intensity and repetition rate, the integrated luminosity could be extended up to $1\,\text{ab}^{-1}$, making the $\mu$TRISTAN a promising platform for probing rare processes and performing precision measurements. A follow-up stage envisions symmetric same-sign anti-muon collisions at a center-of-mass energy of $\sqrt{s} = 2$ TeV.

Several BSM phenomenological studies have been conducted for both stages of the $\mu$TRISTAN~\cite{Calibbi:2024rcm,Calibbi:2025fzi,Kriewald:2024cnt,Batell:2024cdl,Ding:2024zaj,Santiago:2024zpc,Heeck:2024uiz,Goudelis:2023yni,Lichtenstein:2023iut,Dehghani:2025xkd,Bhattacharya:2025xwv,Bolton:2025tqw,deLima:2024ohf,Hamada:2024ojj,Chen:2024tqh,Okabe:2023esr,Fukuda:2023yui,Dev:2023nha}. In particular, charged LFV has been explored in the context of $Z'$ models, heavy neutrino models, axion-like particle (ALP) searches, and effective operator frameworks, focusing predominantly on the lepton sector. In this work, we focus on the process $\mu^{+} e^{-} \to tq$ (with $q = u, c$), induced by charged lepton flavor violating top quark FCNC interactions. We construct simplified scalar, vector, and tensor four-fermion operators extracted from Standard Model Effective Field Theory (SMEFT)~\cite{Buchmuller:1985jz,Grzadkowski:2010es} operators, following the modeling strategy used in~\cite{CMS:2022ztx,ATLAS:2024njy}, as detailed later. Our analysis constrains these EFT operators and, in turn, the rare FCNC decay modes of the top quark involving charged LFV. These bounds can be mapped onto a broad class of BSM scenarios that link the lepton and quark sectors, such as $Z'$ and leptoquark models.
 
The paper is organized as follows: In Sec.~\ref{sec:operator}, we present the EFT framework adopted for this study and review existing constraints on the relevant operators. Sec.~\ref{sec:collider} details the collider analysis of the process $\mu^{+} e^{-} \to tq$ at the $\mu$TRISTAN, based on a cut-based strategy, along with a discussion on the impact of beam polarization. In Sec.~\ref{sec:sensitivity}, we provide the projected sensitivities for the EFT operators and the corresponding limits on rare top quark decay branching ratios. Finally, we summarize our findings and conclusions in Sec.~\ref{sec:conclusion}.
\section{Effective operators}
\label{sec:operator}
In the absence of direct evidence for new physics at current experiments, effective field theories have emerged as powerful tools for interpreting observed deviations and projecting sensitivities in a model-independent manner. Among these, SMEFT provides a systematic extension of the SM by incorporating higher-dimensional operators built from SM fields and respecting its gauge symmetries, without introducing additional degrees of freedom. The SMEFT Lagrangian is given by:
\begin{equation}
    \mathcal{L}_{\rm SMEFT} = \mathcal{L}_{\rm SM} + \sum_{i,d} \frac{C^{d}_{i}}{\Lambda^{d-4}}\,\mathcal{O}^{d}_{i}\;,
\end{equation}
where $\mathcal{O}^{d}_{i}$ are dimension-$d$ operators constructed from SM fields, $C^{d}_{i}$ are their corresponding dimensionless Wilson coefficients (WCs), and $\Lambda$ denotes the effective scale. While the leading contribution in SMEFT arises at dimension-5 through the Weinberg operator~\cite{Weinberg:1979sa}, its phenomenological impact is limited primarily to neutrino mass generation. In practice, the dominant and most widely studied effects begin at dimension-6, which have significant implications across various sectors of particle physics. A complete and non-redundant basis of dimension-6 operators is presented in~\cite{Grzadkowski:2010es}.

\begin{table*}[htb!]
    \centering
    \renewcommand{\arraystretch}{1.2}{
    \begin{tabular}{>{\centering\arraybackslash}p{2.25cm}
                     >{\centering\arraybackslash}p{1.75cm}
                     >{\centering\arraybackslash}p{4cm}
                     >{\centering\arraybackslash}p{0.1cm}
                     >{\centering\arraybackslash}p{1.5cm}
                     >{\centering\arraybackslash}p{3cm}}
        \hline \hline
        \multirow{2}*{Operator} & \multicolumn{2}{c}{SMEFT operators} && \multicolumn{2}{c}{Simplified operators} \\ \cline{2-3} \cline{5-6}
        & Symbol & Structure && Symbol & Structure \\ 
        \hline \hline
        \multirow{2}*{Scalar} & \multirow{2}*{$[\mathcal{O}^{(1)}_{LeQu}]_{ijkl}$} & \multirow{2}*{$(\overline{L}^{a}_{i} e_{j}) \epsilon^{ab} (\overline{Q}^{b}_{k} u_{l})$} && $\mathcal{O}^{S}_{e \mu t u}$ & $(\overline{e} \mu)(\overline{t}u)$ \\
        & & && $\mathcal{O}^{S}_{e \mu t c}$ & $(\overline{e} \mu)(\overline{t}c)$ \\ \hline \hline
        \multirow{5}*{Vector} & $[\mathcal{O}^{(1)}_{LQ}]_{ijkl}$ & $(\overline{L}_{i} \gamma_{\alpha} L_{j}) (\overline{Q}_{k} \gamma^{\alpha} Q_{l})$ && \multirow{5}*{\shortstack[c]{$\mathcal{O}^{V}_{e \mu t u}$ \\ $\mathcal{O}^{V}_{e \mu t c}$}} & \multirow{5}*{\shortstack[c]{$(\overline{e} \gamma_{\alpha} \mu)(\overline{t} \gamma^{\alpha} u)$ \\ $(\overline{e} \gamma_{\alpha} \mu)(\overline{t} \gamma^{\alpha} c)$}} \\
        & $[\mathcal{O}^{(3)}_{LQ}]_{ijkl}$ & $(\overline{L}_{i} \gamma_{\alpha} \tau^{a} L_{j}) (\overline{Q}_{k} \gamma^{\alpha} \tau^{a} Q_{l})$ && & \\
        & $[\mathcal{O}_{eu}]_{ijkl}$ & $(\overline{e}_{i} \gamma_{\alpha} e_{j}) (\overline{u}_{k} \gamma^{\alpha} u_{l})$ && & \\
        & $[\mathcal{O}_{Lu}]_{ijkl}$ & $(\overline{L}_{i} \gamma_{\alpha} L_{j}) (\overline{u}_{k} \gamma^{\alpha} u_{l})$ && & \\
        & $[\mathcal{O}_{Qe}]_{ijkl}$ & $(\overline{e}_{i} \gamma_{\alpha} e_{j}) (\overline{Q}_{k} \gamma^{\alpha} Q_{l})$ && & \\ \hline \hline
        \multirow{2}*{Tensor} & \multirow{2}*{$[\mathcal{O}^{(3)}_{LeQu}]_{ijkl}$} & \multirow{2}*{$(\overline{L}^{a}_{i} \sigma_{\alpha \beta} e_{j}) \epsilon^{ab} (\overline{Q}^{b}_{k} \sigma^{\alpha \beta} u_{l})$} && $\mathcal{O}^{T}_{e \mu t u}$ & $(\overline{e} \sigma_{\alpha \beta} \mu)(\overline{t} \sigma^{\alpha \beta} u)$ \\
        & & && $\mathcal{O}^{T}_{e \mu t c}$ & $(\overline{e} \sigma_{\alpha \beta} \mu)(\overline{t} \sigma^{\alpha \beta} c)$ \\ \hline \hline
    \end{tabular}}
    \caption{SMEFT and corresponding simplified four-fermion operators contributing to the process $\mu^+ e^- \rightarrow tq$ $(q = u, c)$ at the $\mu$TRISTAN. The scalar and tensor operators include their Hermitian conjugates, although they are not shown explicitly in the table. Indices $i,j$ represent lepton flavors, while $k,l$ denote quark flavors. The simplified operators incorporate all relevant flavor combinations necessary for the considered processes.}
    \label{tab:smeft-op}
\end{table*}

The leading contributions to the process $\mu^+ e^- \rightarrow t q$ $(q = u, c)$ arise from three classes of four-fermion SMEFT operators: scalar ($\mathcal{O}^{(1)}_{LeQu}$, including its Hermitian conjugate), vector ($\mathcal{O}^{(1)}_{LQ}$, $\mathcal{O}^{(3)}_{LQ}$, $\mathcal{O}_{eu}$, $\mathcal{O}_{Lu}$, and $\mathcal{O}_{Qe}$), and tensor ($\mathcal{O}^{(3)}_{LeQu}$, also including its Hermitian conjugate), as summarized in Table~\ref{tab:smeft-op}. These operators are flavor-dependent, with lepton flavor indices denoted by $i,j$ and quark flavor indices by $k,l$. Following the approach in~\cite{CMS:2022ztx}, we define simplified operators that capture the relevant flavor structures contributing to the process, abstracted from the full SMEFT basis. The mapping between the simplified and SMEFT operators is given by:
\begin{equation}\label{eq:simop}
\begin{split}
    \mathcal{O}^{S}_{e\mu tq} &= [\mathcal{O}^{(1)}_{LeQu}]_{e\mu tq}\,,\\
    \mathcal{O}^{V}_{e\mu tq} &= [\mathcal{O}^{(1)}_{LQ}]_{e\mu tq} + [\mathcal{O}^{(3)}_{LQ}]_{e\mu tq} + [\mathcal{O}_{eu}]_{e\mu tq} \\ &+ [\mathcal{O}_{Lu}]_{e\mu tq} + [\mathcal{O}_{Qe}]_{e\mu tq}\,,\\
    \mathcal{O}^{T}_{e\mu tq} &= [\mathcal{O}^{(3)}_{LeQu}]_{e\mu tq}\,.
\end{split}
\end{equation}
It should be noted that upon flavor expansion, the operators $[\mathcal{O}^{(1)}_{LQ}]_{e\mu tq}$ and $[\mathcal{O}^{(3)}_{LQ}]_{e\mu tq}$ yield identical Lorentz structures. However, due to the presence of SU(2) generators $\tau^{a}$ in $[\mathcal{O}^{(3)}_{LQ}]_{e\mu tq}$, the charged lepton interactions acquire an overall negative sign, while the neutrino interactions acquire a positive sign~\cite{Durieux:2014xla,Aebischer:2025qhh,ATLAS:2024njy}. Consequently, the effective WCs contributing to the $\mathcal{O}_{LQ}$-like Lorentz structure is given by:
\begin{equation}
    \left(\frac{[C^{(-)}_{LQ}]_{e\mu tq}}{\Lambda^{2}}\right) = \left(\frac{[C^{(1)}_{LQ}]_{e\mu tq}}{\Lambda^{2}}\right) - \left(\frac{[C^{(3)}_{LQ}]_{e\mu tq}}{\Lambda^{2}}\right)\,.
\end{equation}

\subsection{Constraints on operators}
Before presenting our sensitivity projections at the $\mu$TRISTAN, it is important to review the current experimental constraints on charged lepton flavor violating top quark FCNC interactions. The most stringent bounds to date are provided by a recent CMS analysis at $\sqrt{s} = 13$ TeV with an integrated luminosity of 138 fb$^{-1}$~\cite{CMS:2022ztx,Delzanno:2024ooj}. This analysis combines searches for anomalous production via $qg \to e\mu t$ and rare decay processes $t \to e\mu q$, interpreting the results in the framework of effective four-fermion interactions involving scalar, vector, and tensor operators. 

The limits are derived under the assumption that only one operator is active at a time and are quoted in terms of the effective WCs, $C^{X}_{e\mu tq}/\Lambda^{2}$. These bounds can be translated into constraints on the branching ratio of the rare top decay $t \to e\mu q$ $(q = u, c)$ using the following expression~\cite{ATLAS:2024njy,Chala:2018agk}:
\begin{equation} \label{eq:brtop}
\begin{split}
    \mathcal{B}(t \to e\mu q) &= \frac{m^{5}_{t}}{3072\, \pi^{3}\, \Gamma_{t}} \Bigg[\; \left( \frac{C^{S}_{e\mu tq}}{\Lambda^{2}} \right)^{2} \\ &+ 8\, \left( \frac{C^{V}_{e\mu tq}}{\Lambda^{2}} \right)^{2} + 48\, \left( \frac{C^{T}_{e\mu tq}}{\Lambda^{2}} \right)^{2} \;\Bigg],
\end{split}
\end{equation}
where $m_{t}$ and $\Gamma_{t}$ are the top quark mass and total decay width, respectively. {It is worth emphasizing that our analysis is performed strictly within the SMEFT framework, where $SU(2)_L$ gauge invariance restricts the allowed chiral structures of the semileptonic operators. In particular, the scalar and tensor operators $\mathcal{O}^{(1)}_{LeQu}$ and $\mathcal{O}^{(3)}_{LeQu}$ generate fixed chirality combinations. Independent chiral structures that would arise in a general low-energy EFT are not present, unlike~\cite{Dorsner:2016wpm}. Consequently, for the fully integrated unpolarized decay width considered here, the scalar-tensor interference term vanishes due to the antisymmetric Lorentz structure of the tensor operator, and is therefore not included in our expressions.} Since the contribution of these operators to $\Gamma_{t}$ is negligible for small values of the WCs, it is treated as a fixed SM parameter. The resulting bounds are summarized in Tab.~\ref{tab:bounds}.

\begin{table}[htb!]
    \centering
    \renewcommand{\arraystretch}{1.2}{
    \begin{tabular}{>{\centering\arraybackslash}p{2cm}
                     >{\centering\arraybackslash}p{3cm}
                     >{\centering\arraybackslash}p{3cm}}
    \hline \hline
    Operator & $C^{X}_{e\mu tq}/\Lambda^{2}$ (TeV$^{-2}$) & $\mathcal{B} (t \rightarrow e \mu q)$ \\ \hline \hline
    $\mathcal{O}^{S}_{e\mu tu}$ & $0.24$ & $7.00 \times 10^{-8}$ \\ 
    $\mathcal{O}^{V}_{e\mu tu}$ & $0.12$ & $1.30 \times 10^{-7}$ \\
    $\mathcal{O}^{T}_{e\mu tu}$ & $0.06$ & $2.50 \times 10^{-7}$ \\ \hline \hline
    $\mathcal{O}^{S}_{e\mu tc}$ & $0.86$ & $8.90 \times 10^{-7}$ \\
    $\mathcal{O}^{V}_{e\mu tc}$ & $0.37$ & $1.31 \times 10^{-6}$ \\
    $\mathcal{O}^{T}_{e\mu tc}$ & $0.21$ & $2.59 \times 10^{-6}$ \\ \hline \hline
    \end{tabular}}
    \caption{95\% C.L. exclusion limits on charged lepton flavor violating top quark FCNC operators and corresponding bounds on top quark rare decay branching ratios from the LHC at 13 TeV with 138 fb$^{-1}$~\cite{CMS:2022ztx}.}
    \label{tab:bounds}
\end{table}

As shown in Tab.~\ref{tab:bounds}, the existing CMS bounds constrain the WCs down to the level of $\mathcal{O}(10^{-2})\;\text{TeV}^{-2}$, translating to upper limits on the rare top quark branching ratios as low as $\mathcal{O}(10^{-8})$. The constraints are generally stronger for the $u$-type operators compared to $c$-type, reflecting the parton distribution function (PDF) enhancement of the up quark in proton collisions. Among the three operator classes, the tensor operator coefficients are the most tightly constrained, as both the production cross section for $qg \to e\mu t$ and the decay branching ratio of $t \to e\mu q$ exhibit a stronger sensitivity to variations in tensor interactions compared to their scalar and vector counterparts. {This is because, tensor operators tend to contribute more strongly than vector operators, which in turn are larger than scalar operators, because the way their Lorentz structures contract with fermion spinors and affect the kinematics leads to progressively larger decay rates or cross sections. This pattern is evident in Eq.~\ref{eq:brtop}, where the tensor contribution carries the largest weight, followed by vector and then scalar.}{For completeness and for comparison purposes, we also summarize existing bounds on other flavor structures, namely, scenarios with LFV but no quark FCNC, and vice versa, in Appendix~\ref{app:1}.}

\section{Collider analysis}
\label{sec:collider}
\begin{figure}[htb!]
    \begin{center}
        \begin{tikzpicture}[baseline={(current bounding box.center)},style={scale=1, transform shape}]
            \begin{feynman}
                \vertex [blob] (a) {};
                \vertex [above left = 1.5cm and 2.5cm of a] (b) {$e^{-}$};
                \vertex [below left = 1.5cm and 2.5cm of a] (c) {$\mu^{+}$};
                \vertex [above right = 1.5cm and 2.5cm of a] (e) {$t,u$};
                \vertex [below right = 1.5cm and 2.5cm of a] (f){$\overline{u},\overline{t}$};
                
                \diagram* {
                    (b) -- [ultra thick, fermion, arrow size = 1.5 pt] (a) -- [ultra thick, fermion, arrow size = 1.5 pt] (c),
                    (a) -- [ultra thick, fermion, arrow size = 1.5 pt] (e),
                    (f) -- [ultra thick, fermion, arrow size = 1.5 pt] (a)
                };
            \end{feynman}
        \end{tikzpicture} \\
        \begin{tikzpicture}[baseline={(current bounding box.center)},style={scale=1, transform shape}]
            \begin{feynman}
                \vertex [blob] (a) {};
                \vertex [above left = 1.5cm and 2.5cm of a] (b) {$e^{-}$};
                \vertex [below left = 1.5cm and 2.5cm of a] (c) {$\mu^{+}$};
                \vertex [above right = 1.5cm and 2.5cm of a] (e) {$t,c$};
                \vertex [below right = 1.5cm and 2.5cm of a] (f){$\overline{c},\overline{t}$};
                
                \diagram* {
                    (b) -- [ultra thick, fermion, arrow size = 1.5 pt] (a) -- [ultra thick, fermion, arrow size = 1.5 pt] (c),
                    (a) -- [ultra thick, fermion, arrow size = 1.5 pt] (e),
                    (f) -- [ultra thick, fermion, arrow size = 1.5 pt] (a)
                };
            \end{feynman}
        \end{tikzpicture}
        \caption{Feynman diagrams corresponding to $\mu^+ e^- \rightarrow tq$, $q=u,c$ production at the $\mu$TRISTAN. \label{fig:feyn}}
    \end{center}
\end{figure}
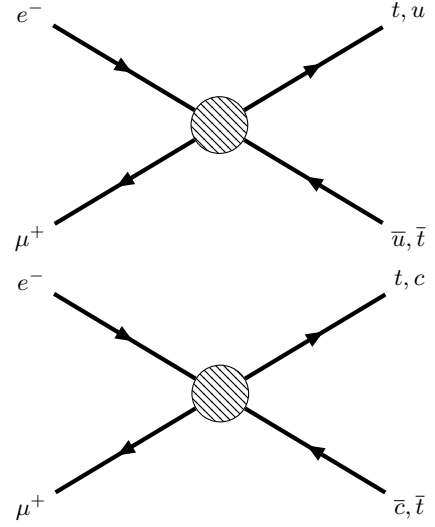

The Feynman diagrams corresponding to the process $\mu^{+} e^{-} \to tq$ $(q = u, c)$ are shown in Fig.~\ref{fig:feyn}. For the collider analysis, we focus on the leptonic decay channel of the top quark, $t \to b \ell \nu$, which leads to a final state consisting of two jets (one being a $b$-jet), a charged lepton, and missing transverse energy (MET). The dominant SM background arises from $\mu^{+} e^{-} \to \ell \nu jj$, primarily via the process $\mu^{+} e^{-} \to W \ell \nu$, where the $W$ boson decays hadronically. This background is reducible and can be significantly suppressed with appropriate kinematic selections, as discussed later.

While the hadronic decay mode of the top quark ($t \to bjj$) benefits from a larger branching ratio and relatively lower background contamination, it poses practical challenges in this analysis. Since our study involves comparing operator structures from different effective classes, angular observables, particularly the separation between the light quark ($q$) and the top quark, play a crucial role in distinguishing operator contributions. In the leptonic channel, this angular separation can be effectively approximated using the separation between the light jet and the $b$-jet. In contrast, the hadronic channel suffers from combinatorial ambiguities due to the presence of multiple light jets, making the reconstruction of the correct jet pairing and extraction of angular correlations considerably more difficult.

For our analysis, we define the following individual operator benchmark points, with all other WCs set to zero:
\begin{widetext}
\begin{equation}
\begin{split}
    \text{BP}(S_u): \hspace{0.2cm} \left(\frac{C^{S}_{e \mu t u}}{\Lambda^{2}}\right) = 0.01\;{\rm TeV}^{-2}, \hspace{1cm}
    \text{BP}(S_c): \hspace{0.2cm} \left(\frac{C^{S}_{e \mu t c}}{\Lambda^{2}}\right) = 0.01\;{\rm TeV}^{-2}, \\
    \text{BP}(V_u): \hspace{0.2cm} \left(\frac{C^{V}_{e \mu t u}}{\Lambda^{2}}\right) = 0.01\;{\rm TeV}^{-2}, \hspace{1cm}
    \text{BP}(V_c): \hspace{0.2cm} \left(\frac{C^{V}_{e \mu t c}}{\Lambda^{2}}\right) = 0.01\;{\rm TeV}^{-2}, \\
    \text{BP}(T_u): \hspace{0.2cm} \left(\frac{C^{T}_{e \mu t u}}{\Lambda^{2}}\right) = 0.01\;{\rm TeV}^{-2}, \hspace{1cm}
    \text{BP}(T_c): \hspace{0.2cm} \left(\frac{C^{T}_{e \mu t c}}{\Lambda^{2}}\right) = 0.01\;{\rm TeV}^{-2}.
\end{split}
\end{equation}
\end{widetext}
All benchmark points respect the current exclusion limits summarized in Tab.~\ref{tab:bounds}. Although some of the operators are subject to weaker experimental constraints, we choose a uniform benchmark value across all cases to facilitate a consistent and direct comparison during the cut-based analysis, which will be discussed in subsequent sections.

\subsection{Beam polarization effects}
Beam polarization is a powerful handle at lepton colliders that can substantially improve the sensitivity to new physics. By tuning the polarization of the initial-state leptons, one can reduce the SM backgrounds and selectively enhance signal contributions from operators with specific chiral structures. This not only boosts the overall signal-to-background ratio but also provides a means to disentangle the effects of different operator classes based on their chirality. As a result, beam polarization serves both as a background suppression technique and a diagnostic tool for probing the Lorentz structure of effective operators.

At the $\mu$TRISTAN, substantial beam polarization is anticipated for both $\mu^{+}$ and $e^{-}$ beams~\cite{Hamada:2022mua}. Surface muons, originating from $\pi^{+}$ decays, are naturally produced with nearly 100\% polarization due to the $V$–$A$ nature of weak interactions. When subjected to a longitudinal magnetic field of $\sim 0.3$ T, and accounting for possible beam emittance effects, a polarization of $P_{\mu^+} \sim \pm 0.80$ is expected. In less favorable scenarios, where the longitudinal magnetic field is absent, the polarization may reduce to $P_{\mu^+} \sim \pm 0.25$, which we adopt as the conservative benchmark for our analysis. On the electron side, technological developments at SuperKEKB~\cite{Roney:2021pwz} suggest that beam polarizations of $P_{e^-} = \pm 0.70$ are realistically achievable.

The production cross section in the presence of beam polarizations translates to the following expression:
\begin{equation}\label{eq:xs}
\begin{split}
\sigma (P_{\mu^+},P_{e^-}) &= \frac{(1-\overline{P_{\mu^+}})(1-P_{e^-})}{4}\,\sigma_{LL} \\ &+ \frac{(1-\overline{P_{\mu^+}})(1+P_{e^-})}{4}\,\sigma_{LR} \\
&+ \frac{(1+\overline{P_{\mu^+}})(1-P_{e^-})}{4}\,\sigma_{RL}\\ &+ \frac{(1+\overline{P_{\mu^+}})(1+P_{e^-})}{4}\,\sigma_{RR}\,,
\end{split}
\end{equation}
where $\overline{P}$ denotes the flipped sign of $P$. {In addition, $\sigma_{XY}$ denotes the polarized cross section defined as $\sigma_{XY} = \sigma(\mu^{+}_{X}\, e^{-}_{Y} \to t q)$, where $X,Y = L,R$, specify the chiralities of the initial state leptons.} The scalar and tensor operators couple fields of opposite chiralities, whereas the vector operator couples fields of the same chirality. This chiral structure directly affects their response to beam polarization and underlies the polarization dependence observed in the signal rates.

The production cross sections for the signal processes $\mu^+ e^- \to t(b \ell \nu)u$, mediated by scalar, vector, and tensor operators, along with the dominant SM background $\mu^+ e^- \to \ell \nu jj$, are presented in Tab.~\ref{tab:pol-eff} for various polarization configurations. Throughout this section, we denote beam polarization configurations using the shorthand $P_{\rm AB} : (P_{\mu^+}, P_{e^-})$, where ${\rm A} = +, -$ corresponds to $P_{\mu^+} = +0.25, -0.25$, and ${\rm B} = +, -$ corresponds to $P_{e^-} = +0.70, -0.70$. The unpolarized case $(P_{\mu^+}, P_{e^-}) = (0.00, 0.00)$ is denoted as $P_{\times \times}$. {Using Eq.~\eqref{eq:xs}, the cross section can be written as
\begin{equation}
    \sigma(P_{AB}) =  \sum_{XY} M^{AB}_{XY}\; \sigma_{XY}\,,
\end{equation}
where the matrix $M$ is
\begin{equation}
M = \left[\;\;
\begin{matrix}
0.31875 & 0.05625 & 0.53125 & 0.09375 \\
0.05625 & 0.31875 & 0.09375 & 0.53125 \\
0.53125 & 0.09375 & 0.31875 & 0.05625 \\
0.09375 & 0.53125 & 0.05625 & 0.31875
\end{matrix}
\;\;\right].
\end{equation}
The $AB$ and $XY$ sequences correspond to $\{++,+-,-+,--\}$ and $\{LL,LR,RL,RR\}$, respectively. For scalar and tensor operators, only $\sigma_{LR}$ and $\sigma_{RL}$ are non-zero. Furthermore, $\sigma_{LR} = \sigma_{RL}$, since the Hermitian conjugate operator carries the same coupling strength. In contrast, for the vector operator only $\sigma_{LL}$ and $\sigma_{RR}$ are non-zero, and the simplified operator definition in Eq.~\eqref{eq:simop} implies $\sigma_{LL} = \sigma_{RR}$. Therefore, the chiral structure of the operators leads to the relations $\sigma(P_{++}) = \sigma(P_{--})$ and $\sigma(P_{+-}) = \sigma(P_{-+})$ for the signal cross sections.} Among these, $P_{++}$ yields lower SM background than $P_{--}$, making it a preferred benchmark configuration. Similarly, $P_{+-}$ offers reduced background compared to $P_{-+}$ and is thus also selected. Hence, our benchmark polarization choices are $P_{\times \times}$ (unpolarized), $P_{++}$, and $P_{+-}$. Importantly, polarization configurations affect operator sensitivity: $P_{++}/P_{--}$ enhance scalar and tensor operator contributions while suppressing the vector ones. Conversely, $P_{+-}/P_{-+}$ configurations enhance vector operator sensitivity while reducing scalar and tensor contributions. This polarization dependence is crucial for distinguishing different operator structures in experimental analyses.

\begin{table*}[htb!]
    \centering
    \renewcommand{\arraystretch}{1.2}{
    \begin{tabular}{>{\centering\arraybackslash}p{4cm}
                     >{\centering\arraybackslash}p{2cm}
                     >{\centering\arraybackslash}p{2cm}
                     >{\centering\arraybackslash}p{2cm}
                     >{\centering\arraybackslash}p{2.5cm}}
    \hline \hline
    \rule{0pt}{2.0em}
     \multirow{2}*{\shortstack[c]{Polarization \\ $(P_{\mu^{+}}, P_{e^{-}})$}} & \multicolumn{3}{c}{Signal} & \multirow{2}*{\shortstack[c]{Background \\ $\sigma_{\ell \nu jj}$ (fb)}} \\ \cline{2-4} \rule{0pt}{1.5em}
      & $\sigma^{{\rm BP}(S_u)}_{t(\ell \nu b)q}$ (fb) & $\sigma^{{\rm BP}(V_u)}_{t(\ell \nu b)q}$ (fb) & $\sigma^{{\rm BP}(T_u)}_{t(\ell \nu b)q}$ (fb) & \\ [0.5em] \hline \hline
     $P_{\times \times}:$ Unpolarized & 0.035 & 0.105 & 0.276 & 148.4 \\ \hline \hline
     $P_{++}:(+0.25,+0.70)$ & 0.041 & 0.086 & 0.323 & 123.4 \\
     $P_{+-}:(+0.25,-0.70)$ & 0.029 & 0.123 & 0.228 & 242.1 \\
     $P_{-+}:(-0.25,+0.70)$ & 0.029 & 0.123 & 0.228 & 74.8 \\
     $P_{--}:(-0.25,-0.70)$ & 0.041 & 0.086 & 0.323 & 150.5 \\ \hline \hline
    \end{tabular}}
    \caption{Production cross sections for the signal process $\mu^{+} e^{-} \rightarrow t(\ell \nu b)u$, mediated by scalar, vector, and tensor operators, along with the dominant SM background $\mu^{+} e^{-} \rightarrow \ell \nu jj$, are presented for various beam polarization configurations.}
    \label{tab:pol-eff}
\end{table*}

{In addition to differentiating operator structures: scalar, vector, and tensor, beam polarization can also be exploited to disentangle the various SMEFT vector structures listed in Tab.~\ref{tab:smeft-op}. A detailed discussion of the role of polarization in distinguishing these vector operators is provided in Appendix~\ref{app:2}.}

\subsection{Cut-based analysis}
The model implementation is carried out using \texttt{FeynRules}~\cite{Alloul:2013bka}. The corresponding universal FeynRules Output (UFO)~\cite{Darme:2023jdn} model files are exported and passed to \texttt{MG5\_aMC}~\cite{Alwall:2011uj} to generate Monte Carlo (MC) events for both the EFT signal and SM background processes. In addition to the dominant background process $\mu^{+} e^{-} \rightarrow \ell \nu jj$, subleading contributions can arise from processes such as $\mu^{+} e^{-} \rightarrow \ell \ell jj$ {and $\mu^{+} e^{-} \rightarrow \ell \ell bb$, particularly due to non-identification of lepton or jet flavor misidentification.} However, such contributions are highly suppressed and do not pass the event selection criteria; therefore, they are neglected in the analysis.

The generated events are passed to \texttt{Pythia8}~\cite{Bierlich:2022pfr} for parton showering, followed by detector simulation using \texttt{Delphes3}~\cite{deFavereau:2013fsa}. In the absence of a dedicated detector card for the $\mu$TRISTAN, we adopt the \texttt{ILCgen}~\cite{Behnke:2013lya} card, which provides a reasonable approximation of the expected detector performance at the $\mu$TRISTAN. For $b$-jet tagging, we employ the ``tight'' $b$-tagging configuration available in the \texttt{ILCgen} card. This model yields a $b$-tagging efficiency of approximately $53\%$, with mistagging rates of around $5\%$ for charm jets and $0.5\%$ for light-flavor jets ($u,d,s,g$). This tight tagging configuration effectively suppresses SM backgrounds, which rarely contain genuine $b$-jets. While the implementation of a $c$-tagging algorithm could in principle aid in distinguishing between $u$- and $c$-type operators, we refrain from doing so due to the limited discriminating power of current $c$-tagging techniques.

\begin{figure}[htb!]
    \centering
    \includegraphics[width=0.95\linewidth]{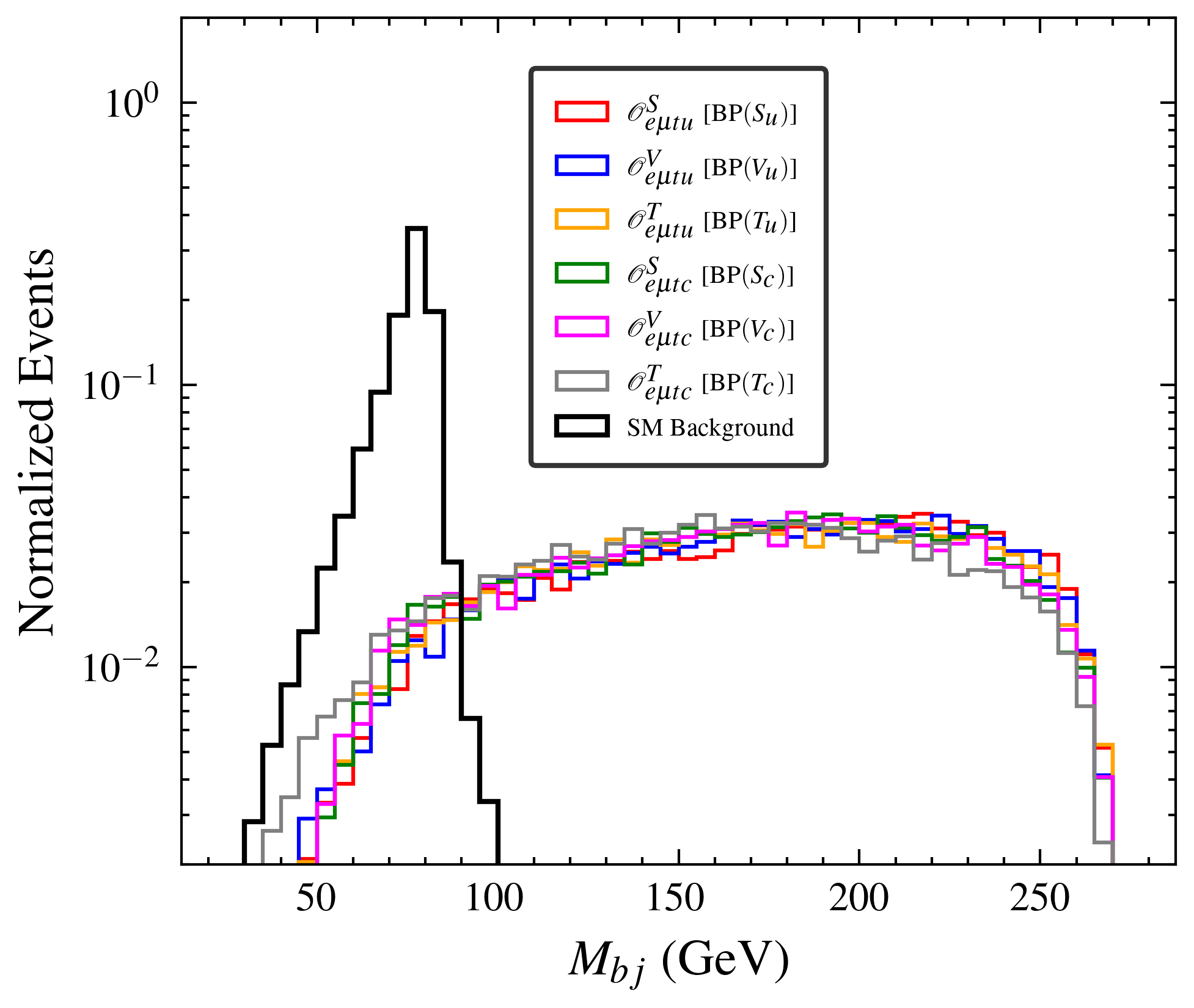} \\
    \includegraphics[width=0.95\linewidth]{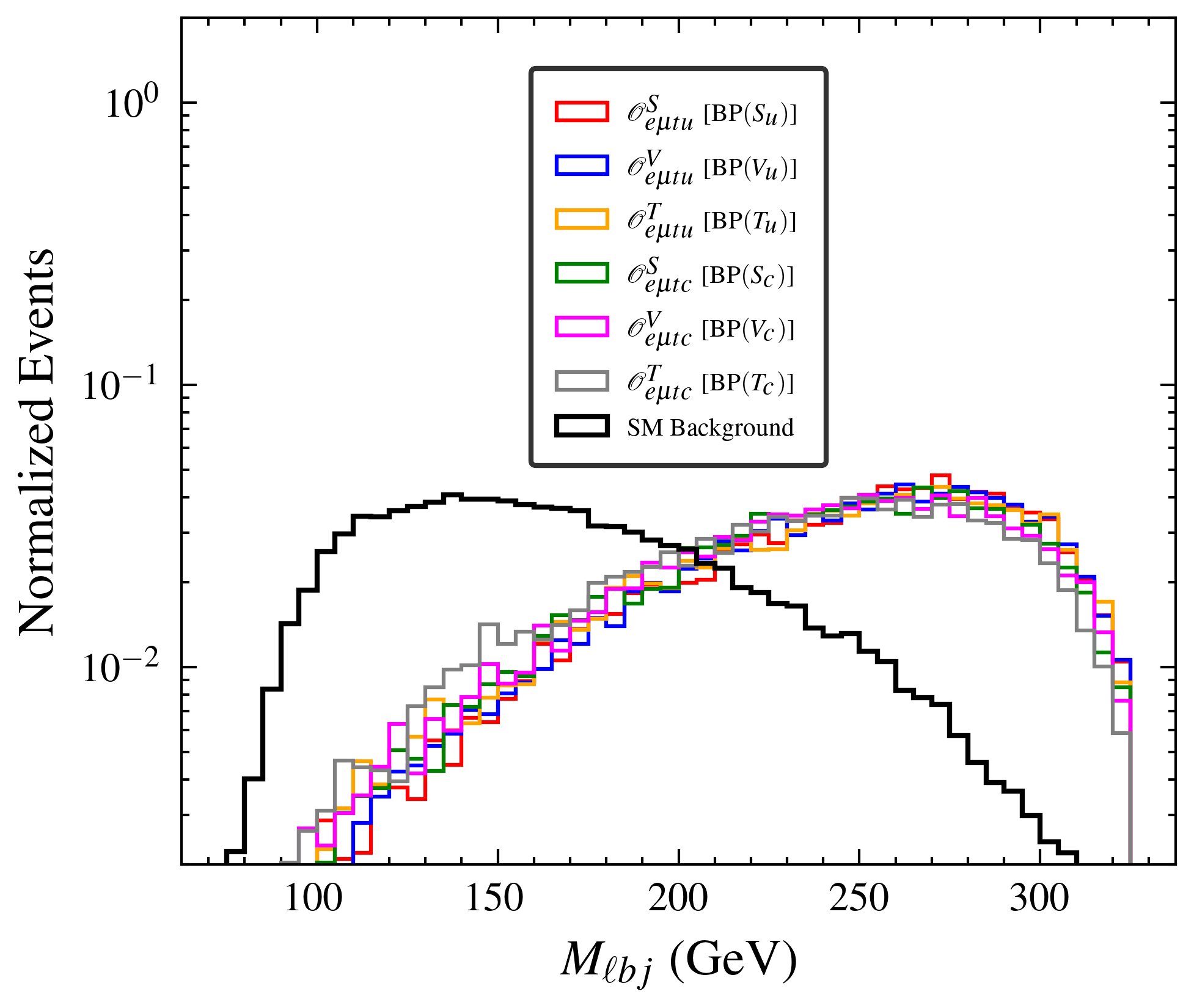}
    \caption{Invariant mass distributions after the baseline cut $\mathcal{C}_{0}$ for signal benchmarks and SM background in the process $\mu^{+} e^{-} \rightarrow t(\ell \nu b)q$ with $q = u, c$ at the $\mu$TRISTAN. $M_{bj}$ (\textit{left}) denotes the invariant mass of the $b$-jet and light jet system, while $M_{\ell bj}$ (\textit{right}) corresponds to the invariant mass of the lepton, $b$-jet, and light jet system.}
    \label{fig:dist}
\end{figure}

For signal event selection, we require events to contain exactly one charged lepton and two jets, with one $b$-tagged and one non-$b$-tagged jet:
\begin{equation} \label{eq:esel}
    N_{\ell} = 1, \hspace{0.5cm} N_{b} = 1, \hspace{0.5cm} N_{j} = 1.
\end{equation}
Due to the asymmetric beam configuration at the $\mu$TRISTAN with $\mu^{+}$ at 1 TeV and $e^{-}$ at 30 GeV the final-state system is strongly boosted along the beam ($z$) direction. In such setups, two primary analysis strategies are typically adopted: (i) reconstructing observables in the partonic center-of-mass frame, or (ii) relying on transverse and boost-invariant observables directly in the lab frame. We follow the latter approach, using kinematic variables such as the invariant mass of visible final-state particles, the transverse momentum of both visible ($p_{T}$) and invisible (MET) particles, rapidity differences, and azimuthal angular separations. Among these, we highlight two key invariant mass observables:
\begin{equation}
    M_{bj} = \sqrt{(p_{b} + p_{j})^{2}}\;, \hspace{1cm}
    M_{\ell bj} = \sqrt{(p_{\ell} + p_{b} + p_{j})^{2}}\;,
\end{equation}
where $p_{\ell}$, $p_{b}$, and $p_{j}$ denote the four-momenta of the lepton, $b$-jet, and light jet, respectively. Fig.~\ref{fig:dist} shows the distributions of $M_{bj}$ (\textit{left}) and $M_{\ell bj}$ (\textit{right}) for the signal benchmarks and SM background in the process $\mu^{+} e^{-} \rightarrow t(\ell \nu b)q$, with $q = u, c$, at the $\mu$TRISTAN. The distributions are obtained after applying detector-level cuts along with the event selection criteria specified in Eq.~\eqref{eq:esel}, collectively referred to as the baseline cut, $\mathcal{C}_{0}$.

Building on the baseline selection, we apply the following sequential kinematic cuts to further enhance the signal-to-background separation:
\begin{equation} \label{eq:ecut}
    \mathcal{C}_{1} : \; M_{bj} > 100\; \text{GeV}, \hspace{1cm}
    \mathcal{C}_{2} : \; M_{\ell bj} > 200\; \text{GeV}.
\end{equation}
The dominant SM background arises from the process $\mu^{+} e^{-} \rightarrow W (jj)\, \ell \nu$, where the invariant mass of the jet pair peaks around the $W$ boson mass. The cut $\mathcal{C}_{1}$ effectively suppresses this background by requiring $M_{bj}$ to lie above 100 GeV. Furthermore, the signal process $\mu^{+} e^{-} \rightarrow tq$ originates from a contact interaction, leading to a harder kinematic spectrum with visible final-state invariant masses peaking closer to the center-of-mass energy. In contrast, the background involves intermediate propagators, resulting in softer distributions. The additional cut $\mathcal{C}_{2}$, imposing $M_{\ell bj} > 200$ GeV, further discriminates against the background.

The cutflow of cross sections for the signal benchmarks and the dominant SM background is summarized in Tab.~\ref{tab:cutflow}. To quantify the observability of the signal, we define the statistical significance as:
\begin{equation}
    \mathcal{Z}_{\mathfrak{L}_{\rm int}} = \frac{\sigma^{\rm S} \times \sqrt{\mathfrak{L}_{\rm int}}}{\sqrt{\sigma^{\rm S} + \sigma^{\rm B}}}\,,
\end{equation}
where $\sigma^{\rm S}$ and $\sigma^{\rm B}$ denote the signal and background cross sections after cuts, respectively, and $\mathfrak{L}_{\rm int}$ is the integrated luminosity. The resulting signal significances for $\mathfrak{L}_{\rm int} = 1\,\text{ab}^{-1}$ are also reported in Tab.~\ref{tab:cutflow}. Despite the conservative choice of benchmark WCs, we find that the signal remains clearly distinguishable from the background. Owing to the strong suppression of the background through the optimized cutflow, even a modest increase in the EFT couplings leads to a marked improvement in significance.
 
\begin{table*}[htb!]
    \centering
    \renewcommand{\arraystretch}{1.2}{
    \begin{tabular}{>{\centering\arraybackslash}p{3cm}
                     >{\centering\arraybackslash}p{2cm}
                     >{\centering\arraybackslash}p{2cm}
                     >{\centering\arraybackslash}p{2cm}
                     >{\centering\arraybackslash}p{2.5cm}}
    \hline \hline
    \multirow{2}*{Benchmarks} & \multicolumn{3}{c}{Cutflow} & \multirow{2}*{$\mathcal{Z}_{1\,{\rm ab}^{-1}}$} \\ \cline{2-4}
     & $\sigma_{0}$ (fb) & $\sigma_{1}$ (fb) & $\sigma_{2}$ (fb) &  \\ \hline \hline
    ${\rm BP}(S_u)$ & 0.014 & 0.012 & 0.011 & $2.008$ \\
    ${\rm BP}(V_u)$ & 0.042 & 0.037 & 0.032 & $4.438$ \\
    ${\rm BP}(T_u)$ & 0.108 & 0.096 & 0.080 & $8.000$ \\ \hline \hline 
    ${\rm BP}(S_c)$ & 0.014 & 0.012 & 0.010 & $1.826$ \\
    ${\rm BP}(V_c)$ & 0.040 & 0.035 & 0.029 & $4.143$ \\
    ${\rm BP}(T_c)$ & 0.103 & 0.087 & 0.070 & $7.379$ \\
    \hline \hline
    SM background & 5.070 & 0.036 & 0.020 & $-$ \\ \hline \hline
    \end{tabular}}
    \caption{Cutflow table showing the signal and background cross sections ($\sigma_i$) after each sequential cut $\mathcal{C}_i$, for various signal benchmarks and the dominant SM background. The final column, $\mathcal{Z}_{1\,\mathrm{ab}^{-1}}$, denotes the signal significance at the $\mu$TRISTAN collider with $\mathfrak{L}_{\rm int} = 1\,\mathrm{ab}^{-1}$.}
    \label{tab:cutflow}
\end{table*}

Tab.~\ref{tab:cutflow1} presents the cross sections after the final selection cut ($\sigma_2$) and the corresponding signal significances, $\mathcal{Z}_{1\,\mathrm{ab}^{-1}}$, for two polarization configurations: $P_{++}$ and $P_{-+}$. As anticipated from the chiral structure of the effective operators, the scalar and tensor benchmarks yield higher significances under the $P_{++}$ configuration, while the vector operator shows enhanced signal significance for the $P_{-+}$ setup.

\begin{table*}[htb!]
    \centering
    \renewcommand{\arraystretch}{1.2}{
    \begin{tabular}{>{\centering\arraybackslash}p{3cm}
                     >{\centering\arraybackslash}p{2cm}
                     >{\centering\arraybackslash}p{2cm}
                     >{\centering\arraybackslash}p{0.25cm}
                     >{\centering\arraybackslash}p{2cm}
                     >{\centering\arraybackslash}p{2cm}}
    \hline \hline
    \multirow{2}*{Benchmarks} & \multicolumn{2}{c}{$P_{++}:(+0.25,+0.70)$} && \multicolumn{2}{c}{$P_{-+}:(-0.25,+0.70)$} \\ \cline{2-3} \cline{5-6}
    & $\sigma_{2}$ (fb) & $\mathcal{Z}_{1\,{\rm ab}^{-1}}$ && $\sigma_{2}$ (fb) & $\mathcal{Z}_{1\,{\rm ab}^{-1}}$ \\ \hline \hline
    ${\rm BP}(S_u)$ & 0.013 & 2.373 && 0.009 & 2.065 \\
    ${\rm BP}(V_u)$ & 0.026 & 3.965 && 0.037 & 5.397 \\
    ${\rm BP}(T_u)$ & 0.094 & 8.922 && 0.066 & 7.571 \\ \hline \hline 
    ${\rm BP}(S_c)$ & 0.012 & 2.228 && 0.008 & 1.886 \\
    ${\rm BP}(V_c)$ & 0.024 & 3.748 && 0.034 & 5.126 \\
    ${\rm BP}(T_c)$ & 0.082 & 8.241 && 0.058 & 7.034 \\ \hline \hline 
    SM background & 0.017 & $-$ && 0.010 & $-$ \\ \hline \hline
    \end{tabular}}
    \caption{Cross sections after the final selection cut ($\sigma_2$) and corresponding signal significance, $\mathcal{Z}_{1\,\mathrm{ab}^{-1}}$, for various signal benchmarks and the SM background under polarization configurations $P_{++}$ and $P_{-+}$.}
    \label{tab:cutflow1}
\end{table*}

\section{Projected sensitivity}
\label{sec:sensitivity}
In this section, we present the projected sensitivity to the effective operator coefficients, $C^{X}_{e\mu tq}/\Lambda^{2}$, using a binned likelihood analysis. By leveraging the shape information of suitably chosen differential distributions, we go beyond total rate measurements and extract stronger constraints. This approach captures not only the overall event yield but also the distinct kinematic features of signal and background processes, thereby enhancing the sensitivity to new physics. We outline the statistical framework used to quantify these projections below.

We estimate the projected sensitivity to the operator coefficients by constructing a binned likelihood using a differential observable, say $\varphi$. The expected number of events in the ${\tt r}^{\rm th}$ bin of $\varphi$ is given by:
\begin{equation}
    \mu_{\tt r} \left(C^{X}_{e\mu tq}/\Lambda^{2}\right) = \mathfrak{L}_{\rm int} \left\{ \sigma^{\rm B}_{\tt r} + \sigma^{\rm S}_{\tt r} \left(C^{X}_{e\mu tq}/\Lambda^{2}\right) \right\}\,,
\end{equation}
where $\mathfrak{L}_{\rm int}$ is the integrated luminosity, $\sigma^{\rm B}_{\tt r}$ is the background cross section in bin ${\tt r}$, and $\sigma^{\rm S}_{\tt r}$ is the signal cross section which depends on the operator coefficient, $C^{X}_{e\mu tq}/\Lambda^{2}$. The corresponding likelihood function \cite{ParticleDataGroup:2024cfk} is defined as a product of Poisson probabilities over all bins:
\begin{equation}
    \mathscr{L} \left(C^{X}_{e\mu tq}/\Lambda^{2}\right)  = \prod^{N}_{{\tt r}=1} \frac{\left\{\mu_{\tt r} \left(C^{X}_{e\mu tq}/\Lambda^{2}\right) \right\}^{n_{\tt r}}e^{-\mu_{\tt r} \left(C^{X}_{e\mu tq}/\Lambda^{2}\right)}}{n_{\tt r}!}\,,
\end{equation}
where, the observable $\varphi$ is divided into $N$ bins, and $n_{\tt r}$ is the observed number of events in the ${\tt r}^{\rm th}$ bin. Since no observed data exist for future colliders, the observed event counts are assumed to equal the expected background yields. So, the observed events are taken as the background expectation:
\begin{equation}
    n_{\tt r} = \mathfrak{L}_{\rm int}\,\sigma^{\rm B}_{\tt r}\,.
\end{equation}
The log-likelihood is then given by:
\begin{equation}
\begin{split}
    \log{\mathscr{L} \left(C^{X}_{e\mu tq}/\Lambda^{2}\right)} &= \sum^{N}_{{\tt r}=1} \Bigg\{n_{\tt r} \log{\mu_{\tt r} \left(C^{X}_{e\mu tq}/\Lambda^{2}\right)} \\ &- \mu_{\tt r} \left(C^{X}_{e\mu tq}/\Lambda^{2}\right) \Bigg\} + {\rm constant}\,.
\end{split}
\end{equation}
To quantify the sensitivity, we define the profile likelihood ratio:
\begin{equation} \label{eq:profile}
\begin{split}
    \lambda \left(C^{X}_{e\mu tq}/\Lambda^{2}\right) &= \frac{\mathscr{L} \left(C^{X}_{e\mu tq}/\Lambda^{2}\right)}{\mathscr{L} \left(\hat{C}^{X}_{e\mu tq}/\Lambda^{2}\right)}\,, \\
    \mathscr{Q} \left(C^{X}_{e\mu tq}/\Lambda^{2}\right) &= -2 \log{\lambda \left(C^{X}_{e\mu tq}/\Lambda^{2}\right)}\,,
\end{split}
\end{equation}
where, $\hat{C}^{X}_{e\mu tq}/\Lambda^{2}$ denotes the value of the operator coefficient that maximizes the likelihood. Since we are evaluating the sensitivity to new physics against the SM background, we perform the analysis around the null hypothesis, i.e., $\hat{C}^{X}_{e\mu tq}/\Lambda^{2} = 0$.

The test statistic $\mathscr{Q}$ defined in Eq.~\eqref{eq:profile} is used to derive projected bounds on the operator coefficient $C^{X}_{e\mu tq}/\Lambda^{2}$ at a given confidence level (C.L.). According to Wilks' theorem~\cite{Wilks:1938dza}, in the asymptotic limit, the distribution of $\mathscr{Q}$ approaches a chi-squared ($\chi^2$) distribution with degrees of freedom equal to the number of parameters being tested. It is important to note that in the low-statistics regime, $\mathscr{Q}$ may deviate from the ideal $\chi^2$ behavior, in which case the true distribution should be estimated using Monte Carlo simulations. However, such simulations are computationally demanding. For projection studies focused on assessing the reach of future colliders, the use of Wilks' theorem provides a reasonable and commonly adopted approximation. Using the Wilks' theorem, we translate the value of $\mathscr{Q}$ into confidence intervals on $C^{X}_{e\mu tq}/\Lambda^{2}$. For instance, the critical values of the test statistic $\mathscr{Q}$ corresponding to the 68\% and 95\% C.L. are given by:
\begin{equation}
\begin{split}
    \mathscr{Q} \left(C^{X}_{e\mu tq}/\Lambda^{2}\right) \leq \chi^2_{1,\,68\%} &= 1.00\,, \\
    \mathscr{Q} \left(C^{X}_{e\mu tq}/\Lambda^{2}\right) \leq \chi^2_{1,\,95\%} &= 3.84\,, \\
    \mathscr{Q} \left(C^{X}_{e\mu tq}/\Lambda^{2}\right) \leq \chi^2_{2,\,68\%} &= 2.30\,, \\
    \mathscr{Q} \left(C^{X}_{e\mu tq}/\Lambda^{2}\right) \leq \chi^2_{2,\,95\%} &= 5.99\,,
\end{split}
\end{equation}
where $\chi^2_{f,\,p\%}$ denotes the $p^{\rm th}$ percentile of the chi-squared distribution with $f$ degrees of freedom. {In the present analysis, the $f=1$ thresholds are used when varying a single WC at a time, while the $f=2$ thresholds are adopted when two WCs are simultaneously allowed to vary in the fit.} These thresholds define the regions of parameter space that are consistent with the background-only hypothesis at the corresponding confidence levels.

\begin{figure}[htb!]
    \centering
    \includegraphics[width=\linewidth]{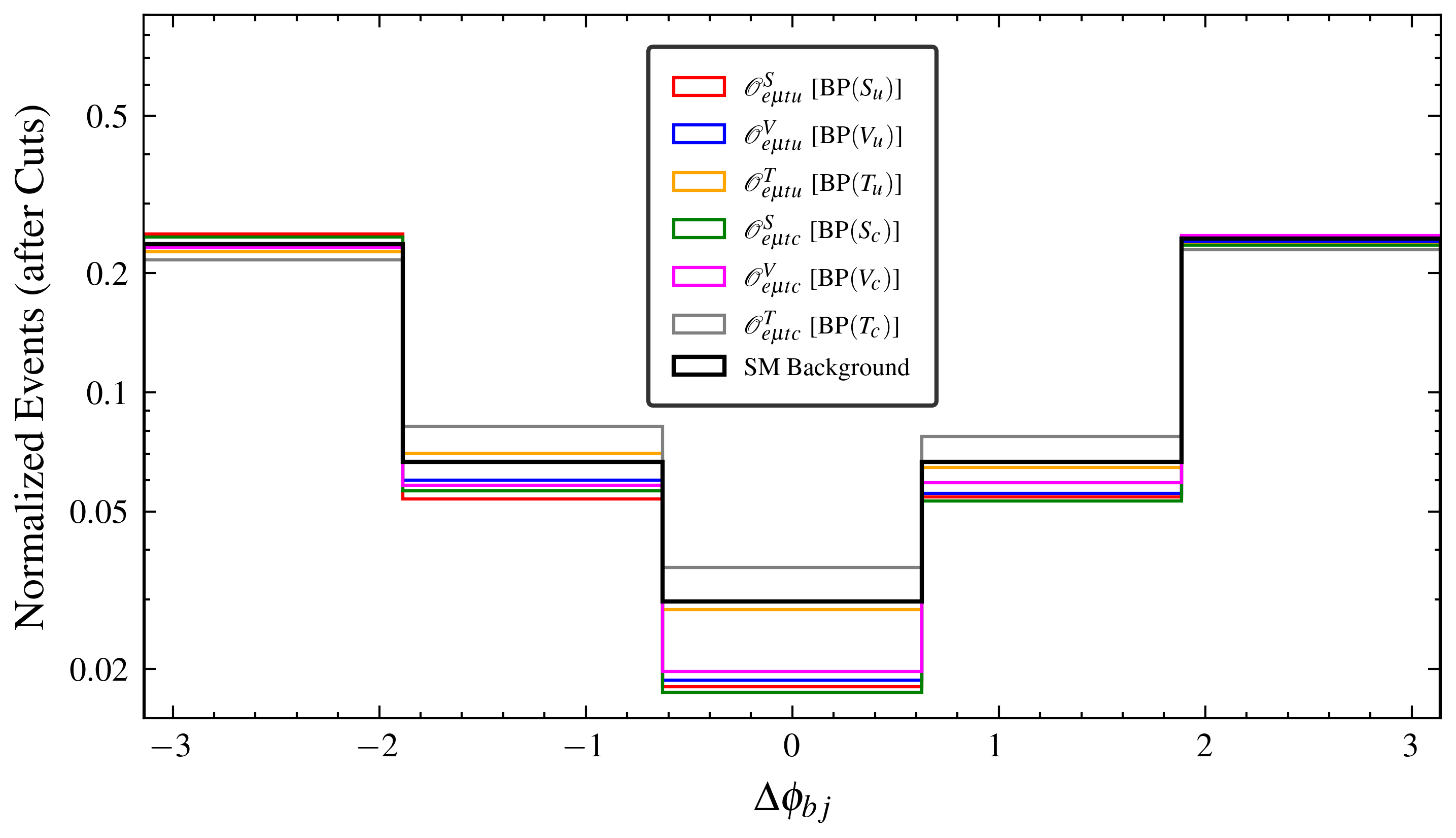}
    \caption{Normalized $\Delta \phi_{bj}$ distribution for signal benchmarks and the SM background, after applying the selection cuts in Eqs. ~\eqref{eq:esel} and~\eqref{eq:ecut}.}
    \label{fig:distx}
\end{figure}
For the sensitivity analysis, we employ the differential distribution $\Delta \phi_{bj}$, illustrated in Fig.~\ref{fig:distx}, defined as:
\begin{equation}
    \Delta \phi_{bj} = \phi_{b} - \phi_{j}\,,
\end{equation}
where $\phi_{b}$ and $\phi_{j}$ denote the azimuthal angles of the $b$-jet and the light jet, respectively. As evident from Fig.~\ref{fig:distx}, the azimuthal angle separation between jets exhibits notable sensitivity to the Lorentz structure of the effective operators, making it a powerful discriminator. We divide the $\Delta \phi_{bj}$ distribution into five equally spaced bins and perform a binned likelihood analysis to extract projected bounds on the operator coefficients, $C^{X}_{e\mu tq}/\Lambda^{2}$. By scanning over these parameters and evaluating the test statistic $\mathscr{Q}$ for each value, we determine the allowed parameter space consistent with the background-only hypothesis.
\subsection{Bounds on EFT coefficients}

\begin{figure*}[htb!]
    \centering
    \includegraphics[width=0.325\linewidth]{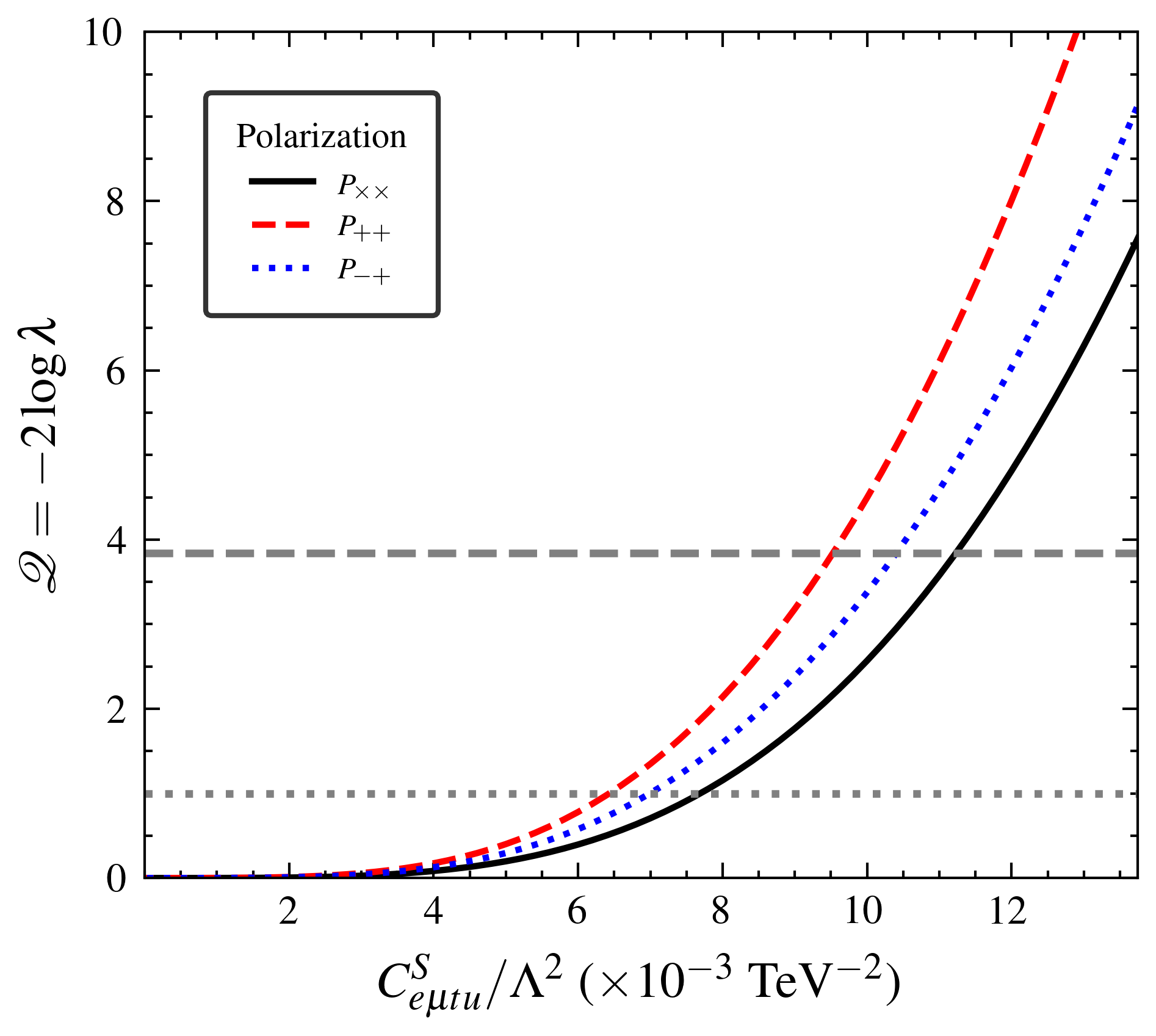}
    \includegraphics[width=0.325\linewidth]{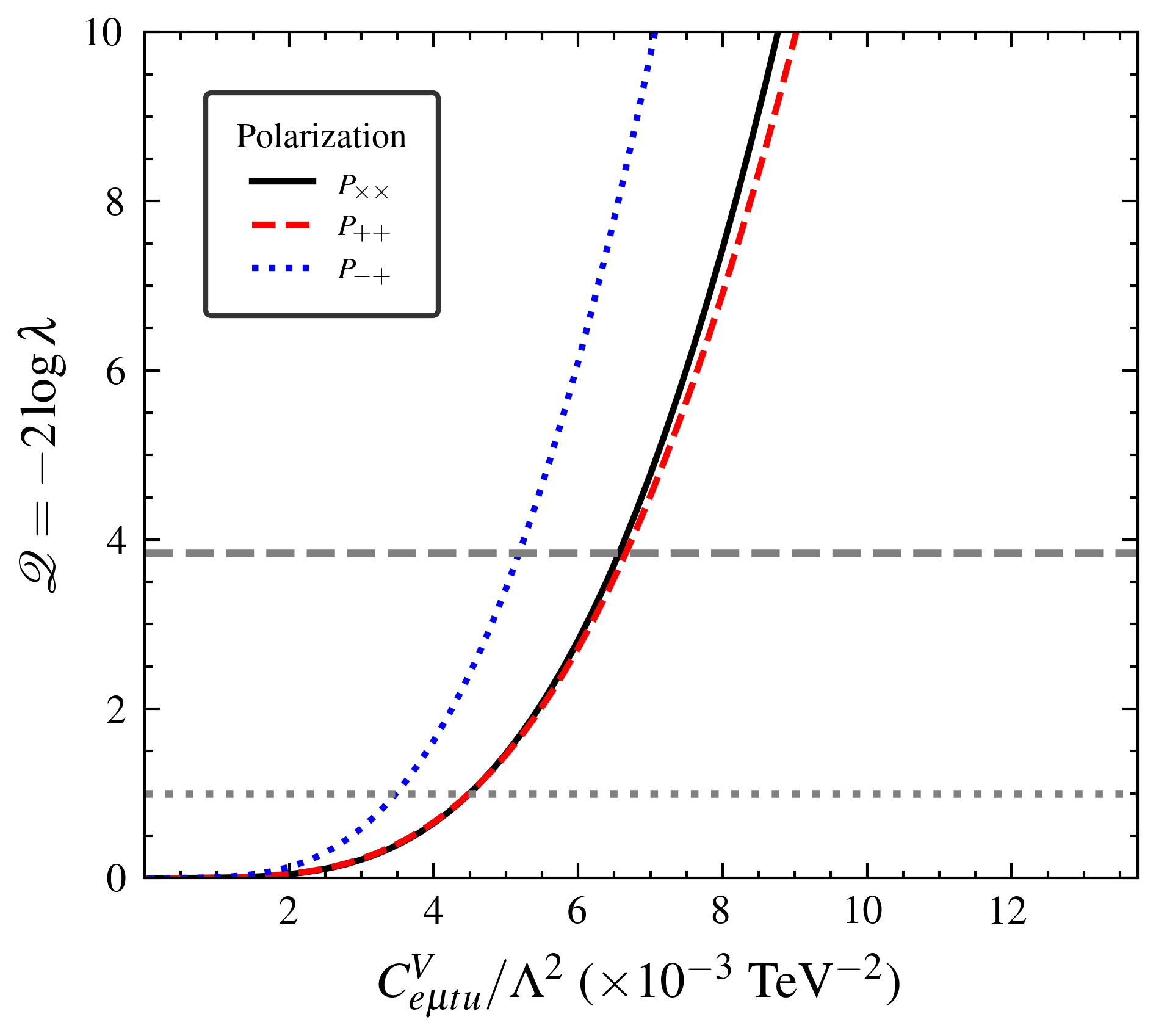}
    \includegraphics[width=0.325\linewidth]{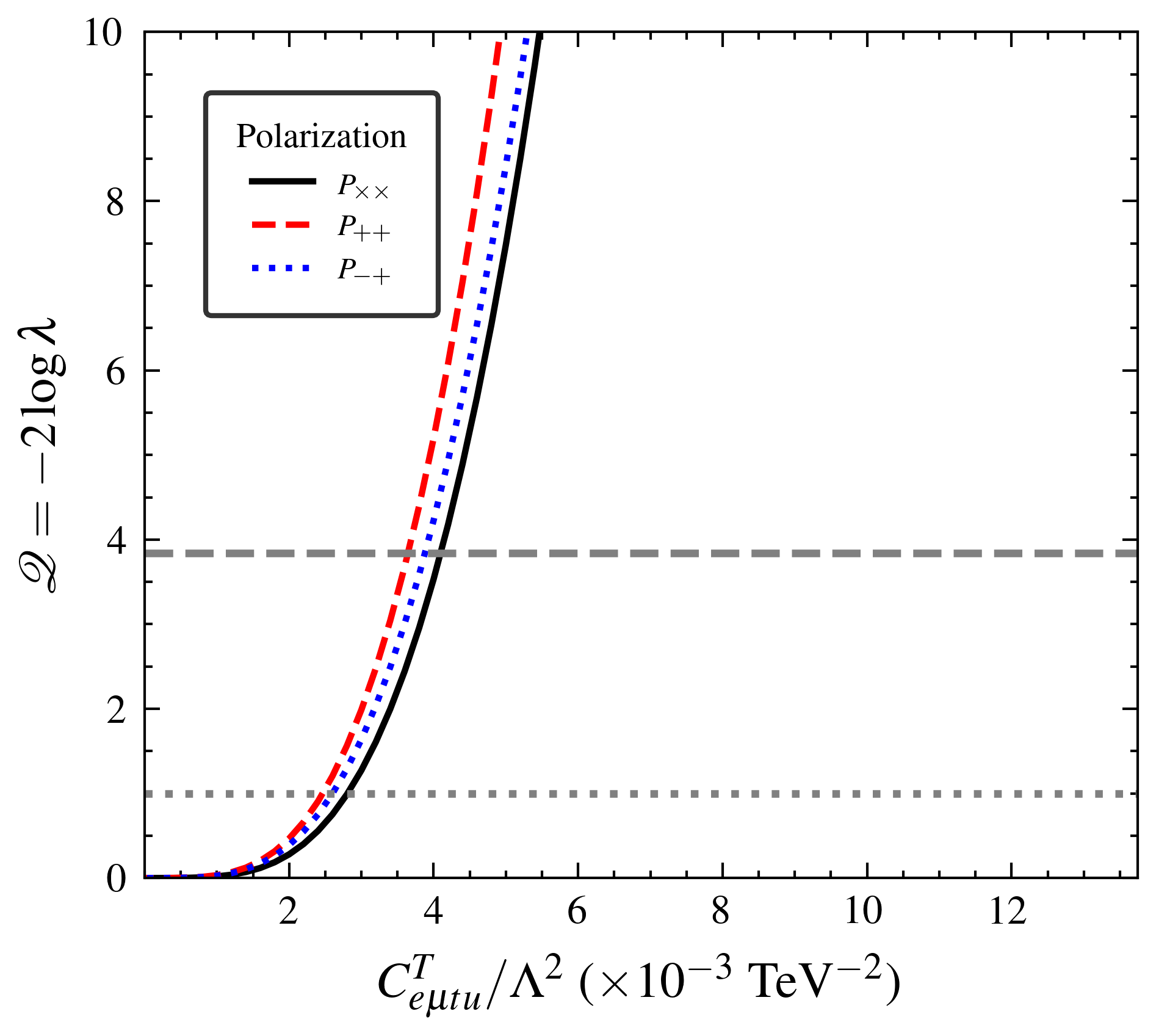} \\
    \includegraphics[width=0.325\linewidth]{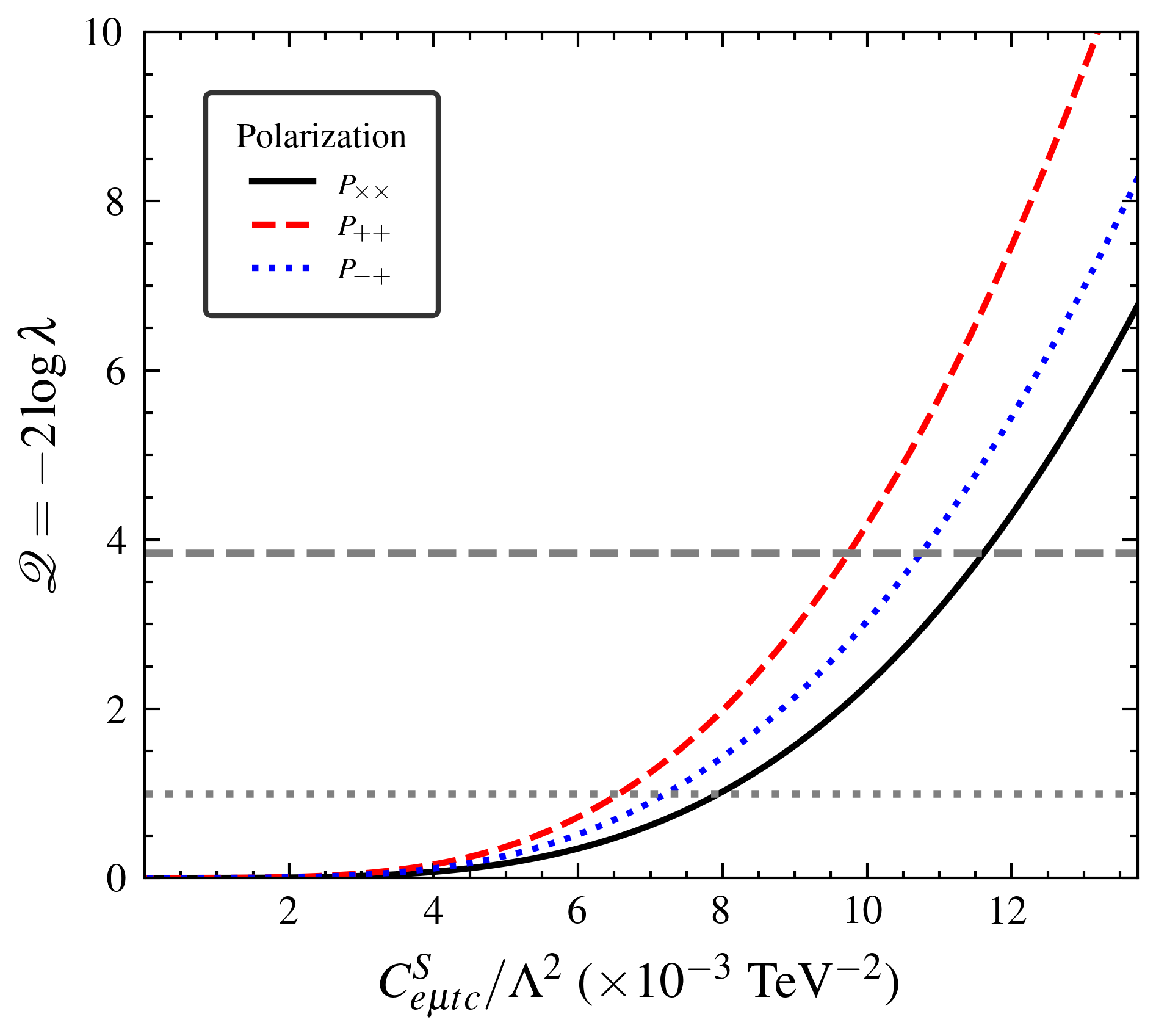}
    \includegraphics[width=0.325\linewidth]{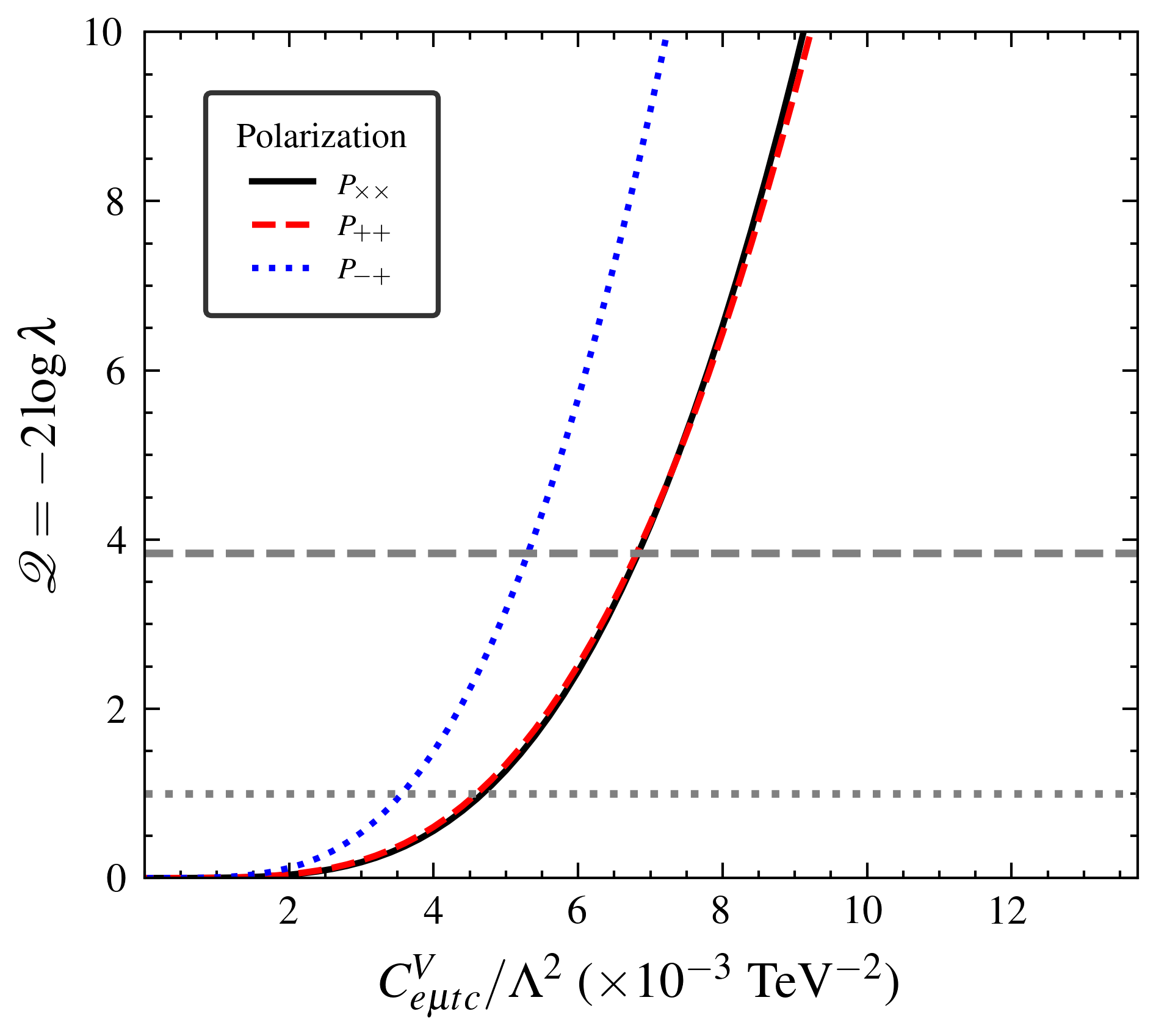}
    \includegraphics[width=0.325\linewidth]{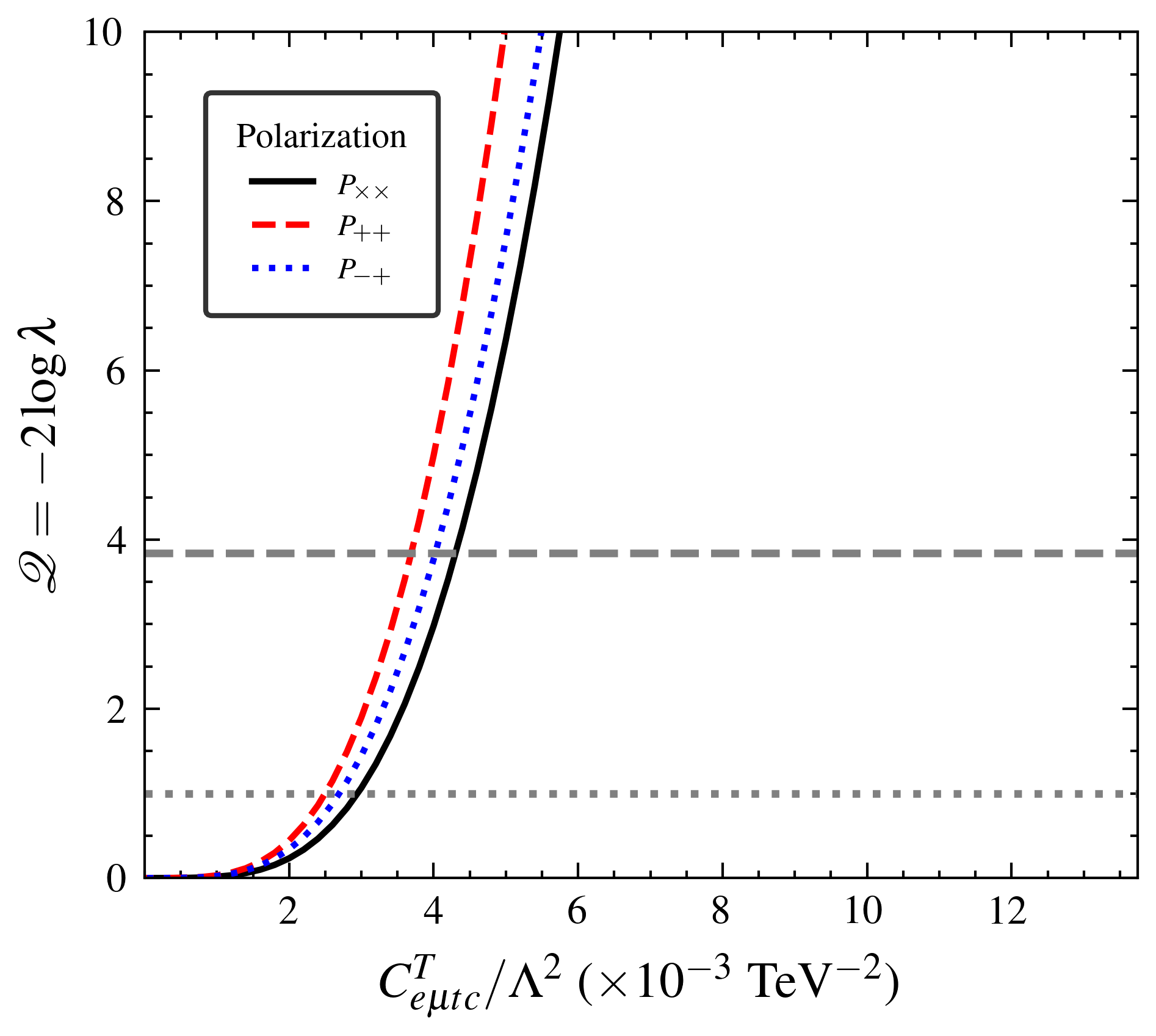}
    \caption{Single-parameter log-likelihood scans as functions of the WCs for scalar (\textit{left}), vector (\textit{middle}), and tensor (\textit{right}) operators under different polarization configurations at the $\mu$TRISTAN with $\mathfrak{L}_{\rm int} = 1\;\text{ab}^{-1}$. The {\textit{top} and \textit{bottom}} panels correspond to $u$- and $c$-type operators, respectively. Grey dashed (dotted) lines indicate the 95\% (68\%) C.L. limits.}
    \label{fig:1dlike}
\end{figure*}

We present the variation of the test statistic $\mathscr{Q}$ as a function of the WCs in Fig.~\ref{fig:1dlike}, corresponding to $tq$ production at the $\mu$TRISTAN with an integrated luminosity of $1\;\text{ab}^{-1}$. The projected 95\% and 68\% C.L. limits are indicated by grey dashed and dotted lines, respectively. The results are shown for three polarization configurations: unpolarized ($P_{\times \times}$), $P_{++}$, and $P_{+-}$.

For the scalar and tensor operators, both polarized setups: $P_{++}$ and $P_{+-}$ offer improved sensitivity over the unpolarized case, with $P_{++}$ yielding the strongest constraints. In contrast, for the vector operator, the $P_{+-}$ configuration enhances sensitivity compared to the unpolarized scenario, while $P_{++}$ does not improve and in some cases slightly worsens the reach. It is also noteworthy that, unlike at the LHC where up-quark PDFs significantly boost sensitivity to $u$-type operators over $c$-type ones, the lepton collider setup at the $\mu$TRISTAN ensures that both $u$- and $c$-type operator sensitivities are nearly comparable.

\begin{table*}[htb!]
    \centering
    \renewcommand{\arraystretch}{1.2}{
    \begin{tabular}{>{\centering\arraybackslash}p{2cm}
                     >{\centering\arraybackslash}p{2cm}
                     >{\centering\arraybackslash}p{2cm}
                     >{\centering\arraybackslash}p{2cm}
                     >{\centering\arraybackslash}p{2cm}
                     >{\centering\arraybackslash}p{2.5cm}}
    \hline \hline
    \multirow{2}*{Operator} & \multirow{2}*{Luminosity} & \multicolumn{3}{c}{Polarization} & \multirow{2}*{LHC} \\ \cline{3-5}
    && $P_{\times \times}$ & $P_{++}$ & $P_{-+}$ & \\
    \hline \hline
    \multirow{2}*{$\mathcal{O}^{S}_{e\mu tu}$} & 100 fb$^{-1}$ & 0.0235 & \textbf{0.0207} & 0.0231 & \multirow{2}*{0.24} \\
    & 1 ab$^{-1}$ & 0.0112 & \textbf{0.0095} & 0.0104 & \\ \hline
    \multirow{2}*{$\mathcal{O}^{V}_{e\mu tu}$} & 100 fb$^{-1}$ & 0.0137 & 0.0144 & \textbf{0.0115} & \multirow{2}*{0.12} \\
    & 1 ab$^{-1}$ & 0.0066 & 0.0067 & \textbf{0.0052} & \\ \hline
    \multirow{2}*{$\mathcal{O}^{T}_{e\mu tu}$} & 100 fb$^{-1}$ & 0.0085 & \textbf{0.0079} & 0.0086 & \multirow{2}*{0.06} \\
    & 1 ab$^{-1}$ & 0.0041 & \textbf{0.0036} & 0.0039 & \\
    \hline \hline
    \multirow{2}*{$\mathcal{O}^{S}_{e\mu tc}$} & 100 fb$^{-1}$ & 0.0243 & \textbf{0.0212} & 0.0238 & \multirow{2}*{0.86} \\
    & 1 ab$^{-1}$ & 0.0116 & \textbf{0.0097} & 0.0107 & \\ \hline
    \multirow{2}*{$\mathcal{O}^{V}_{e\mu tc}$} & 100 fb$^{-1}$ & 0.0143 & 0.0148 & \textbf{0.0118} & \multirow{2}*{0.37} \\
    & 1 ab$^{-1}$ & 0.0068 & 0.0068 & \textbf{0.0053} & \\ \hline
    \multirow{2}*{$\mathcal{O}^{T}_{e\mu tc}$} & 100 fb$^{-1}$ & 0.0090 & \textbf{0.0079} & 0.0089 & \multirow{2}*{0.21} \\
    & 1 ab$^{-1}$ & 0.0043 & \textbf{0.0036} & 0.0040 & \\
    \hline \hline
    \end{tabular}}
    \caption{95\% C.L. limits on the WCs, $C^{X}_{e\mu tq}/\Lambda^{2}$ (in TeV$^{-2}$) from $tq$ production at the $\mu$TRISTAN for integrated luminosities of a $100\;\text{fb}^{-1}$ and $1\;\text{ab}^{-1}$, for different polarization setups. The last columns corresponds to the LHC bounds, detailed in Tab.~\ref{tab:bounds}. Best-case sensitivities in each row are highlighted in bold for clarity.}
    \label{tab:sensitivity1}
\end{table*}

In Tab.~\ref{tab:sensitivity1}, we present the projected 95\% confidence level bounds on the WCs, $C^{X}_{e\mu tq}/\Lambda^{2}$, derived from $tq$ production at the $\mu$TRISTAN for integrated luminosities of $100\;\text{fb}^{-1}$ and $1\;\text{ab}^{-1}$, across different polarization configurations. Even with a relatively modest luminosity of $100\;\text{fb}^{-1}$, achievable within the early years of the $\mu$TRISTAN operation, the sensitivity surpasses current LHC limits by roughly an order of magnitude. With an integrated luminosity of $1\;\text{ab}^{-1}$, the bounds improve by more than a factor of two across all operator types and polarization configurations, highlighting the enhanced sensitivity and precision achievable at the $\mu$TRISTAN facility.

\subsection{Bounds on branching ratios}
The projected bounds on the WCs, $C^{X}_{e\mu tq}/\Lambda^{2}$, presented in Tab.~\ref{tab:sensitivity1}, can be translated into upper limits on the branching ratios of rare top quark decays using Eq.~\eqref{eq:brtop}. These projections are summarized in Tab.~\ref{tab:sensitivity2} for the process $\mu^{+} e^{-} \rightarrow tq$ at the $\mu$TRISTAN, assuming integrated luminosities of $100\;\text{fb}^{-1}$ and $1\;\text{ab}^{-1}$ under different polarization configurations. The resulting bounds improve significantly over current LHC constraints, with branching ratio sensitivities reaching 2-3 orders of magnitude better, depending on the operator class and flavor structure. Notably, even at the early-stage luminosity of $100\;\text{fb}^{-1}$, the sensitivity outperforms existing limits, while the $1\;\text{ab}^{-1}$ dataset offers substantially stronger exclusion potential. These improvements highlight the capability of the $\mu$TRISTAN collider to explore extremely rare flavor-violating top decays with high precision, offering a promising avenue to uncover signals of new physics in the charged lepton and top quark sectors.

\begin{table*}[htb!]
    \centering
    \renewcommand{\arraystretch}{1.2}{
    \begin{tabular}{>{\centering\arraybackslash}p{2cm}
                    >{\centering\arraybackslash}p{2cm}
                    >{\centering\arraybackslash}p{2cm}
                    >{\centering\arraybackslash}p{2cm}
                    >{\centering\arraybackslash}p{2cm}
                    >{\centering\arraybackslash}p{2.5cm}}
    \hline \hline
    \multirow{2}*{Operator} & \multirow{2}*{Luminosity} & \multicolumn{3}{c}{Polarization} & \multirow{2}*{LHC} \\ \cline{3-5}
    && $P_{\times \times}$ & $P_{++}$ & $P_{-+}$ & \\
    \hline \hline
    \multirow{2}*{$\mathcal{O}^{S}_{e\mu tu}$} & 100 fb$^{-1}$ & $6.3 \times 10^{-10}$ & {\boldmath $4.9 \times 10^{-10}$} & $6.1 \times 10^{-10}$ & \multirow{2}*{$7.0 \times 10^{-8}$} \\
    & 1 ab$^{-1}$ & $1.4 \times 10^{-10}$ & {\boldmath $1.0 \times 10^{-10}$} & $1.2 \times 10^{-10}$ &  \\ \hline
    \multirow{2}*{$\mathcal{O}^{V}_{e\mu tu}$} & 100 fb$^{-1}$ & $1.7 \times 10^{-9}$ & $1.9 \times 10^{-9}$ & {\boldmath $1.2 \times 10^{-9}$} & \multirow{2}*{$1.3 \times 10^{-7}$} \\
    & 1 ab$^{-1}$ & $4.0 \times 10^{-10}$ & $4.1 \times 10^{-10}$ & {\boldmath $2.5 \times 10^{-10}$} &  \\ \hline
    \multirow{2}*{$\mathcal{O}^{T}_{e\mu tu}$} & 100 fb$^{-1}$ & $4.0 \times 10^{-9}$ & {\boldmath $3.4 \times 10^{-9}$} & $4.1 \times 10^{-9}$ & \multirow{2}*{$2.5 \times 10^{-7}$} \\
    & 1 ab$^{-1}$ & $9.2 \times 10^{-10}$ & {\boldmath $7.1 \times 10^{-10}$} & $8.4 \times 10^{-10}$ &  \\
    \hline \hline
    \multirow{2}*{$\mathcal{O}^{S}_{e\mu tc}$} & 100 fb$^{-1}$ &
    $6.8 \times 10^{-10}$ & {\boldmath $5.1 \times 10^{-10}$} & $6.5 \times 10^{-10}$ & \multirow{2}*{$8.9 \times 10^{-7}$} \\
    & 1 ab$^{-1}$ & $1.5 \times 10^{-10}$ & {\boldmath $1.1 \times 10^{-10}$} & $1.3 \times 10^{-10}$ &  \\ \hline
    \multirow{2}*{$\mathcal{O}^{V}_{e\mu tc}$} & 100 fb$^{-1}$ & $1.9 \times 10^{-9}$ & $2.0 \times 10^{-9}$ & {\boldmath $1.3 \times 10^{-9}$} & \multirow{2}*{$1.3 \times 10^{-6}$} \\
    & 1 ab$^{-1}$ & $4.2 \times 10^{-10}$ & $4.2 \times 10^{-10}$ & {\boldmath $2.6 \times 10^{-10}$} &  \\ \hline
    \multirow{2}*{$\mathcal{O}^{T}_{e\mu tc}$} & 100 fb$^{-1}$ & $4.4 \times 10^{-9}$ & {\boldmath $3.4 \times 10^{-9}$} & $4.4 \times 10^{-9}$ & \multirow{2}*{$2.6 \times 10^{-6}$} \\
    & 1 ab$^{-1}$ & $1.0 \times 10^{-9}$ & {\boldmath $7.1 \times 10^{-10}$} & $8.8 \times 10^{-10}$ &  \\
    \hline \hline
    \end{tabular}}
    \caption{Projected limits on the branching ratios of rare top quark decay, $\mathcal{B}(t \to e \mu q)$, from $tq$ production at the $\mu$TRISTAN for integrated luminosities of $100\;\text{fb}^{-1}$ and $1\;\text{ab}^{-1}$, for different polarization setups. The last column shows the current LHC bounds as listed in Tab.~\ref{tab:bounds}. Best-case sensitivities in each row are highlighted in bold for clarity.}
    \label{tab:sensitivity2}
\end{table*}

\subsection{Correlated sensitivities}
\begin{figure*}[htb!]
    \centering
    \includegraphics[width=0.325\linewidth]{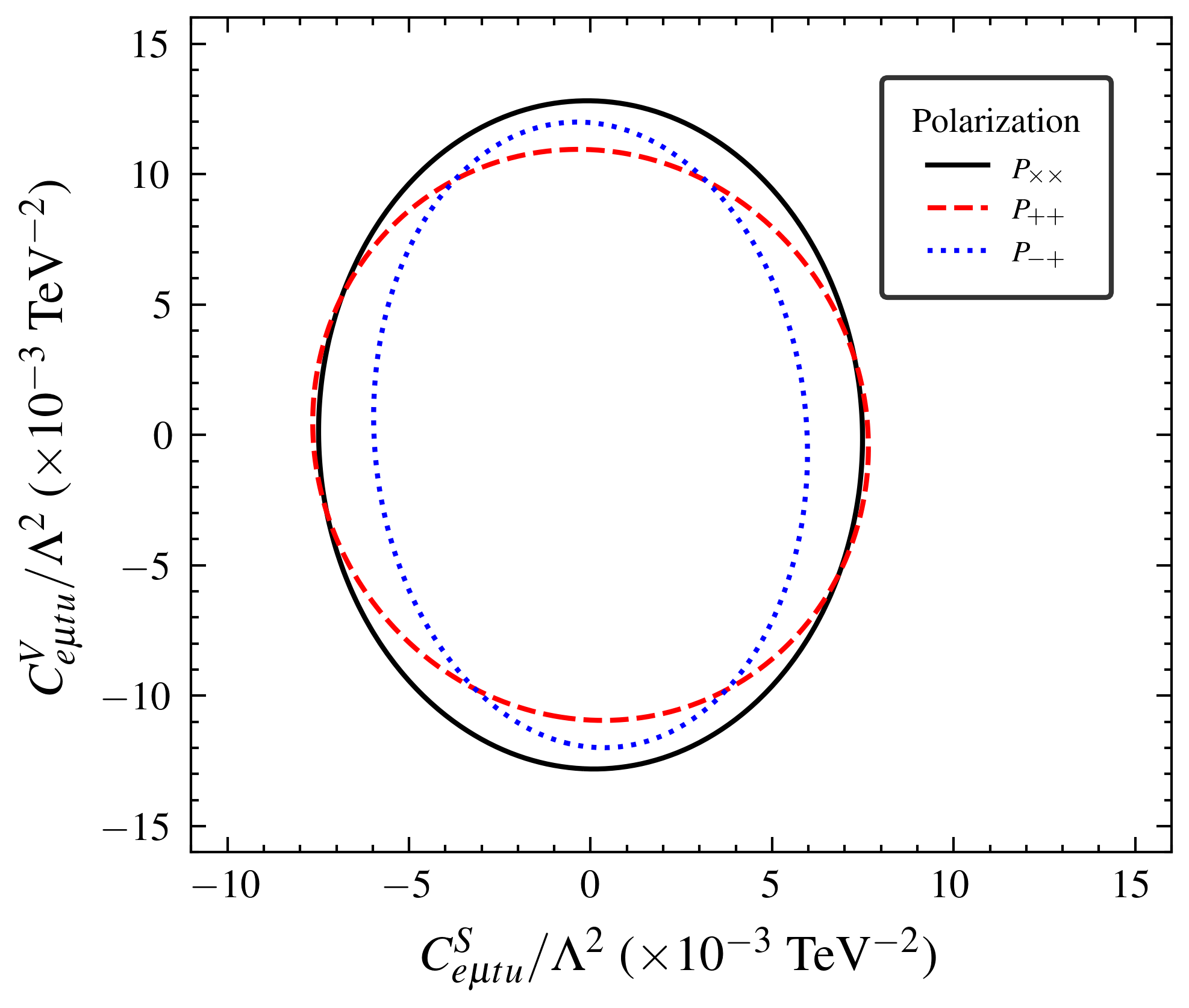}
    \includegraphics[width=0.325\linewidth]{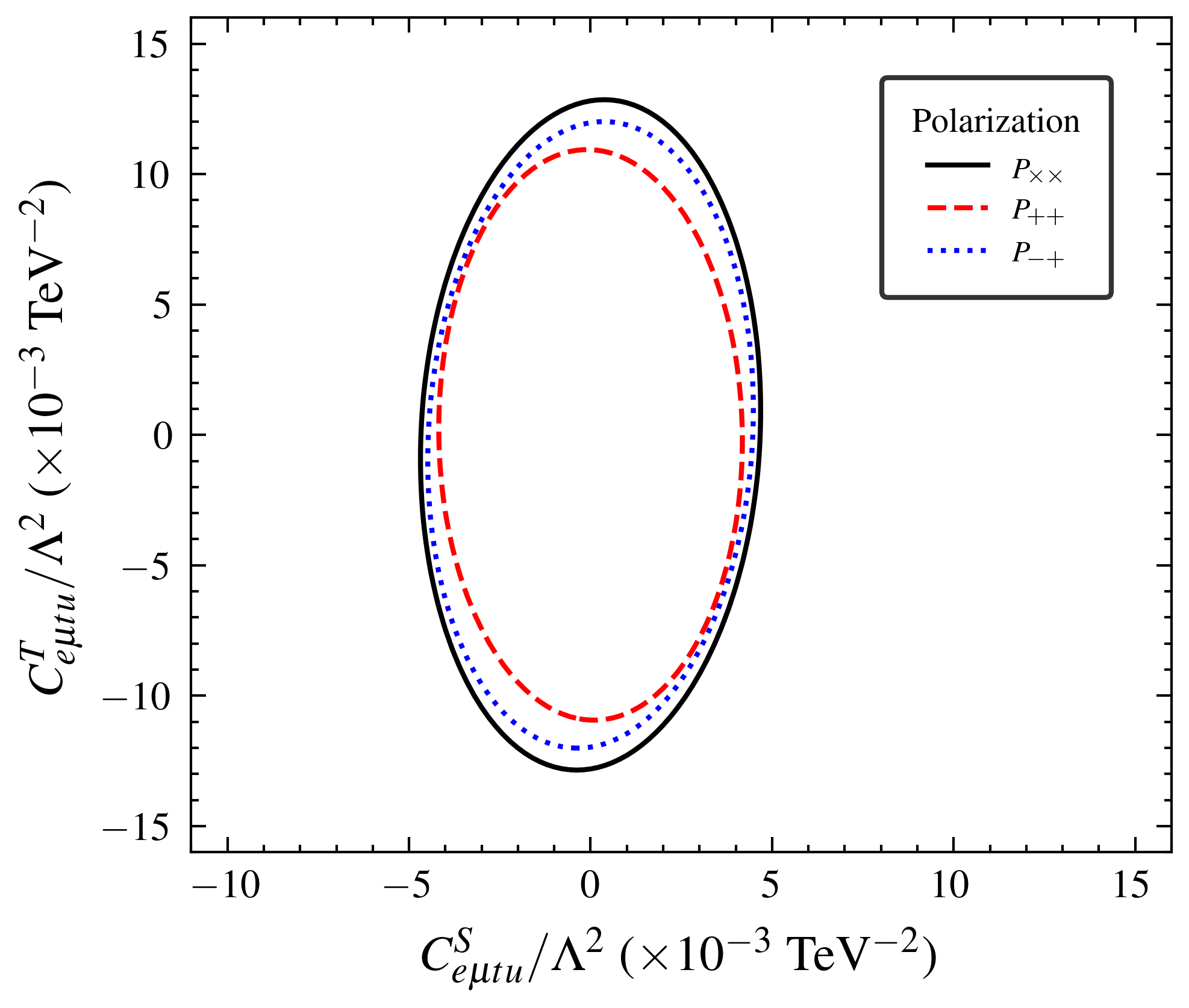}
    \includegraphics[width=0.325\linewidth]{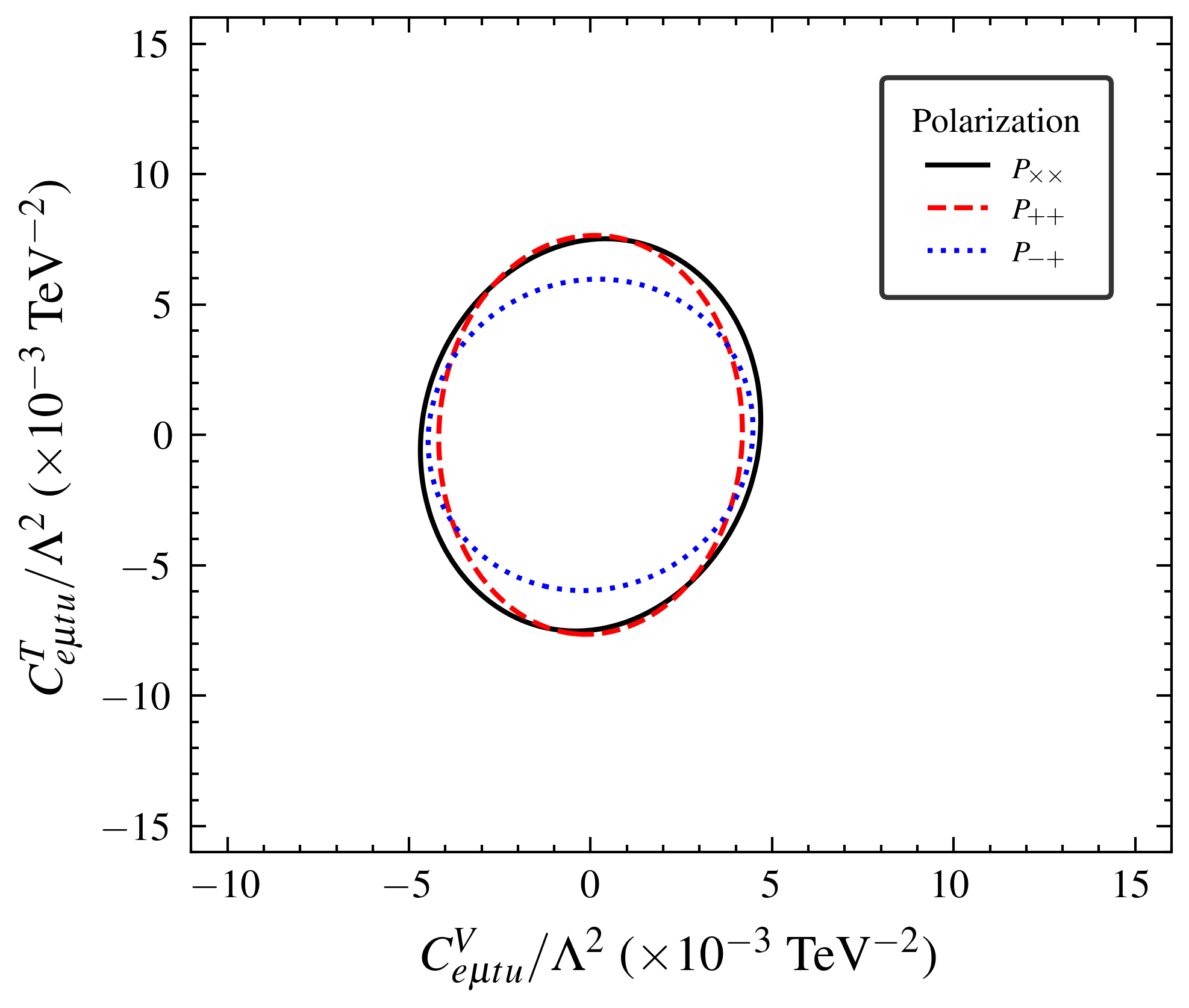} \\
    \includegraphics[width=0.325\linewidth]{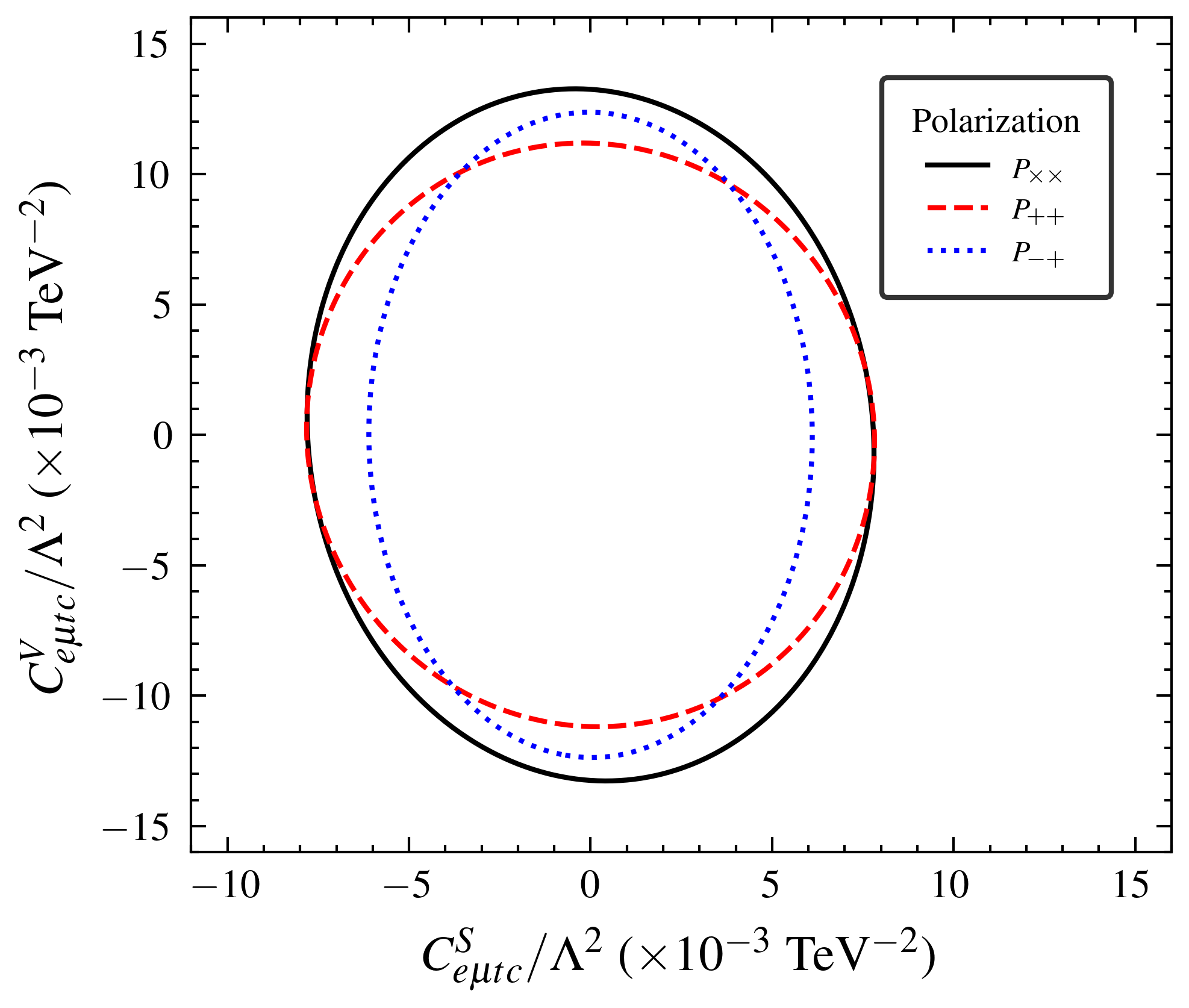}
    \includegraphics[width=0.325\linewidth]{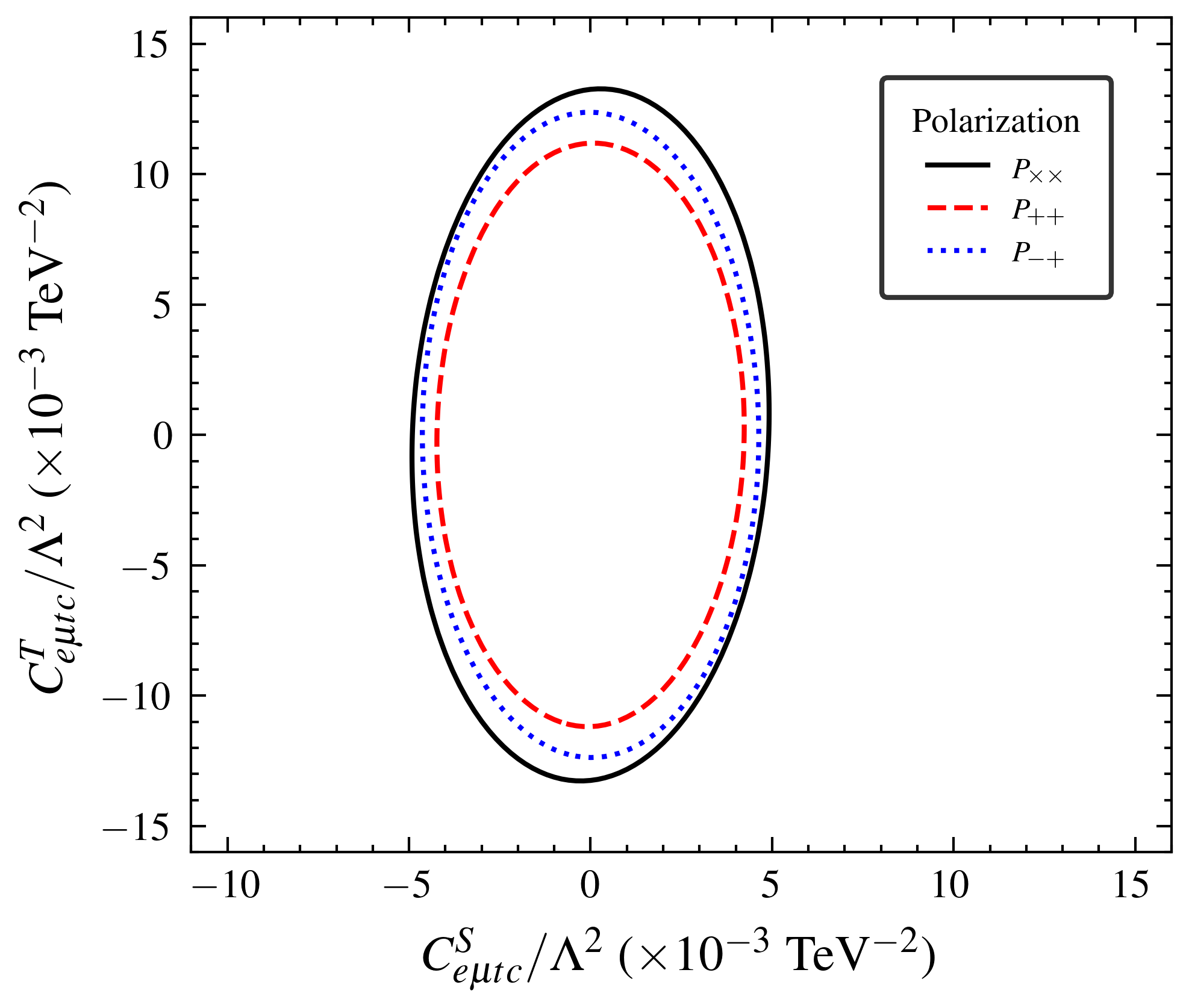}
    \includegraphics[width=0.325\linewidth]{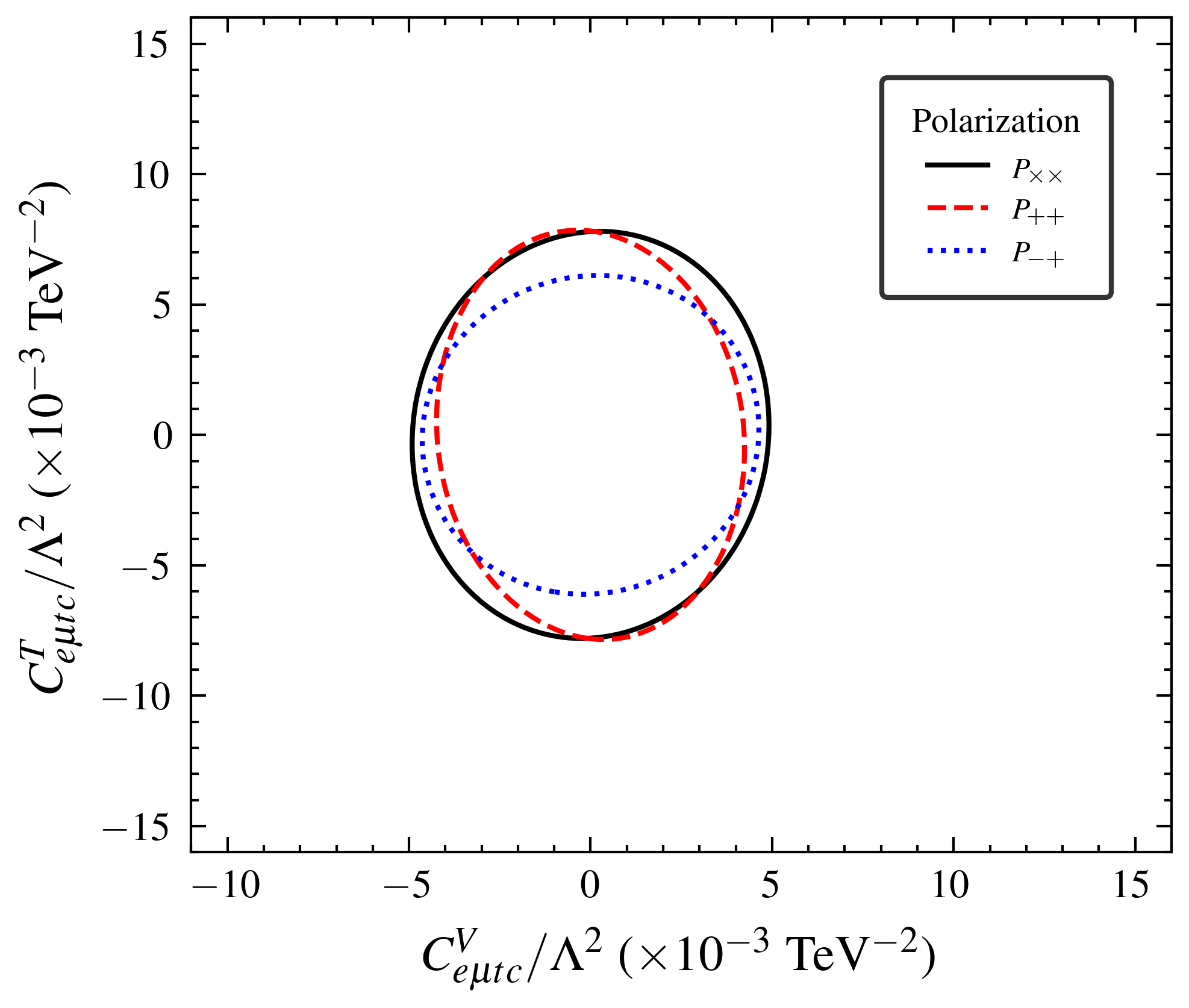}
    \caption{Correlated 95\% C.L. exclusion contours for EFT operator parameter spaces of $u$-type (\textit{top}) and $c$-type (\textit{bottom}) classes from $tq$ production at the $\mu$TRISTAN, assuming an integrated luminosity of $1\;\text{ab}^{-1}$, for different beam polarization configurations.}
    \label{fig:2dcorr}
\end{figure*}

In Fig.~\ref{fig:2dcorr}, we present the projected 95\% C.L. correlated exclusion limits on the parameter spaces of different EFT operator classes, shown separately for $u$-type (left panels) and $c$-type (right panels) interactions. These bounds are derived from $tq$ production at the $\mu$TRISTAN, assuming an integrated luminosity of $1\;\text{ab}^{-1}$ under various beam polarization configurations. The observed variations in the shape, orientation, and extent of the exclusion contours across different polarization setups reflect the chiral selectivity of the operators and the beam polarization. By appropriately tuning the beam polarization, one can enhance sensitivity in specific directions of parameter space, thereby isolating dominant contributions from particular operator classes. This demonstrates the utility of polarization in boosting signal significance and providing a more refined probe of the underlying new physics dynamics.

\section{Summary and conclusion}
\label{sec:conclusion}
In this work, we have studied charged lepton flavor violating flavor-changing neutral current (FCNC) interactions involving the top quark at the proposed high-energy $\mu^{+} e^{-}$ collider stage of the $\mu$TRISTAN facility, operating at a center-of-mass energy of $\sqrt{s} = 346$ GeV. Focusing on the process $\mu^{+} e^{-} \to tq$ with $q = u, c$, we analyzed simplified scalar, vector, and tensor four-fermion operators derived from the Standard Model Effective Field Theory (SMEFT) framework. These operators offer a model-independent parameterization of potential new physics responsible for charged lepton flavor violation (LFV) and FCNC effects.

Our collider analysis targeted the leptonic decay channel of the top quark and utilized a cut-based strategy to suppress the dominant SM backgrounds. We demonstrated that kinematic observables such as invariant mass and azimuthal angular separation between jets play a crucial role in enhancing signal sensitivity. The effect of initial-state beam polarization was studied in detail, revealing that it can significantly boost the sensitivity to specific operator structures and aid in disentangling their chiral nature.

We employed a binned likelihood analysis using the distribution of $\Delta\phi_{bj}$ to derive projected 95\% C.L. bounds on the Wilson coefficients, $C^{X}_{e\mu tq}/\Lambda^2$. Our results show that, even in the unpolarized configuration and with a modest integrated luminosity of $100\;\text{fb}^{-1}$, attainable within the early operational phase, the $\mu$TRISTAN can outperform existing LHC bounds by nearly an order of magnitude. With $1\;\text{ab}^{-1}$ of data, the projected sensitivity improves by a further factor of 2–3. When translated into constraints on the rare decay branching ratios $\mathcal{B}(t \to e \mu q)$, this corresponds to improvements of up to three orders of magnitude over existing limits.

Additionally, we presented correlated sensitivity projections in the Wilson coefficient parameter space, highlighting how different beam polarization configurations can be used to selectively enhance sensitivity in certain directions. This illustrates the unique potential of a polarized $\mu^{+} e^{-}$ collider in disentangling the contributions of different EFT operator classes.

In summary, the $\mu$TRISTAN collider offers a clean and powerful environment to probe flavor-violating interactions with high precision. Our analysis establishes that such a facility can play a leading role in exploring rare top FCNC processes involving charged LFV, significantly extending the reach beyond current and near-future hadron coider capabilities. Furthermore, although we have framed our study in the language of effective operators, these interactions can arise in a broad range of BSM scenarios, including models with extended gauge sectors (e.g., $Z'$), scalar or vector leptoquarks, and compositeness frameworks. The bounds obtained in our analysis can thus be interpreted as constraints on these underlying theories, highlighting the utility of SMEFT-based collider studies in mapping out the viable parameter space of a wide class of new physics models.

\acknowledgments
AS thanks Subhaditya Bhattacharya for his useful suggestions during this work and Shailesh Pincha for insightful discussions on asymmetric colliders.

\appendix
{
\section{BSM scenarios generating $e\mu tq$ operators}
\label{app:uv}
In this section, we provide two specific BSM scenarios that generate sizeable $e\mu tq$ interactions and thereby source the EFT operators studied in this paper. We assume the presence of new flavour structures arising from BSM spurions that can naturally generate large off-diagonal entries in the lepton and quark sectors while suppressing diagonal flavour structures. A simple theoretical motivation for such flavour structures arises from the presence of additional flavour spurions associated with the BSM sector. In this work we assume that the heavy states couple to SM fermions through new flavour matrices that are not aligned with the SM Yukawa structures. These spurions can naturally contain sizeable off-diagonal entries while keeping flavour-diagonal components suppressed. As a result, the effective flavour matrices $X_\ell$ and $X_q$ may preferentially generate $(e\mu)$ and $(tq)$ transitions, providing a natural origin for the $e\mu tq$ operators considered in this work.

\subsection*{Scalar leptoquark with flavourful couplings}
We consider a heavy scalar leptoquark $S \sim (3,1,-1/3)$ under the SM gauge symmetry. Its most general renormalizable interactions with SM fermions, consistent with gauge invariance, are
\begin{equation}
\mathcal{L}_S = -\, \overline{Q}_i\,(\mathcal{C}_q)_{ij}\,S\,L_j 
               -\, \overline{u}_i\,(\mathcal{C}_u)_{ij}\,S\, e_j 
               + \mathrm{h.c.}\,,
\label{eq:L_S_general_app}
\end{equation}
where $\mathcal{C}_q$ and $\mathcal{C}_u$ are flavour matrices. We assume that these matrices originate from BSM flavour spurions that are not aligned with the SM Yukawa structures and can contain sizeable off-diagonal entries. The flavour structure of the couplings can be parametrized as
\begin{align}
\mathcal{C}_q &= \lambda\, X_q , \\
\mathcal{C}_u &= \tilde{\lambda}\, X_u^\dagger ,
\end{align}
where $X_q$ and $X_u$ denote BSM flavour spurions. These spurions are not required to follow the SM Yukawa hierarchy and may contain comparatively large off-diagonal entries, particularly in the $tq$ sector. Integrating out $S$ at tree level generates the effective operator
\begin{equation}
\mathcal{L}_{\rm EFT} \supset 
-\,\frac{1}{M_S^2}\,
(\overline{e}\,\mathcal{Y}_\ell\,\mu)
(\overline{t}\,\mathcal{Y}_q\,q)
+ \mathrm{h.c.}\,,
\label{eq:S_app}
\end{equation}
where $\mathcal{Y}_q$ inherits its flavour structure from the underlying spurions appearing in $\mathcal{C}_{q,u}$. The dominant flavour entry corresponds to the component generating the $tq$ transition,
\begin{equation}
(\mathcal{Y}_q)_{tq} \sim (X_q)_{tq},
\label{eq:Yq_scaling_app}
\end{equation}
while entries involving purely light generations remain suppressed according to the assumed flavour hierarchy. The tree-level matching coefficient therefore takes the schematic form
\begin{equation}
\mathbb{C}_{e\mu tq} 
\sim 
\frac{1}{M_S^2}\;
(\mathcal{Y}_\ell)_{e\mu}\,
(\mathcal{Y}_q)_{tq}\,,
\label{eq:C_app}
\end{equation}
highlighting the origin of the $e-\mu$ and $t-q$ flavour violation from the underlying spurion structure.

\subsection*{Flavorful $Z'$ model with BSM flavour currents}
As an alternative, we consider a heavy neutral vector $Z'$ coupled to flavourful fermion currents,
\begin{equation}
\mathcal{L}_{Z'} =
Z'_\mu \left(
g_q\, \overline{Q}_{i}\gamma^\mu (\mathcal{G}_q)_{ij} Q_j
+
g_\ell\, \overline{L}_{i}\gamma^\mu (\mathcal{G}_\ell)_{ij} L_j
\right).
\label{eq:L_Zp_app}
\end{equation}
We assume that the flavour structure of these currents is controlled by BSM spurions,
\begin{align}
\mathcal{G}_q &= X_q X_q^\dagger ,\\
\mathcal{G}_\ell &= X_\ell X_\ell^\dagger ,
\end{align}
where $X_q$ and $X_\ell$ denote flavour matrices that can contain sizeable off-diagonal entries such as $(\mathcal{G}_q)_{tq}$ and $(\mathcal{G}_\ell)_{e\mu}$. Integrating out $Z'$ yields at tree level
\begin{equation}
\mathcal{L}_{\rm EFT} \supset
-\frac{g_q g_\ell}{M_{Z'}^2}\,
(\overline{e}\gamma_\mu (\mathcal{G}_\ell)_{e\mu}\mu)\,
(\overline{t}\gamma^\mu (\mathcal{G}_q)_{tq} q)
+ \mathrm{h.c.}\,,
\label{eq:match_Zp_app}
\end{equation}
where the effective coefficient is controlled by the off-diagonal spurion entries generating the $e-\mu$ and $t-q$ transitions, while flavour-diagonal operators remain suppressed by the assumed structure of the BSM spurions.
}
{
\section{Constraints on other flavor combinations}
\label{app:1}
In this section, we summarize existing constraints on semi-leptonic four-fermion operators contributing to flavor-violating scenarios not directly considered in this work, namely: (i) charged lepton flavor violation (LFV) with quark flavor conservation, and (ii) quark FCNC with lepton flavor conservation. Since our analysis focuses exclusively on top quark FCNC effects in combination with charged LFV, we do not include bounds arising from quark flavor-changing charged-current (FCCC) observables. For completeness, we also report bounds on down-type semi-leptonic operators, both vector and scalar, to facilitate comparison with their up-type counterparts. These include:
\begin{equation}
\begin{split}
[\mathcal{O}_{ed}]_{ijkl}\hspace{0.4cm} &= (\overline{e}_{i} \gamma_{\alpha} e_{j}) \, (\overline{u}_{k} \gamma^{\alpha} u_{l}) \,,\\
[\mathcal{O}_{Le}]_{ijkl}\hspace{0.35cm} &= (\overline{L}_{i} \gamma_{\alpha} L_{j}) \, (\overline{d}_{k} \gamma^{\alpha} d_{l}) \,,\\
[\mathcal{O}_{LeQd}]_{ijkl} &= (\overline{L}^{a}_{i} e_{j}) \, \epsilon^{ab} \, (\overline{Q}^{b}_{k} d_{l}) \,,
\end{split}    
\end{equation}
Tables~\ref{tab:ext_1a} and \ref{tab:ext_1b} summarize existing constraints on quark FCNC operators without LFV for scalar/tensor and vector structures, respectively. Similarly, Tables~\ref{tab:ext_2a} and \ref{tab:ext_2b} present constraints on LFV operators without quark FCNC. As evident, these existing bounds are significantly stronger than those on lepton flavor violating top FCNC operators, primarily due to stringent limits from $B$-physics observables and the non-observation of charged LFV processes. 

\begin{table*}[htb!]
    \centering
    \renewcommand{\arraystretch}{1.2}{
    \begin{tabular}{>{\centering\arraybackslash}p{2cm}
                    >{\centering\arraybackslash}p{3cm}
                    >{\centering\arraybackslash}p{3cm}
                    >{\centering\arraybackslash}p{5cm}}
    \hline \hline
    \multicolumn{4}{c}{\textbf{FCNC ($\checkmark$) LFV ($\times$)}} \\
    \hline
    Operator & Flavor structure & Observable & Constraint (TeV$^{-2}$) \\ \hline \hline
    \multirow{3}*{$\mathcal{O}^{(1)}_{LeQu}$} & $\mu\mu tc, eetc$ & $B \to D^{(*)}\ell \nu$ & $(0.09 \pm 0.58) \times 10^{-7}$~\cite{Bhattacharya:2023beo} \\ 
    & $\tau\tau tu$ & $B^{\pm} \to \tau \nu_{\tau}$ & $-(6.64 \pm 0.21) \times 10^{-9}$~\cite{Bhattacharya:2023beo} \\ 
    & $\mu\mu cu$ & $D \to \mu \nu_{\mu}$ & $-(8.61 \pm 0.18) \times 10^{-10}$~\cite{Bhattacharya:2023beo} \\ 
    \hline
    \multirow{2}*{$\mathcal{O}_{LeQd}$} & $\mu\mu db$ & $b \to d \mu\mu$ & $[-3.48,0.67] \times 10^{-5}$~\cite{Greljo:2022jac} \\
    & $\mu\mu bs$ & $b \to s \mu\mu$ & $[-6.80,16.5] \times 10^{-6}$~\cite{Greljo:2022jac} \\
    \hline 
    \multirow{2}*{$\mathcal{O}^{(3)}_{LeQu}$} & $\mu\mu tc, eetc$ & $B \to D^{(*)}\ell \nu$ & $-(0.33 \pm 6.61) \times 10^{-10}$~\cite{Bhattacharya:2023beo} \\ 
    & $\mu\mu tu, eetu$ & $B \to \pi \ell \nu$ & $(0.07 \pm 0.16) \times 10^{-7}$~\cite{Bhattacharya:2023beo} \\
     \hline \hline
    \end{tabular}}
    \caption{Existing constraints on scalar and tensor operators along with the respective flavor combination and relevant observable. Here, $\ell = \mu, e$.}
    \label{tab:ext_1a}
\end{table*}

\begin{table*}[htb!]
    \centering
    \renewcommand{\arraystretch}{1.2}{
    \begin{tabular}{>{\centering\arraybackslash}p{2cm}
                    >{\centering\arraybackslash}p{3cm}
                    >{\centering\arraybackslash}p{3cm}
                    >{\centering\arraybackslash}p{5cm}}
    \hline \hline
    \multicolumn{4}{c}{\textbf{FCNC ($\checkmark$) LFV ($\times$)}} \\
    \hline
    Operator & Flavor structure & Observable & Constraint (TeV$^{-2}$) \\ \hline \hline
    \multirow{3}*{$\mathcal{O}^{(1)}_{LQ}$} & $\mu\mu bs, \mu\mu tc$ & $b \to s\mu\mu$ & $-(3.98 \pm 1.20) \times 10^{-4}$~\cite{Bhattacharya:2023beo} \\
     & $\mu\mu bd, \mu\mu tu$ & $b \to d\mu\mu$ & $-\left(1.90^{+0.65}_{-0.68}\right)\times 10^{-4}$~\cite{Bhattacharya:2023beo} \\
     & $\mu\mu sd, \mu\mu cu$ & $\mathcal{B}\left(K_{L} \to \mu\mu\right)$ & $-\left(1.21^{+0.33}_{-0.31}\right)\times 10^{-4}$~\cite{Bhattacharya:2023beo} \\ \hline
    \multirow{3}*{$\mathcal{O}^{(3)}_{LQ}$} & $\mu\mu bs, \mu\mu tc$ & $b \to s\mu\mu$ & $-(3.98 \pm 1.20) \times 10^{-4}$~\cite{Bhattacharya:2023beo} \\
     & $\mu\mu bd, \mu\mu tu$ & $b \to d\mu\mu$ & $-\left(1.90^{+0.65}_{-0.68}\right)\times 10^{-4}$~\cite{Bhattacharya:2023beo} \\
     & $\mu\mu sd, \mu\mu cu$ & $\mathcal{B}\left(K_{L} \to \mu\mu\right)$ & $-\left(1.21^{+0.33}_{-0.31}\right)\times 10^{-4}$~\cite{Bhattacharya:2023beo} \\ \hline
     \multirow{2}*{$\mathcal{O}_{ed}$} & $\mu\mu bd$ & $b \to d \mu \mu$ & $[-7.03, 3.76] \times 10^{-4}$~\cite{Greljo:2022jac} \\ 
     & $\mu\mu bs$ & $b \to s\mu\mu$ & $[-4.05, 4.37] \times 10^{-4}$~\cite{Greljo:2022jac} \\
     \hline
     \multirow{2}*{$\mathcal{O}_{Ld}$} & $\mu\mu bd$ & $b \to d \mu \mu$ & $[-2.79, 2.43] \times 10^{-4}$~\cite{Greljo:2022jac} \\ 
     & $eebs$ & $b \to see$ & $[-4.04, 1.09] \times 10^{-4}$~\cite{Greljo:2022jac} \\
     \hline
    \multirow{3}*{$\mathcal{O}_{Qe}$} & $\mu\mu bs, \mu\mu tc$ & $\mathcal{B} \left(B^{0}_{s} \to \mu\mu\right)$ & $-\left(7.39^{+0.42}_{-0.44} \right) \times 10^{-4}$~\cite{Bhattacharya:2023beo} \\
     & $\mu\mu bd, \mu\mu tu$ & $b \to d\mu\mu$ & $\left(0.68^{+2.72}_{-2.79}\right)\times 10^{-4}$~\cite{Bhattacharya:2023beo} \\
     & $\mu\mu sd, \mu\mu cu$ & $\mathcal{B}\left(K_{L} \to \mu\mu\right)$ & $-\left(1.21^{+0.33}_{-0.31}\right)\times 10^{-4}$~\cite{Bhattacharya:2023beo} \\
     \hline \hline
    \end{tabular}}
    \caption{Existing constraints on vector operators along with the respective flavor combination and relevant observable. Here, $\ell = \mu, e$.}
    \label{tab:ext_1b}
\end{table*}

\begin{table*}[htb!]
    \centering
    \renewcommand{\arraystretch}{1.2}{
    \begin{tabular}{>{\centering\arraybackslash}p{2cm}
                    >{\centering\arraybackslash}p{3cm}
                    >{\centering\arraybackslash}p{3cm}
                    >{\centering\arraybackslash}p{5cm}}
    \hline \hline
    \multicolumn{4}{c}{\textbf{FCNC ($\times$) LFV ($\checkmark$)}} \\
    \hline
    Operator & Flavor structure & Observable & Constraint (TeV$^{-2}$) \\ \hline \hline
    \multirow{6}*{$\mathcal{O}^{(1)}_{LeQu}$} & $\mu ecc$ & $\mu \to e\gamma$ & $2.6 \times 10^{-4}$~\cite{Pruna:2014asa} \\ 
    & $\mu ett$ & $\mu \to e\gamma$ & $1.9 \times 10^{-6}$~\cite{Pruna:2014asa} \\
    & $\tau ecc$ & $\tau \to e\gamma$ & $2.4 \times 10^{0}$~\cite{Pruna:2014asa} \\
    & $\tau ett$ & $\tau \to e\gamma$ & $1.8 \times 10^{-2}$~\cite{Pruna:2014asa} \\
    & $\tau\mu cc$ & $\tau \to \mu\gamma$ & $2.8 \times 10^{0}$~\cite{Pruna:2014asa} \\
    & $\tau\mu tt$ & $\tau \to \mu\gamma$ & $2.1 \times 10^{-2}$~\cite{Pruna:2014asa} \\
    \hline
    \multirow{6}*{$\mathcal{O}^{(3)}_{LeQu}$} & $\mu ecc$ & $\mu \to e\gamma$ & $4.8 \times 10^{-7}$~\cite{Pruna:2014asa} \\
    & $\mu ett$ & $\mu \to e\gamma$ & $3.6 \times 10^{-9}$~\cite{Pruna:2014asa} \\
    & $\tau ecc$ & $\tau \to e\gamma$ & $4.6 \times 10^{-3}$~\cite{Pruna:2014asa} \\
    & $\tau ett$ & $\tau \to e\gamma$ & $3.4 \times 10^{-5}$~\cite{Pruna:2014asa} \\
    & $\tau\mu cc$ & $\tau \to \mu\gamma$ & $5.4 \times 10^{-3}$~\cite{Pruna:2014asa} \\
    & $\tau\mu tt$ & $\tau \to \mu\gamma$ & $4.0 \times 10^{-5}$~\cite{Pruna:2014asa} \\
     \hline \hline
    \end{tabular}}
    \caption{Existing constraints on scalar and tensor operators along with the respective flavor combination and relevant observable. Here, $\ell = \mu, e$.}
    \label{tab:ext_2a}
\end{table*}

\begin{table*}[htb!]
    \centering
    \renewcommand{\arraystretch}{1.2}{
    \begin{tabular}{>{\centering\arraybackslash}p{2cm}
                    >{\centering\arraybackslash}p{3cm}
                    >{\centering\arraybackslash}p{3cm}
                    >{\centering\arraybackslash}p{5cm}}
    \hline \hline
    \multicolumn{4}{c}{\textbf{FCNC ($\times$) LFV ($\checkmark$)}} \\
    \hline
    Operator & Flavor structure & Observable & Constraint (TeV$^{-2}$) \\ \hline \hline
    \multirow{2}*{$\mathcal{O}^{(1)}_{LQ}$} & $\mu euu, \mu edd, \mu ess, \mu ecc$ & SF & $(1.82 \pm 0.69) \times 10^{-3}$~\cite{Kumar:2021yod} \\
    & $\mu ebb, \mu ett$ & SF & $(2.69 \pm 1.01) \times 10^{-4}$~\cite{Kumar:2021yod} \\ \hline
    \multirow{2}*{$\mathcal{O}^{(3)}_{LQ}$} & $\mu euu, \mu edd, \mu ess, \mu ecc$ & SF & $(6.39 \pm 2.39) \times 10^{-4}$~\cite{Kumar:2021yod} \\ 
    & $\mu ebb, \mu ett$ & SF & $(2.30 \pm 0.86) \times 10^{-4}$~\cite{Kumar:2021yod} \\ \hline
     \multirow{2}*{$\mathcal{O}_{eu}$} & $\mu euu, \mu ecc$ & SF & $(9.68 \pm 3.68) \times 10^{-4}$~\cite{Kumar:2021yod} \\
     & $\mu ett$ & SF & $(2.69 \pm 1.01) \times 10^{-4}$~\cite{Kumar:2021yod} \\ \hline
     \multirow{2}*{$\mathcal{O}_{ed}$} & $\mu edd, \mu ess$ & SF & $(2.00 \pm 0.75) \times 10^{-3}$~\cite{Kumar:2021yod} \\
     & $\mu ebb$ & SF & $(2.25 \pm 0.96) \times 10^{-3}$~\cite{Kumar:2021yod} \\ \hline
     \multirow{2}*{$\mathcal{O}_{Lu}$} & $\mu euu, \mu ecc$ & SF & $(1.00 \pm 0.38) \times 10^{-3}$~\cite{Kumar:2021yod} \\
     & $\mu ett$ & SF & $(2.81 \pm 1.06) \times 10^{-4}$~\cite{Kumar:2021yod} \\ \hline
     \multirow{2}*{$\mathcal{O}_{Ld}$} & $\mu edd, \mu ess$ & SF & $(1.96 \pm 0.74) \times 10^{-3}$~\cite{Kumar:2021yod} \\
     & $\mu ebb$ & SF & $(2.49 \pm 0.93) \times 10^{-3}$~\cite{Kumar:2021yod} \\ \hline
    \multirow{2}*{$\mathcal{O}_{Qe}$} & $\mu euu, \mu edd, \mu ess, \mu ecc$ & SF & $(2.02 \pm 0.76) \times 10^{-3}$~\cite{Kumar:2021yod} \\
     & $\mu ebb, \mu ett$ & SF & $(2.81 \pm 1.04) \times 10^{-4}$~\cite{Kumar:2021yod} \\
     \hline \hline
    \end{tabular}}
    \caption{Existing constraints on vector operators along with the respective flavor combination. The constraints correspond to simultaneous fit (SF) on observables: $Z \to \ell\ell$, $\tau \to 3\ell$, $\mu \to 3e$, and $\tau \to (\rho, \phi, \pi)\, \ell$~\cite{Kumar:2021yod}. Here, $\ell = \mu, e$.}
    \label{tab:ext_2b}
\end{table*}

\section{Disentangling the vector chiral structures}
\label{app:2}
In this section, we explore the role of beam polarization in disentangling the different chiral structures of the vector operators in the SMEFT framework. While the simplified operator structures listed in Tab.~\ref{tab:smeft-op} already exhibit kinematic distinctions from the scalar and tensor operators, the use of polarized beams provides an additional handle to separate the individual components of the vector operators. We select the following vector operator benchmarks relevant for $\mu^{+}e^{-} \to tu$ production:
\begin{equation}
\begin{split}
    &\text{BP}_{LQ}^{(u)}: \hspace{0.2cm} \left(\frac{[C^{(-)}_{LQ}]_{e \mu t u}}{\Lambda^{2}}\right) = 0.01\;{\rm TeV}^{-2}\,, \\
    &\text{BP}_{Lu}^{(u)}: \hspace{0.2cm} \left(\frac{[C_{Lu}]_{e \mu t u}}{\Lambda^{2}}\right) = 0.01\;{\rm TeV}^{-2}\,, \\
    &\text{BP}_{eu}^{(u)}: \hspace{0.2cm} \left(\frac{[C_{eu}]_{e \mu t u}}{\Lambda^{2}}\right) = 0.01\;{\rm TeV}^{-2}\,, \\
    &\text{BP}_{Qe}^{(u)}: \hspace{0.2cm} \left(\frac{[C_{Qe}]_{e \mu t c}}{\Lambda^{2}}\right) = 0.01\;{\rm TeV}^{-2}\,. \\
\end{split}
\end{equation}
For each benchmark, all other Wilson coefficients are set to zero. In this section, we focus exclusively on the $u$-type benchmarks, since an analogous analysis for $c$-type benchmarks would yield qualitatively similar results. The corresponding signal cross sections for $\mu^{+} e^{-} \rightarrow t(\ell \nu b) u$ production are presented in Tab.~\ref{tab:pol-eff1} for both the unpolarized case and various beam polarization configurations.
\begin{table*}[htb!]
    \centering
    \renewcommand{\arraystretch}{1.2}{
    \begin{tabular}{>{\centering\arraybackslash}p{3.5cm}
                     >{\centering\arraybackslash}p{2.25cm}
                     >{\centering\arraybackslash}p{2.25cm}
                     >{\centering\arraybackslash}p{2.25cm}
                     >{\centering\arraybackslash}p{2.25cm}}
    \hline \hline
    \rule{0pt}{2.0em}
     \multirow{2}*{\shortstack[c]{Polarization \\ $(P_{\mu^{+}}, P_{e^{-}})$}} & \multicolumn{4}{c}{Signal} \\ \cline{2-5} \rule{0pt}{1.5em}
      & $\sigma^{{\rm BP}_{LQ}^{(u)}}_{t(\ell \nu b)q}$ (fb) & $\sigma^{{\rm BP}_{eu}^{(u)}}_{t(\ell \nu b)q}$ (fb) & $\sigma^{{\rm BP}_{Lu}^{(u)}}_{t(\ell \nu b)q}$ (fb) & $\sigma^{{\rm BP}_{Qe}^{(u)}}_{t(\ell \nu b)q}$ (fb) \\ [0.5em] \hline \hline
     $P_{\times \times}:$ Unpolarized & 0.026 & 0.026 & 0.026 & 0.026 \\ \hline \hline
     $P_{++}:(+0.25,+0.70)$ & 0.010 & 0.034 & 0.010 & 0.034 \\
     $P_{+-}:(+0.25,-0.70)$ & 0.056 & 0.006 & 0.056 & 0.006 \\
     $P_{-+}:(-0.25,+0.70)$ & 0.006 & 0.056 & 0.006 & 0.056 \\
     $P_{--}:(-0.25,-0.70)$ & 0.034 & 0.010 & 0.034 & 0.010 \\ \hline \hline
    \end{tabular}}
    \caption{Production cross sections for the signal process $\mu^{+} e^{-} \rightarrow t(\ell \nu b)u$, mediated by different vector operators are presented for various beam polarization configurations.}
    \label{tab:pol-eff1}
\end{table*}

We observe that all operator structures contribute similarly in the unpolarized setup. The polarization configuration $P_{+-}$ enhances the $\mathcal{O}^{(-)}_{LQ}$ and $\mathcal{O}_{Lu}$ contributions while suppressing $\mathcal{O}_{eu}$ and $\mathcal{O}_{Qe}$, whereas the $P_{-+}$ configuration exhibits the opposite behavior. We perform the collider simulation and likelihood-based sensitivity analysis following the same procedure outlined in Sec.~\ref{sec:collider} and \ref{sec:sensitivity}. The resulting 95\% C.L. limits on the Wilson coefficients of the vector operators for $tu$ production at the $\mu$TRISTAN, for integrated luminosities of $100\;\text{fb}^{-1}$ and $1\;\text{ab}^{-1}$ under different beam polarization configurations, are presented in Tab.~\ref{tab:sensitivityv}. Observations from Tab.~\ref{tab:sensitivity1} indicate that the $c$-type operator structures are expected to exhibit similar sensitivity levels. This section highlights that beam polarization remains a crucial tool for distinguishing the chiral components of the vector operators.

\begin{table*}[htb!]
    \centering
    \renewcommand{\arraystretch}{1.2}{
    \begin{tabular}{>{\centering\arraybackslash}p{3cm}
                     >{\centering\arraybackslash}p{2.5cm}
                     >{\centering\arraybackslash}p{2.5cm}
                     >{\centering\arraybackslash}p{2.5cm}
                     >{\centering\arraybackslash}p{2.5cm}}
    \hline \hline
    \multirow{2}*{Operator} & \multirow{2}*{Luminosity} & \multicolumn{3}{c}{Polarization} \\ \cline{3-5}
    && $P_{\times \times}$ & $P_{+-}$ & $P_{-+}$ \\
    \hline \hline
    \multirow{2}*{$[\mathcal{O}^{(-)}_{LQ}]_{e\mu tu}$} & 100 fb$^{-1}$ & 0.0274 & \textbf{0.0184} & 0.0522 \\
    & 1 ab$^{-1}$ & 0.0131 & \textbf{0.0077} & 0.0235 \\ \hline
    \multirow{2}*{$[\mathcal{O}_{eu}]_{e\mu tu}$} & 100 fb$^{-1}$ & 0.0277 & 0.0566 & \textbf{0.0171} \\
    & 1 ab$^{-1}$ & 0.0132 & 0.0235 & \textbf{0.0077} \\ \hline
    \multirow{2}*{$[\mathcal{O}_{Lu}]_{e\mu tu}$} & 100 fb$^{-1}$ & 0.0272 & \textbf{0.0188} & 0.0511 \\
    & 1 ab$^{-1}$ & 0.0130 & \textbf{0.0079} & 0.0231 \\ \hline
    \multirow{2}*{$[\mathcal{O}_{Qe}]_{e\mu tu}$} & 100 fb$^{-1}$ & 0.0276 & 0.0569 & \textbf{0.0168} \\
    & 1 ab$^{-1}$ & 0.0132 & 0.0239 & \textbf{0.0076} \\
    \hline \hline
    \end{tabular}}
    \caption{95\% C.L. limits on the WCs of vector operators (up-type), $C/\Lambda^{2}$ (in TeV$^{-2}$) from $tu$ production at the $\mu$TRISTAN for integrated luminosities of a $100\;\text{fb}^{-1}$ and $1\;\text{ab}^{-1}$, for different polarization setups. Best-case sensitivities in each row are highlighted in bold for clarity.}
    \label{tab:sensitivityv}
\end{table*}
}

\clearpage
\bibliography{biblio}

\begin{thebibliography}{124}%
\makeatletter
\providecommand \@ifxundefined [1]{%
 \@ifx{#1\undefined}
}%
\providecommand \@ifnum [1]{%
 \ifnum #1\expandafter \@firstoftwo
 \else \expandafter \@secondoftwo
 \fi
}%
\providecommand \@ifx [1]{%
 \ifx #1\expandafter \@firstoftwo
 \else \expandafter \@secondoftwo
 \fi
}%
\providecommand \natexlab [1]{#1}%
\providecommand \enquote  [1]{``#1''}%
\providecommand \bibnamefont  [1]{#1}%
\providecommand \bibfnamefont [1]{#1}%
\providecommand \citenamefont [1]{#1}%
\providecommand \href@noop [0]{\@secondoftwo}%
\providecommand \href [0]{\begingroup \@sanitize@url \@href}%
\providecommand \@href[1]{\@@startlink{#1}\@@href}%
\providecommand \@@href[1]{\endgroup#1\@@endlink}%
\providecommand \@sanitize@url [0]{\catcode `\\12\catcode `\$12\catcode
  `\&12\catcode `\#12\catcode `\^12\catcode `\_12\catcode `\%12\relax}%
\providecommand \@@startlink[1]{}%
\providecommand \@@endlink[0]{}%
\providecommand \url  [0]{\begingroup\@sanitize@url \@url }%
\providecommand \@url [1]{\endgroup\@href {#1}{\urlprefix }}%
\providecommand \urlprefix  [0]{URL }%
\providecommand \Eprint [0]{\href }%
\providecommand \doibase [0]{https://doi.org/}%
\providecommand \selectlanguage [0]{\@gobble}%
\providecommand \bibinfo  [0]{\@secondoftwo}%
\providecommand \bibfield  [0]{\@secondoftwo}%
\providecommand \translation [1]{[#1]}%
\providecommand \BibitemOpen [0]{}%
\providecommand \bibitemStop [0]{}%
\providecommand \bibitemNoStop [0]{.\EOS\space}%
\providecommand \EOS [0]{\spacefactor3000\relax}%
\providecommand \BibitemShut  [1]{\csname bibitem#1\endcsname}%
\let\auto@bib@innerbib\@empty
\bibitem [{\citenamefont {Glashow}\ \emph {et~al.}(1970)\citenamefont
  {Glashow}, \citenamefont {Iliopoulos},\ and\ \citenamefont
  {Maiani}}]{Glashow:1970gm}%
  \BibitemOpen
  \bibfield  {author} {\bibinfo {author} {\bibfnamefont {S.~L.}\ \bibnamefont
  {Glashow}}, \bibinfo {author} {\bibfnamefont {J.}~\bibnamefont
  {Iliopoulos}},\ and\ \bibinfo {author} {\bibfnamefont {L.}~\bibnamefont
  {Maiani}},\ }\bibfield  {title} {\bibinfo {title} {{Weak Interactions with
  Lepton-Hadron Symmetry}},\ }\href {https://doi.org/10.1103/PhysRevD.2.1285}
  {\bibfield  {journal} {\bibinfo  {journal} {Phys. Rev. D}\ }\textbf {\bibinfo
  {volume} {2}},\ \bibinfo {pages} {1285} (\bibinfo {year} {1970})}\BibitemShut
  {NoStop}%
\bibitem [{\citenamefont {Saito}\ \emph {et~al.}(2015)\citenamefont {Saito}
  \emph {et~al.}}]{Belle:2014nmp}%
  \BibitemOpen
  \bibfield  {author} {\bibinfo {author} {\bibfnamefont {T.}~\bibnamefont
  {Saito}} \emph {et~al.} (\bibinfo {collaboration} {Belle}),\ }\bibfield
  {title} {\bibinfo {title} {{Measurement of the $\bar{B} \rightarrow X_s
  \gamma$ Branching Fraction with a Sum of Exclusive Decays}},\ }\href
  {https://doi.org/10.1103/PhysRevD.91.052004} {\bibfield  {journal} {\bibinfo
  {journal} {Phys. Rev. D}\ }\textbf {\bibinfo {volume} {91}},\ \bibinfo
  {pages} {052004} (\bibinfo {year} {2015})},\ \Eprint
  {https://arxiv.org/abs/1411.7198} {arXiv:1411.7198 [hep-ex]} \BibitemShut
  {NoStop}%
\bibitem [{\citenamefont {Lees}\ \emph
  {et~al.}(2012{\natexlab{a}})\citenamefont {Lees} \emph
  {et~al.}}]{BaBar:2012eja}%
  \BibitemOpen
  \bibfield  {author} {\bibinfo {author} {\bibfnamefont {J.~P.}\ \bibnamefont
  {Lees}} \emph {et~al.} (\bibinfo {collaboration} {BaBar}),\ }\bibfield
  {title} {\bibinfo {title} {{Exclusive Measurements of $b \to s\gamma$
  Transition Rate and Photon Energy Spectrum}},\ }\href
  {https://doi.org/10.1103/PhysRevD.86.052012} {\bibfield  {journal} {\bibinfo
  {journal} {Phys. Rev. D}\ }\textbf {\bibinfo {volume} {86}},\ \bibinfo
  {pages} {052012} (\bibinfo {year} {2012}{\natexlab{a}})},\ \Eprint
  {https://arxiv.org/abs/1207.2520} {arXiv:1207.2520 [hep-ex]} \BibitemShut
  {NoStop}%
\bibitem [{\citenamefont {Lees}\ \emph
  {et~al.}(2012{\natexlab{b}})\citenamefont {Lees} \emph
  {et~al.}}]{BaBar:2012fqh}%
  \BibitemOpen
  \bibfield  {author} {\bibinfo {author} {\bibfnamefont {J.~P.}\ \bibnamefont
  {Lees}} \emph {et~al.} (\bibinfo {collaboration} {BaBar}),\ }\bibfield
  {title} {\bibinfo {title} {{Precision Measurement of the $B \to X_s \gamma$
  Photon Energy Spectrum, Branching Fraction, and Direct CP Asymmetry $A_{CP}(B
  \to X_{s+d}\gamma)$}},\ }\href
  {https://doi.org/10.1103/PhysRevLett.109.191801} {\bibfield  {journal}
  {\bibinfo  {journal} {Phys. Rev. Lett.}\ }\textbf {\bibinfo {volume} {109}},\
  \bibinfo {pages} {191801} (\bibinfo {year} {2012}{\natexlab{b}})},\ \Eprint
  {https://arxiv.org/abs/1207.2690} {arXiv:1207.2690 [hep-ex]} \BibitemShut
  {NoStop}%
\bibitem [{\citenamefont {Horiguchi}\ \emph {et~al.}(2017)\citenamefont
  {Horiguchi} \emph {et~al.}}]{Belle:2017hum}%
  \BibitemOpen
  \bibfield  {author} {\bibinfo {author} {\bibfnamefont {T.}~\bibnamefont
  {Horiguchi}} \emph {et~al.} (\bibinfo {collaboration} {Belle}),\ }\bibfield
  {title} {\bibinfo {title} {{Evidence for Isospin Violation and Measurement of
  $CP$ Asymmetries in $B \to K^{\ast}(892) \gamma$}},\ }\href
  {https://doi.org/10.1103/PhysRevLett.119.191802} {\bibfield  {journal}
  {\bibinfo  {journal} {Phys. Rev. Lett.}\ }\textbf {\bibinfo {volume} {119}},\
  \bibinfo {pages} {191802} (\bibinfo {year} {2017})},\ \Eprint
  {https://arxiv.org/abs/1707.00394} {arXiv:1707.00394 [hep-ex]} \BibitemShut
  {NoStop}%
\bibitem [{\citenamefont {Aubert}\ \emph
  {et~al.}(2009{\natexlab{a}})\citenamefont {Aubert} \emph
  {et~al.}}]{BaBar:2009byi}%
  \BibitemOpen
  \bibfield  {author} {\bibinfo {author} {\bibfnamefont {B.}~\bibnamefont
  {Aubert}} \emph {et~al.} (\bibinfo {collaboration} {BaBar}),\ }\bibfield
  {title} {\bibinfo {title} {{Measurement of Branching Fractions and CP and
  Isospin Asymmetries in $B \to K^*(892)\gamma$ Decays}},\ }\href
  {https://doi.org/10.1103/PhysRevLett.103.211802} {\bibfield  {journal}
  {\bibinfo  {journal} {Phys. Rev. Lett.}\ }\textbf {\bibinfo {volume} {103}},\
  \bibinfo {pages} {211802} (\bibinfo {year} {2009}{\natexlab{a}})},\ \Eprint
  {https://arxiv.org/abs/0906.2177} {arXiv:0906.2177 [hep-ex]} \BibitemShut
  {NoStop}%
\bibitem [{\citenamefont {Choudhury}\ \emph {et~al.}(2021)\citenamefont
  {Choudhury} \emph {et~al.}}]{BELLE:2019xld}%
  \BibitemOpen
  \bibfield  {author} {\bibinfo {author} {\bibfnamefont {S.}~\bibnamefont
  {Choudhury}} \emph {et~al.} (\bibinfo {collaboration} {BELLE}),\ }\bibfield
  {title} {\bibinfo {title} {{Test of lepton flavor universality and search for
  lepton flavor violation in $B \rightarrow K\ell \ell$ decays}},\ }\href
  {https://doi.org/10.1007/JHEP03(2021)105} {\bibfield  {journal} {\bibinfo
  {journal} {JHEP}\ }\textbf {\bibinfo {volume} {03}},\ \bibinfo {pages}
  {105}},\ \Eprint {https://arxiv.org/abs/1908.01848} {arXiv:1908.01848
  [hep-ex]} \BibitemShut {NoStop}%
\bibitem [{\citenamefont {Aubert}\ \emph
  {et~al.}(2009{\natexlab{b}})\citenamefont {Aubert} \emph
  {et~al.}}]{BaBar:2008jdv}%
  \BibitemOpen
  \bibfield  {author} {\bibinfo {author} {\bibfnamefont {B.}~\bibnamefont
  {Aubert}} \emph {et~al.} (\bibinfo {collaboration} {BaBar}),\ }\bibfield
  {title} {\bibinfo {title} {{Direct CP, Lepton Flavor and Isospin Asymmetries
  in the Decays $B \to K^{(*)} \ell^{+} \ell^{-}$}},\ }\href
  {https://doi.org/10.1103/PhysRevLett.102.091803} {\bibfield  {journal}
  {\bibinfo  {journal} {Phys. Rev. Lett.}\ }\textbf {\bibinfo {volume} {102}},\
  \bibinfo {pages} {091803} (\bibinfo {year} {2009}{\natexlab{b}})},\ \Eprint
  {https://arxiv.org/abs/0807.4119} {arXiv:0807.4119 [hep-ex]} \BibitemShut
  {NoStop}%
\bibitem [{\citenamefont {Wei}\ \emph {et~al.}(2009)\citenamefont {Wei} \emph
  {et~al.}}]{Belle:2009zue}%
  \BibitemOpen
  \bibfield  {author} {\bibinfo {author} {\bibfnamefont {J.~T.}\ \bibnamefont
  {Wei}} \emph {et~al.} (\bibinfo {collaboration} {Belle}),\ }\bibfield
  {title} {\bibinfo {title} {{Measurement of the Differential Branching
  Fraction and Forward-Backward Asymmetry for $B \to K^{(*)}\ell^+\ell^-$}},\
  }\href {https://doi.org/10.1103/PhysRevLett.103.171801} {\bibfield  {journal}
  {\bibinfo  {journal} {Phys. Rev. Lett.}\ }\textbf {\bibinfo {volume} {103}},\
  \bibinfo {pages} {171801} (\bibinfo {year} {2009})},\ \Eprint
  {https://arxiv.org/abs/0904.0770} {arXiv:0904.0770 [hep-ex]} \BibitemShut
  {NoStop}%
\bibitem [{\citenamefont {Aaij}\ \emph {et~al.}(2021)\citenamefont {Aaij} \emph
  {et~al.}}]{LHCb:2021zwz}%
  \BibitemOpen
  \bibfield  {author} {\bibinfo {author} {\bibfnamefont {R.}~\bibnamefont
  {Aaij}} \emph {et~al.} (\bibinfo {collaboration} {LHCb}),\ }\bibfield
  {title} {\bibinfo {title} {{Branching Fraction Measurements of the Rare
  $B^0_s\rightarrow\phi\mu^+\mu^-$ and $B^0_s\rightarrow
  f_2^\prime(1525)\mu^+\mu^-$- Decays}},\ }\href
  {https://doi.org/10.1103/PhysRevLett.127.151801} {\bibfield  {journal}
  {\bibinfo  {journal} {Phys. Rev. Lett.}\ }\textbf {\bibinfo {volume} {127}},\
  \bibinfo {pages} {151801} (\bibinfo {year} {2021})},\ \Eprint
  {https://arxiv.org/abs/2105.14007} {arXiv:2105.14007 [hep-ex]} \BibitemShut
  {NoStop}%
\bibitem [{\citenamefont {Abudin\'en}\ \emph {et~al.}(2023)\citenamefont
  {Abudin\'en} \emph {et~al.}}]{Belle-II:2023bps}%
  \BibitemOpen
  \bibfield  {author} {\bibinfo {author} {\bibfnamefont {F.}~\bibnamefont
  {Abudin\'en}} \emph {et~al.} (\bibinfo {collaboration} {Belle-II}),\
  }\bibfield  {title} {\bibinfo {title} {{Measurement of the B0 lifetime and
  flavor-oscillation frequency using hadronic decays reconstructed in
  2019\textendash{}2021 Belle II data}},\ }\href
  {https://doi.org/10.1103/PhysRevD.107.L091102} {\bibfield  {journal}
  {\bibinfo  {journal} {Phys. Rev. D}\ }\textbf {\bibinfo {volume} {107}},\
  \bibinfo {pages} {L091102} (\bibinfo {year} {2023})},\ \Eprint
  {https://arxiv.org/abs/2302.12791} {arXiv:2302.12791 [hep-ex]} \BibitemShut
  {NoStop}%
\bibitem [{\citenamefont {Aaij}\ \emph {et~al.}(2022)\citenamefont {Aaij} \emph
  {et~al.}}]{LHCb:2021moh}%
  \BibitemOpen
  \bibfield  {author} {\bibinfo {author} {\bibfnamefont {R.}~\bibnamefont
  {Aaij}} \emph {et~al.} (\bibinfo {collaboration} {LHCb}),\ }\bibfield
  {title} {\bibinfo {title} {{Precise determination of the
  $B_{\mathrm{s}}^0$\textendash{}$\overline B_{\mathrm{s}}^0$ oscillation
  frequency}},\ }\href {https://doi.org/10.1038/s41567-021-01394-x} {\bibfield
  {journal} {\bibinfo  {journal} {Nature Phys.}\ }\textbf {\bibinfo {volume}
  {18}},\ \bibinfo {pages} {1} (\bibinfo {year} {2022})},\ \Eprint
  {https://arxiv.org/abs/2104.04421} {arXiv:2104.04421 [hep-ex]} \BibitemShut
  {NoStop}%
\bibitem [{\citenamefont {Aaij}\ \emph {et~al.}(2013)\citenamefont {Aaij} \emph
  {et~al.}}]{LHCb:2013zpr}%
  \BibitemOpen
  \bibfield  {author} {\bibinfo {author} {\bibfnamefont {R.}~\bibnamefont
  {Aaij}} \emph {et~al.} (\bibinfo {collaboration} {LHCb}),\ }\bibfield
  {title} {\bibinfo {title} {{Measurement of $D^0–\bar D^0$ Mixing Parameters
  and Search for $CP$ Violation Using $D^0 \to K^+ \pi^-$ Decays}},\ }\href
  {https://doi.org/10.1103/PhysRevLett.111.251801} {\bibfield  {journal}
  {\bibinfo  {journal} {Phys. Rev. Lett.}\ }\textbf {\bibinfo {volume} {111}},\
  \bibinfo {pages} {251801} (\bibinfo {year} {2013})},\ \Eprint
  {https://arxiv.org/abs/1309.6534} {arXiv:1309.6534 [hep-ex]} \BibitemShut
  {NoStop}%
\bibitem [{\citenamefont {Cortina~Gil}\ \emph {et~al.}(2021)\citenamefont
  {Cortina~Gil} \emph {et~al.}}]{NA62:2021zjw}%
  \BibitemOpen
  \bibfield  {author} {\bibinfo {author} {\bibfnamefont {E.}~\bibnamefont
  {Cortina~Gil}} \emph {et~al.} (\bibinfo {collaboration} {NA62}),\ }\bibfield
  {title} {\bibinfo {title} {{Measurement of the very rare
  K$^{+}$\textrightarrow{}$ {\pi}^{+}\nu \overline{\nu} $ decay}},\ }\href
  {https://doi.org/10.1007/JHEP06(2021)093} {\bibfield  {journal} {\bibinfo
  {journal} {JHEP}\ }\textbf {\bibinfo {volume} {06}},\ \bibinfo {pages}
  {093}},\ \Eprint {https://arxiv.org/abs/2103.15389} {arXiv:2103.15389
  [hep-ex]} \BibitemShut {NoStop}%
\bibitem [{\citenamefont {Hayrapetyan}\ \emph {et~al.}(2024)\citenamefont
  {Hayrapetyan} \emph {et~al.}}]{CMS:2024ubt}%
  \BibitemOpen
  \bibfield  {author} {\bibinfo {author} {\bibfnamefont {A.}~\bibnamefont
  {Hayrapetyan}} \emph {et~al.} (\bibinfo {collaboration} {CMS}),\ }\bibfield
  {title} {\bibinfo {title} {{Search for flavor-changing neutral current
  interactions of the top quark mediated by a Higgs boson in proton-proton
  collisions at 13 TeV}},\ }\href@noop {} {\  (\bibinfo {year} {2024})},\
  \Eprint {https://arxiv.org/abs/2407.15172} {arXiv:2407.15172 [hep-ex]}
  \BibitemShut {NoStop}%
\bibitem [{\citenamefont {Aad}\ \emph {et~al.}(2022{\natexlab{a}})\citenamefont
  {Aad} \emph {et~al.}}]{ATLAS:2021amo}%
  \BibitemOpen
  \bibfield  {author} {\bibinfo {author} {\bibfnamefont {G.}~\bibnamefont
  {Aad}} \emph {et~al.} (\bibinfo {collaboration} {ATLAS}),\ }\bibfield
  {title} {\bibinfo {title} {{Search for flavour-changing neutral-current
  interactions of a top quark and a gluon in pp collisions at $\sqrt{s}=13$~TeV
  with the ATLAS detector}},\ }\href
  {https://doi.org/10.1140/epjc/s10052-022-10182-7} {\bibfield  {journal}
  {\bibinfo  {journal} {Eur. Phys. J. C}\ }\textbf {\bibinfo {volume} {82}},\
  \bibinfo {pages} {334} (\bibinfo {year} {2022}{\natexlab{a}})},\ \Eprint
  {https://arxiv.org/abs/2112.01302} {arXiv:2112.01302 [hep-ex]} \BibitemShut
  {NoStop}%
\bibitem [{\citenamefont {Aad}\ \emph {et~al.}(2023{\natexlab{a}})\citenamefont
  {Aad} \emph {et~al.}}]{ATLAS:2022per}%
  \BibitemOpen
  \bibfield  {author} {\bibinfo {author} {\bibfnamefont {G.}~\bibnamefont
  {Aad}} \emph {et~al.} (\bibinfo {collaboration} {ATLAS}),\ }\bibfield
  {title} {\bibinfo {title} {{Search for flavour-changing neutral-current
  couplings between the top quark and the photon with the ATLAS detector at
  s=13 TeV}},\ }\href {https://doi.org/10.1016/j.physletb.2022.137379}
  {\bibfield  {journal} {\bibinfo  {journal} {Phys. Lett. B}\ }\textbf
  {\bibinfo {volume} {842}},\ \bibinfo {pages} {137379} (\bibinfo {year}
  {2023}{\natexlab{a}})},\ \bibinfo {note} {[Erratum: Phys.Lett.B 847, 138286
  (2024)]},\ \Eprint {https://arxiv.org/abs/2205.02537} {arXiv:2205.02537
  [hep-ex]} \BibitemShut {NoStop}%
\bibitem [{\citenamefont {Aad}\ \emph {et~al.}(2023{\natexlab{b}})\citenamefont
  {Aad} \emph {et~al.}}]{ATLAS:2023qzr}%
  \BibitemOpen
  \bibfield  {author} {\bibinfo {author} {\bibfnamefont {G.}~\bibnamefont
  {Aad}} \emph {et~al.} (\bibinfo {collaboration} {ATLAS}),\ }\bibfield
  {title} {\bibinfo {title} {{Search for flavor-changing neutral-current
  couplings between the top quark and the Z boson with proton-proton collisions
  at s=13\,\,TeV with the ATLAS detector}},\ }\href
  {https://doi.org/10.1103/PhysRevD.108.032019} {\bibfield  {journal} {\bibinfo
   {journal} {Phys. Rev. D}\ }\textbf {\bibinfo {volume} {108}},\ \bibinfo
  {pages} {032019} (\bibinfo {year} {2023}{\natexlab{b}})},\ \Eprint
  {https://arxiv.org/abs/2301.11605} {arXiv:2301.11605 [hep-ex]} \BibitemShut
  {NoStop}%
\bibitem [{\citenamefont {Aad}\ \emph {et~al.}(2024)\citenamefont {Aad} \emph
  {et~al.}}]{ATLAS:2024njy}%
  \BibitemOpen
  \bibfield  {author} {\bibinfo {author} {\bibfnamefont {G.}~\bibnamefont
  {Aad}} \emph {et~al.} (\bibinfo {collaboration} {ATLAS}),\ }\bibfield
  {title} {\bibinfo {title} {{Search for charged-lepton-flavor violating
  \ensuremath{\mu}\ensuremath{\tau}qt interactions in top-quark production and
  decay in pp collisions at s=13\,\,TeV with the ATLAS detector at the LHC}},\
  }\href {https://doi.org/10.1103/PhysRevD.110.012014} {\bibfield  {journal}
  {\bibinfo  {journal} {Phys. Rev. D}\ }\textbf {\bibinfo {volume} {110}},\
  \bibinfo {pages} {012014} (\bibinfo {year} {2024})},\ \Eprint
  {https://arxiv.org/abs/2403.06742} {arXiv:2403.06742 [hep-ex]} \BibitemShut
  {NoStop}%
\bibitem [{\citenamefont {Tumasyan}\ \emph {et~al.}(2022)\citenamefont
  {Tumasyan} \emph {et~al.}}]{CMS:2022ztx}%
  \BibitemOpen
  \bibfield  {author} {\bibinfo {author} {\bibfnamefont {A.}~\bibnamefont
  {Tumasyan}} \emph {et~al.} (\bibinfo {collaboration} {CMS}),\ }\bibfield
  {title} {\bibinfo {title} {{Search for charged-lepton flavor violation in top
  quark production and decay in pp collisions at $ \sqrt{s} $ = 13 TeV}},\
  }\href {https://doi.org/10.1007/JHEP06(2022)082} {\bibfield  {journal}
  {\bibinfo  {journal} {JHEP}\ }\textbf {\bibinfo {volume} {06}},\ \bibinfo
  {pages} {082}},\ \Eprint {https://arxiv.org/abs/2201.07859} {arXiv:2201.07859
  [hep-ex]} \BibitemShut {NoStop}%
\bibitem [{\citenamefont {Fukuda}\ \emph {et~al.}(1998)\citenamefont {Fukuda}
  \emph {et~al.}}]{Super-Kamiokande:1998kpq}%
  \BibitemOpen
  \bibfield  {author} {\bibinfo {author} {\bibfnamefont {Y.}~\bibnamefont
  {Fukuda}} \emph {et~al.} (\bibinfo {collaboration} {Super-Kamiokande}),\
  }\bibfield  {title} {\bibinfo {title} {{Evidence for oscillation of
  atmospheric neutrinos}},\ }\href
  {https://doi.org/10.1103/PhysRevLett.81.1562} {\bibfield  {journal} {\bibinfo
   {journal} {Phys. Rev. Lett.}\ }\textbf {\bibinfo {volume} {81}},\ \bibinfo
  {pages} {1562} (\bibinfo {year} {1998})},\ \Eprint
  {https://arxiv.org/abs/hep-ex/9807003} {arXiv:hep-ex/9807003} \BibitemShut
  {NoStop}%
\bibitem [{\citenamefont {Ahmad}\ \emph {et~al.}(2002)\citenamefont {Ahmad}
  \emph {et~al.}}]{SNO:2002tuh}%
  \BibitemOpen
  \bibfield  {author} {\bibinfo {author} {\bibfnamefont {Q.~R.}\ \bibnamefont
  {Ahmad}} \emph {et~al.} (\bibinfo {collaboration} {SNO}),\ }\bibfield
  {title} {\bibinfo {title} {{Direct evidence for neutrino flavor
  transformation from neutral current interactions in the Sudbury Neutrino
  Observatory}},\ }\href {https://doi.org/10.1103/PhysRevLett.89.011301}
  {\bibfield  {journal} {\bibinfo  {journal} {Phys. Rev. Lett.}\ }\textbf
  {\bibinfo {volume} {89}},\ \bibinfo {pages} {011301} (\bibinfo {year}
  {2002})},\ \Eprint {https://arxiv.org/abs/nucl-ex/0204008}
  {arXiv:nucl-ex/0204008} \BibitemShut {NoStop}%
\bibitem [{\citenamefont {Afanaciev}\ \emph {et~al.}(2024)\citenamefont
  {Afanaciev} \emph {et~al.}}]{MEGII:2023ltw}%
  \BibitemOpen
  \bibfield  {author} {\bibinfo {author} {\bibfnamefont {K.}~\bibnamefont
  {Afanaciev}} \emph {et~al.} (\bibinfo {collaboration} {MEG II}),\ }\bibfield
  {title} {\bibinfo {title} {{A search for $\mu ^+ \rightarrow \textrm{e}^+
  \gamma $ with the first dataset of the MEG~II experiment}},\ }\href
  {https://doi.org/10.1140/epjc/s10052-024-12416-2} {\bibfield  {journal}
  {\bibinfo  {journal} {Eur. Phys. J. C}\ }\textbf {\bibinfo {volume} {84}},\
  \bibinfo {pages} {216} (\bibinfo {year} {2024})},\ \bibinfo {note} {[Erratum:
  Eur.Phys.J.C 84, 1042 (2024)]},\ \Eprint {https://arxiv.org/abs/2310.12614}
  {arXiv:2310.12614 [hep-ex]} \BibitemShut {NoStop}%
\bibitem [{\citenamefont {Aubert}\ \emph {et~al.}(2010)\citenamefont {Aubert}
  \emph {et~al.}}]{BaBar:2009hkt}%
  \BibitemOpen
  \bibfield  {author} {\bibinfo {author} {\bibfnamefont {B.}~\bibnamefont
  {Aubert}} \emph {et~al.} (\bibinfo {collaboration} {BaBar}),\ }\bibfield
  {title} {\bibinfo {title} {{Searches for Lepton Flavor Violation in the
  Decays $\tau^\pm \to e^\pm \gamma$ and $\tau^\pm \to \mu^\pm \gamma$}},\
  }\href {https://doi.org/10.1103/PhysRevLett.104.021802} {\bibfield  {journal}
  {\bibinfo  {journal} {Phys. Rev. Lett.}\ }\textbf {\bibinfo {volume} {104}},\
  \bibinfo {pages} {021802} (\bibinfo {year} {2010})},\ \Eprint
  {https://arxiv.org/abs/0908.2381} {arXiv:0908.2381 [hep-ex]} \BibitemShut
  {NoStop}%
\bibitem [{\citenamefont {Abdesselam}\ \emph {et~al.}(2021)\citenamefont
  {Abdesselam} \emph {et~al.}}]{Belle:2021ysv}%
  \BibitemOpen
  \bibfield  {author} {\bibinfo {author} {\bibfnamefont {A.}~\bibnamefont
  {Abdesselam}} \emph {et~al.} (\bibinfo {collaboration} {Belle}),\ }\bibfield
  {title} {\bibinfo {title} {{Search for lepton-flavor-violating tau-lepton
  decays to $\ell\gamma$ at Belle}},\ }\href
  {https://doi.org/10.1007/JHEP10(2021)019} {\bibfield  {journal} {\bibinfo
  {journal} {JHEP}\ }\textbf {\bibinfo {volume} {10}},\ \bibinfo {pages}
  {19}},\ \Eprint {https://arxiv.org/abs/2103.12994} {arXiv:2103.12994
  [hep-ex]} \BibitemShut {NoStop}%
\bibitem [{\citenamefont {Bellgardt}\ \emph {et~al.}(1988)\citenamefont
  {Bellgardt} \emph {et~al.}}]{SINDRUM:1987nra}%
  \BibitemOpen
  \bibfield  {author} {\bibinfo {author} {\bibfnamefont {U.}~\bibnamefont
  {Bellgardt}} \emph {et~al.} (\bibinfo {collaboration} {SINDRUM}),\ }\bibfield
   {title} {\bibinfo {title} {{Search for the Decay $\mu^+ \to e^+ e^+ e^-$}},\
  }\href {https://doi.org/10.1016/0550-3213(88)90462-2} {\bibfield  {journal}
  {\bibinfo  {journal} {Nucl. Phys. B}\ }\textbf {\bibinfo {volume} {299}},\
  \bibinfo {pages} {1} (\bibinfo {year} {1988})}\BibitemShut {NoStop}%
\bibitem [{\citenamefont {Hayasaka}\ \emph {et~al.}(2010)\citenamefont
  {Hayasaka} \emph {et~al.}}]{Hayasaka:2010np}%
  \BibitemOpen
  \bibfield  {author} {\bibinfo {author} {\bibfnamefont {K.}~\bibnamefont
  {Hayasaka}} \emph {et~al.},\ }\bibfield  {title} {\bibinfo {title} {{Search
  for Lepton Flavor Violating Tau Decays into Three Leptons with 719 Million
  Produced Tau+Tau- Pairs}},\ }\href
  {https://doi.org/10.1016/j.physletb.2010.03.037} {\bibfield  {journal}
  {\bibinfo  {journal} {Phys. Lett. B}\ }\textbf {\bibinfo {volume} {687}},\
  \bibinfo {pages} {139} (\bibinfo {year} {2010})},\ \Eprint
  {https://arxiv.org/abs/1001.3221} {arXiv:1001.3221 [hep-ex]} \BibitemShut
  {NoStop}%
\bibitem [{\citenamefont {Badertscher}\ \emph {et~al.}(1980)\citenamefont
  {Badertscher} \emph {et~al.}}]{Badertscher:1980bt}%
  \BibitemOpen
  \bibfield  {author} {\bibinfo {author} {\bibfnamefont {A.}~\bibnamefont
  {Badertscher}} \emph {et~al.},\ }\bibfield  {title} {\bibinfo {title} {{New
  Upper Limits for Muon - Electron Conversion in Sulfur}},\ }\href
  {https://doi.org/10.1007/BF02776193} {\bibfield  {journal} {\bibinfo
  {journal} {Lett. Nuovo Cim.}\ }\textbf {\bibinfo {volume} {28}},\ \bibinfo
  {pages} {401} (\bibinfo {year} {1980})}\BibitemShut {NoStop}%
\bibitem [{\citenamefont {Dohmen}\ \emph {et~al.}(1993)\citenamefont {Dohmen}
  \emph {et~al.}}]{SINDRUMII:1993gxf}%
  \BibitemOpen
  \bibfield  {author} {\bibinfo {author} {\bibfnamefont {C.}~\bibnamefont
  {Dohmen}} \emph {et~al.} (\bibinfo {collaboration} {SINDRUM II}),\ }\bibfield
   {title} {\bibinfo {title} {{Test of lepton flavor conservation in mu
  ---\ensuremath{>} e conversion on titanium}},\ }\href
  {https://doi.org/10.1016/0370-2693(93)91383-X} {\bibfield  {journal}
  {\bibinfo  {journal} {Phys. Lett. B}\ }\textbf {\bibinfo {volume} {317}},\
  \bibinfo {pages} {631} (\bibinfo {year} {1993})}\BibitemShut {NoStop}%
\bibitem [{\citenamefont {Honecker}\ \emph {et~al.}(1996)\citenamefont
  {Honecker} \emph {et~al.}}]{SINDRUMII:1996fti}%
  \BibitemOpen
  \bibfield  {author} {\bibinfo {author} {\bibfnamefont {W.}~\bibnamefont
  {Honecker}} \emph {et~al.} (\bibinfo {collaboration} {SINDRUM II}),\
  }\bibfield  {title} {\bibinfo {title} {{Improved limit on the branching ratio
  of mu ---\ensuremath{>} e conversion on lead}},\ }\href
  {https://doi.org/10.1103/PhysRevLett.76.200} {\bibfield  {journal} {\bibinfo
  {journal} {Phys. Rev. Lett.}\ }\textbf {\bibinfo {volume} {76}},\ \bibinfo
  {pages} {200} (\bibinfo {year} {1996})}\BibitemShut {NoStop}%
\bibitem [{\citenamefont {Bertl}\ \emph {et~al.}(2006)\citenamefont {Bertl}
  \emph {et~al.}}]{SINDRUMII:2006dvw}%
  \BibitemOpen
  \bibfield  {author} {\bibinfo {author} {\bibfnamefont {W.~H.}\ \bibnamefont
  {Bertl}} \emph {et~al.} (\bibinfo {collaboration} {SINDRUM II}),\ }\bibfield
  {title} {\bibinfo {title} {{A Search for muon to electron conversion in
  muonic gold}},\ }\href {https://doi.org/10.1140/epjc/s2006-02582-x}
  {\bibfield  {journal} {\bibinfo  {journal} {Eur. Phys. J. C}\ }\textbf
  {\bibinfo {volume} {47}},\ \bibinfo {pages} {337} (\bibinfo {year}
  {2006})}\BibitemShut {NoStop}%
\bibitem [{\citenamefont {Davidson}\ \emph {et~al.}(2022)\citenamefont
  {Davidson}, \citenamefont {Echenard}, \citenamefont {Bernstein},
  \citenamefont {Heeck},\ and\ \citenamefont {Hitlin}}]{Davidson:2022jai}%
  \BibitemOpen
  \bibfield  {author} {\bibinfo {author} {\bibfnamefont {S.}~\bibnamefont
  {Davidson}}, \bibinfo {author} {\bibfnamefont {B.}~\bibnamefont {Echenard}},
  \bibinfo {author} {\bibfnamefont {R.~H.}\ \bibnamefont {Bernstein}}, \bibinfo
  {author} {\bibfnamefont {J.}~\bibnamefont {Heeck}},\ and\ \bibinfo {author}
  {\bibfnamefont {D.~G.}\ \bibnamefont {Hitlin}},\ }\bibfield  {title}
  {\bibinfo {title} {{Charged Lepton Flavor Violation}},\ }\href@noop {} {\
  (\bibinfo {year} {2022})},\ \Eprint {https://arxiv.org/abs/2209.00142}
  {arXiv:2209.00142 [hep-ex]} \BibitemShut {NoStop}%
\bibitem [{\citenamefont {Hou}\ and\ \citenamefont
  {Modak}(2021)}]{Hou:2020chc}%
  \BibitemOpen
  \bibfield  {author} {\bibinfo {author} {\bibfnamefont {W.-S.}\ \bibnamefont
  {Hou}}\ and\ \bibinfo {author} {\bibfnamefont {T.}~\bibnamefont {Modak}},\
  }\bibfield  {title} {\bibinfo {title} {{Probing Top Changing Neutral Higgs
  Couplings at Colliders}},\ }\href {https://doi.org/10.1142/S0217732321300068}
  {\bibfield  {journal} {\bibinfo  {journal} {Mod. Phys. Lett. A}\ }\textbf
  {\bibinfo {volume} {36}},\ \bibinfo {pages} {2130006} (\bibinfo {year}
  {2021})},\ \Eprint {https://arxiv.org/abs/2012.05735} {arXiv:2012.05735
  [hep-ph]} \BibitemShut {NoStop}%
\bibitem [{\citenamefont {Fuyuto}\ \emph {et~al.}(2016)\citenamefont {Fuyuto},
  \citenamefont {Hou},\ and\ \citenamefont {Kohda}}]{Fuyuto:2015gmk}%
  \BibitemOpen
  \bibfield  {author} {\bibinfo {author} {\bibfnamefont {K.}~\bibnamefont
  {Fuyuto}}, \bibinfo {author} {\bibfnamefont {W.-S.}\ \bibnamefont {Hou}},\
  and\ \bibinfo {author} {\bibfnamefont {M.}~\bibnamefont {Kohda}},\ }\bibfield
   {title} {\bibinfo {title} {{Z' -induced FCNC decays of top, beauty, and
  strange quarks}},\ }\href {https://doi.org/10.1103/PhysRevD.93.054021}
  {\bibfield  {journal} {\bibinfo  {journal} {Phys. Rev. D}\ }\textbf {\bibinfo
  {volume} {93}},\ \bibinfo {pages} {054021} (\bibinfo {year} {2016})},\
  \Eprint {https://arxiv.org/abs/1512.09026} {arXiv:1512.09026 [hep-ph]}
  \BibitemShut {NoStop}%
\bibitem [{\citenamefont {Bhattacharya}\ \emph
  {et~al.}(2025{\natexlab{a}})\citenamefont {Bhattacharya}, \citenamefont
  {Kolay}, \citenamefont {Pradhan},\ and\ \citenamefont
  {Sarkar}}]{Bhattacharya:2025mlg}%
  \BibitemOpen
  \bibfield  {author} {\bibinfo {author} {\bibfnamefont {S.}~\bibnamefont
  {Bhattacharya}}, \bibinfo {author} {\bibfnamefont {L.}~\bibnamefont {Kolay}},
  \bibinfo {author} {\bibfnamefont {D.}~\bibnamefont {Pradhan}},\ and\ \bibinfo
  {author} {\bibfnamefont {A.}~\bibnamefont {Sarkar}},\ }\bibfield  {title}
  {\bibinfo {title} {{Up-type FCNC in presence of Dark Matter}},\ }\href@noop
  {} {\  (\bibinfo {year} {2025}{\natexlab{a}})},\ \Eprint
  {https://arxiv.org/abs/2504.20045} {arXiv:2504.20045 [hep-ph]} \BibitemShut
  {NoStop}%
\bibitem [{\citenamefont {Bhattacharya}\ \emph {et~al.}(2024)\citenamefont
  {Bhattacharya}, \citenamefont {Jahedi}, \citenamefont {Nandi},\ and\
  \citenamefont {Sarkar}}]{Bhattacharya:2023beo}%
  \BibitemOpen
  \bibfield  {author} {\bibinfo {author} {\bibfnamefont {S.}~\bibnamefont
  {Bhattacharya}}, \bibinfo {author} {\bibfnamefont {S.}~\bibnamefont
  {Jahedi}}, \bibinfo {author} {\bibfnamefont {S.}~\bibnamefont {Nandi}},\ and\
  \bibinfo {author} {\bibfnamefont {A.}~\bibnamefont {Sarkar}},\ }\bibfield
  {title} {\bibinfo {title} {{Probing flavor constrained SMEFT operators
  through tc production at the muon collider}},\ }\href
  {https://doi.org/10.1007/JHEP07(2024)061} {\bibfield  {journal} {\bibinfo
  {journal} {JHEP}\ }\textbf {\bibinfo {volume} {07}},\ \bibinfo {pages}
  {061}},\ \Eprint {https://arxiv.org/abs/2312.14872} {arXiv:2312.14872
  [hep-ph]} \BibitemShut {NoStop}%
\bibitem [{\citenamefont {Sun}\ \emph {et~al.}(2023)\citenamefont {Sun},
  \citenamefont {Yan}, \citenamefont {Zhao},\ and\ \citenamefont
  {Zhao}}]{Sun:2023cuf}%
  \BibitemOpen
  \bibfield  {author} {\bibinfo {author} {\bibfnamefont {S.}~\bibnamefont
  {Sun}}, \bibinfo {author} {\bibfnamefont {Q.-S.}\ \bibnamefont {Yan}},
  \bibinfo {author} {\bibfnamefont {X.}~\bibnamefont {Zhao}},\ and\ \bibinfo
  {author} {\bibfnamefont {Z.}~\bibnamefont {Zhao}},\ }\bibfield  {title}
  {\bibinfo {title} {{Constraining rare B decays by
  \ensuremath{\mu}+\ensuremath{\mu}-\textrightarrow{}tc at future lepton
  colliders}},\ }\href {https://doi.org/10.1103/PhysRevD.108.075016} {\bibfield
   {journal} {\bibinfo  {journal} {Phys. Rev. D}\ }\textbf {\bibinfo {volume}
  {108}},\ \bibinfo {pages} {075016} (\bibinfo {year} {2023})},\ \Eprint
  {https://arxiv.org/abs/2302.01143} {arXiv:2302.01143 [hep-ph]} \BibitemShut
  {NoStop}%
\bibitem [{\citenamefont {Ghasemi~Bostanabad}\ and\ \citenamefont
  {Mohammadi~Najafabadi}(2025)}]{GhasemiBostanabad:2025xua}%
  \BibitemOpen
  \bibfield  {author} {\bibinfo {author} {\bibfnamefont {M.}~\bibnamefont
  {Ghasemi~Bostanabad}}\ and\ \bibinfo {author} {\bibfnamefont
  {M.}~\bibnamefont {Mohammadi~Najafabadi}},\ }\bibfield  {title} {\bibinfo
  {title} {{Machine learning approaches to top quark flavor-changing
  four-fermion interactions in trilepton signals at the LHC}},\ }\href
  {https://doi.org/10.1103/mk8x-nrpn} {\bibfield  {journal} {\bibinfo
  {journal} {Phys. Rev. D}\ }\textbf {\bibinfo {volume} {111}},\ \bibinfo
  {pages} {112003} (\bibinfo {year} {2025})},\ \Eprint
  {https://arxiv.org/abs/2502.18667} {arXiv:2502.18667 [hep-ph]} \BibitemShut
  {NoStop}%
\bibitem [{\citenamefont {Atkinson}\ \emph {et~al.}(2025)\citenamefont
  {Atkinson}, \citenamefont {Englert}, \citenamefont {Kirk},\ and\
  \citenamefont {Tetlalmatzi-Xolocotzi}}]{Atkinson:2024hqp}%
  \BibitemOpen
  \bibfield  {author} {\bibinfo {author} {\bibfnamefont {O.}~\bibnamefont
  {Atkinson}}, \bibinfo {author} {\bibfnamefont {C.}~\bibnamefont {Englert}},
  \bibinfo {author} {\bibfnamefont {M.}~\bibnamefont {Kirk}},\ and\ \bibinfo
  {author} {\bibfnamefont {G.}~\bibnamefont {Tetlalmatzi-Xolocotzi}},\
  }\bibfield  {title} {\bibinfo {title} {{Collider-flavour complementarity from
  the bottom to the top}},\ }\href
  {https://doi.org/10.1140/epjc/s10052-024-13739-w} {\bibfield  {journal}
  {\bibinfo  {journal} {Eur. Phys. J. C}\ }\textbf {\bibinfo {volume} {85}},\
  \bibinfo {pages} {258} (\bibinfo {year} {2025})},\ \Eprint
  {https://arxiv.org/abs/2411.00940} {arXiv:2411.00940 [hep-ph]} \BibitemShut
  {NoStop}%
\bibitem [{\citenamefont {Chen}\ and\ \citenamefont
  {Nomura}(2022)}]{Chen:2022dzc}%
  \BibitemOpen
  \bibfield  {author} {\bibinfo {author} {\bibfnamefont {C.-H.}\ \bibnamefont
  {Chen}}\ and\ \bibinfo {author} {\bibfnamefont {T.}~\bibnamefont {Nomura}},\
  }\bibfield  {title} {\bibinfo {title} {{Scotogenic top-quark FCNC decays}},\
  }\href {https://doi.org/10.1103/PhysRevD.106.095005} {\bibfield  {journal}
  {\bibinfo  {journal} {Phys. Rev. D}\ }\textbf {\bibinfo {volume} {106}},\
  \bibinfo {pages} {095005} (\bibinfo {year} {2022})},\ \Eprint
  {https://arxiv.org/abs/2204.01214} {arXiv:2204.01214 [hep-ph]} \BibitemShut
  {NoStop}%
\bibitem [{\citenamefont {Gait\'an}\ \emph {et~al.}(2018)\citenamefont
  {Gait\'an}, \citenamefont {Martinez}, \citenamefont {de~Oca},\ and\
  \citenamefont {Garc\'es}}]{Gaitan:2017tka}%
  \BibitemOpen
  \bibfield  {author} {\bibinfo {author} {\bibfnamefont {R.}~\bibnamefont
  {Gait\'an}}, \bibinfo {author} {\bibfnamefont {R.}~\bibnamefont {Martinez}},
  \bibinfo {author} {\bibfnamefont {J.~H.~M.}\ \bibnamefont {de~Oca}},\ and\
  \bibinfo {author} {\bibfnamefont {E.~A.}\ \bibnamefont {Garc\'es}},\
  }\bibfield  {title} {\bibinfo {title} {{SM Higgs boson and $t\rightarrow cZ$
  decays in the 2HDM type III with CP violation}},\ }\href
  {https://doi.org/10.1103/PhysRevD.98.035031} {\bibfield  {journal} {\bibinfo
  {journal} {Phys. Rev. D}\ }\textbf {\bibinfo {volume} {98}},\ \bibinfo
  {pages} {035031} (\bibinfo {year} {2018})},\ \Eprint
  {https://arxiv.org/abs/1710.04262} {arXiv:1710.04262 [hep-ph]} \BibitemShut
  {NoStop}%
\bibitem [{\citenamefont {Shi}\ and\ \citenamefont
  {Zhang}(2019)}]{Shi:2019epw}%
  \BibitemOpen
  \bibfield  {author} {\bibinfo {author} {\bibfnamefont {L.}~\bibnamefont
  {Shi}}\ and\ \bibinfo {author} {\bibfnamefont {C.}~\bibnamefont {Zhang}},\
  }\bibfield  {title} {\bibinfo {title} {{Probing the top quark flavor-changing
  couplings at CEPC}},\ }\href {https://doi.org/10.1088/1674-1137/43/11/113104}
  {\bibfield  {journal} {\bibinfo  {journal} {Chin. Phys. C}\ }\textbf
  {\bibinfo {volume} {43}},\ \bibinfo {pages} {113104} (\bibinfo {year}
  {2019})},\ \Eprint {https://arxiv.org/abs/1906.04573} {arXiv:1906.04573
  [hep-ph]} \BibitemShut {NoStop}%
\bibitem [{\citenamefont {Arroyo-Ure\~na}\ \emph {et~al.}(2019)\citenamefont
  {Arroyo-Ure\~na}, \citenamefont {Gait\'an}, \citenamefont {Herrera-Chac\'on},
  \citenamefont {Montes~de Oca~Y.},\ and\ \citenamefont
  {Valencia-P\'erez}}]{Arroyo-Urena:2019qhl}%
  \BibitemOpen
  \bibfield  {author} {\bibinfo {author} {\bibfnamefont {M.~A.}\ \bibnamefont
  {Arroyo-Ure\~na}}, \bibinfo {author} {\bibfnamefont {R.}~\bibnamefont
  {Gait\'an}}, \bibinfo {author} {\bibfnamefont {E.~A.}\ \bibnamefont
  {Herrera-Chac\'on}}, \bibinfo {author} {\bibfnamefont {J.~H.}\ \bibnamefont
  {Montes~de Oca~Y.}},\ and\ \bibinfo {author} {\bibfnamefont {T.~A.}\
  \bibnamefont {Valencia-P\'erez}},\ }\bibfield  {title} {\bibinfo {title}
  {{Search for the $t\to ch$ decay at hadron colliders}},\ }\href
  {https://doi.org/10.1007/JHEP07(2019)041} {\bibfield  {journal} {\bibinfo
  {journal} {JHEP}\ }\textbf {\bibinfo {volume} {07}},\ \bibinfo {pages}
  {041}},\ \Eprint {https://arxiv.org/abs/1903.02718} {arXiv:1903.02718
  [hep-ph]} \BibitemShut {NoStop}%
\bibitem [{\citenamefont {Liu}\ and\ \citenamefont
  {Moretti}(2020)}]{Liu:2020kxt}%
  \BibitemOpen
  \bibfield  {author} {\bibinfo {author} {\bibfnamefont {Y.-B.}\ \bibnamefont
  {Liu}}\ and\ \bibinfo {author} {\bibfnamefont {S.}~\bibnamefont {Moretti}},\
  }\bibfield  {title} {\bibinfo {title} {{Probing the top-Higgs boson FCNC
  couplings via the $h\to \gamma\gamma$ channel at the HE-LHC and FCC-hh}},\
  }\href {https://doi.org/10.1103/PhysRevD.101.075029} {\bibfield  {journal}
  {\bibinfo  {journal} {Phys. Rev. D}\ }\textbf {\bibinfo {volume} {101}},\
  \bibinfo {pages} {075029} (\bibinfo {year} {2020})},\ \Eprint
  {https://arxiv.org/abs/2002.05311} {arXiv:2002.05311 [hep-ph]} \BibitemShut
  {NoStop}%
\bibitem [{\citenamefont {Liu}\ \emph {et~al.}(2022)\citenamefont {Liu},
  \citenamefont {Yan},\ and\ \citenamefont {Zhang}}]{Liu:2021crr}%
  \BibitemOpen
  \bibfield  {author} {\bibinfo {author} {\bibfnamefont {Y.}~\bibnamefont
  {Liu}}, \bibinfo {author} {\bibfnamefont {B.}~\bibnamefont {Yan}},\ and\
  \bibinfo {author} {\bibfnamefont {R.}~\bibnamefont {Zhang}},\ }\bibfield
  {title} {\bibinfo {title} {{Loop induced top quark FCNC through top quark and
  dark matter interactions}},\ }\href
  {https://doi.org/10.1016/j.physletb.2022.136964} {\bibfield  {journal}
  {\bibinfo  {journal} {Phys. Lett. B}\ }\textbf {\bibinfo {volume} {827}},\
  \bibinfo {pages} {136964} (\bibinfo {year} {2022})},\ \Eprint
  {https://arxiv.org/abs/2103.07859} {arXiv:2103.07859 [hep-ph]} \BibitemShut
  {NoStop}%
\bibitem [{\citenamefont {Gutierrez}\ \emph {et~al.}(2021)\citenamefont
  {Gutierrez}, \citenamefont {Jain},\ and\ \citenamefont
  {Kao}}]{Gutierrez:2020eby}%
  \BibitemOpen
  \bibfield  {author} {\bibinfo {author} {\bibfnamefont {P.}~\bibnamefont
  {Gutierrez}}, \bibinfo {author} {\bibfnamefont {R.}~\bibnamefont {Jain}},\
  and\ \bibinfo {author} {\bibfnamefont {C.}~\bibnamefont {Kao}},\ }\bibfield
  {title} {\bibinfo {title} {{Flavor changing top decays to charm and a Higgs
  boson with $\tau\tau$ at the LHC}},\ }\href
  {https://doi.org/10.1103/PhysRevD.103.115020} {\bibfield  {journal} {\bibinfo
   {journal} {Phys. Rev. D}\ }\textbf {\bibinfo {volume} {103}},\ \bibinfo
  {pages} {115020} (\bibinfo {year} {2021})},\ \Eprint
  {https://arxiv.org/abs/2012.09209} {arXiv:2012.09209 [hep-ph]} \BibitemShut
  {NoStop}%
\bibitem [{\citenamefont {Bie}\ \emph {et~al.}(2021)\citenamefont {Bie},
  \citenamefont {Liu},\ and\ \citenamefont {Wang}}]{Bie:2020sro}%
  \BibitemOpen
  \bibfield  {author} {\bibinfo {author} {\bibfnamefont {S.-Y.}\ \bibnamefont
  {Bie}}, \bibinfo {author} {\bibfnamefont {G.-L.}\ \bibnamefont {Liu}},\ and\
  \bibinfo {author} {\bibfnamefont {W.}~\bibnamefont {Wang}},\ }\bibfield
  {title} {\bibinfo {title} {{Top rare decays $t\to cV$ in mirror twin Higgs
  models}},\ }\href {https://doi.org/10.1088/1674-1137/abc1d5} {\bibfield
  {journal} {\bibinfo  {journal} {Chin. Phys. C}\ }\textbf {\bibinfo {volume}
  {45}},\ \bibinfo {pages} {013106} (\bibinfo {year} {2021})},\ \Eprint
  {https://arxiv.org/abs/2009.04858} {arXiv:2009.04858 [hep-ph]} \BibitemShut
  {NoStop}%
\bibitem [{\citenamefont {Balaji}(2020)}]{Balaji:2020qjg}%
  \BibitemOpen
  \bibfield  {author} {\bibinfo {author} {\bibfnamefont {S.}~\bibnamefont
  {Balaji}},\ }\bibfield  {title} {\bibinfo {title} {{$CP$ asymmetries in the
  rare top decays $t\to c\gamma$ and $t\to c g$}},\ }\href
  {https://doi.org/10.1103/PhysRevD.102.113010} {\bibfield  {journal} {\bibinfo
   {journal} {Phys. Rev. D}\ }\textbf {\bibinfo {volume} {102}},\ \bibinfo
  {pages} {113010} (\bibinfo {year} {2020})},\ \Eprint
  {https://arxiv.org/abs/2009.03315} {arXiv:2009.03315 [hep-ph]} \BibitemShut
  {NoStop}%
\bibitem [{\citenamefont {Frank}\ and\ \citenamefont
  {Turan}(2006)}]{Frank:2006ku}%
  \BibitemOpen
  \bibfield  {author} {\bibinfo {author} {\bibfnamefont {M.}~\bibnamefont
  {Frank}}\ and\ \bibinfo {author} {\bibfnamefont {I.}~\bibnamefont {Turan}},\
  }\bibfield  {title} {\bibinfo {title} {{Rare decay of the top t
  ---\ensuremath{>} c l anti-l and single top production at ILC}},\ }\href
  {https://doi.org/10.1103/PhysRevD.74.073014} {\bibfield  {journal} {\bibinfo
  {journal} {Phys. Rev. D}\ }\textbf {\bibinfo {volume} {74}},\ \bibinfo
  {pages} {073014} (\bibinfo {year} {2006})},\ \Eprint
  {https://arxiv.org/abs/hep-ph/0609069} {arXiv:hep-ph/0609069} \BibitemShut
  {NoStop}%
\bibitem [{\citenamefont {d'Enterria}\ and\ \citenamefont
  {Le}(2025)}]{dEnterria:2023wjq}%
  \BibitemOpen
  \bibfield  {author} {\bibinfo {author} {\bibfnamefont {D.}~\bibnamefont
  {d'Enterria}}\ and\ \bibinfo {author} {\bibfnamefont {V.~D.}\ \bibnamefont
  {Le}},\ }\bibfield  {title} {\bibinfo {title} {{Rare and exclusive few-body
  decays of the Higgs, Z, W bosons, and the top quark}},\ }\href
  {https://doi.org/10.1088/1361-6471/ad3c59} {\bibfield  {journal} {\bibinfo
  {journal} {J. Phys. G}\ }\textbf {\bibinfo {volume} {52}},\ \bibinfo {pages}
  {053001} (\bibinfo {year} {2025})},\ \Eprint
  {https://arxiv.org/abs/2312.11211} {arXiv:2312.11211 [hep-ph]} \BibitemShut
  {NoStop}%
\bibitem [{\citenamefont {Kala}\ \emph {et~al.}(2025)\citenamefont {Kala},
  \citenamefont {Kolay}, \citenamefont {Mukherjee},\ and\ \citenamefont
  {Nandi}}]{Kala:2025srq}%
  \BibitemOpen
  \bibfield  {author} {\bibinfo {author} {\bibfnamefont {S.}~\bibnamefont
  {Kala}}, \bibinfo {author} {\bibfnamefont {L.}~\bibnamefont {Kolay}},
  \bibinfo {author} {\bibfnamefont {L.}~\bibnamefont {Mukherjee}},\ and\
  \bibinfo {author} {\bibfnamefont {S.}~\bibnamefont {Nandi}},\ }\bibfield
  {title} {\bibinfo {title} {{Constraining anomalous $W tb$ and related SMEFT
  couplings using low-energy and electroweak precision observables}},\
  }\href@noop {} {\  (\bibinfo {year} {2025})},\ \Eprint
  {https://arxiv.org/abs/2505.07926} {arXiv:2505.07926 [hep-ph]} \BibitemShut
  {NoStop}%
\bibitem [{\citenamefont {Khachatryan}\ \emph {et~al.}(2016)\citenamefont
  {Khachatryan} \emph {et~al.}}]{CMS:2016cvq}%
  \BibitemOpen
  \bibfield  {author} {\bibinfo {author} {\bibfnamefont {V.}~\bibnamefont
  {Khachatryan}} \emph {et~al.} (\bibinfo {collaboration} {CMS}),\ }\bibfield
  {title} {\bibinfo {title} {{Search for lepton flavour violating decays of the
  Higgs boson to $e \tau$ and $e \mu$ in proton\textendash{}proton collisions
  at $\sqrt s=$ 8 TeV}},\ }\href
  {https://doi.org/10.1016/j.physletb.2016.09.062} {\bibfield  {journal}
  {\bibinfo  {journal} {Phys. Lett. B}\ }\textbf {\bibinfo {volume} {763}},\
  \bibinfo {pages} {472} (\bibinfo {year} {2016})},\ \Eprint
  {https://arxiv.org/abs/1607.03561} {arXiv:1607.03561 [hep-ex]} \BibitemShut
  {NoStop}%
\bibitem [{\citenamefont {Sirunyan}\ \emph {et~al.}(2018)\citenamefont
  {Sirunyan} \emph {et~al.}}]{CMS:2017con}%
  \BibitemOpen
  \bibfield  {author} {\bibinfo {author} {\bibfnamefont {A.~M.}\ \bibnamefont
  {Sirunyan}} \emph {et~al.} (\bibinfo {collaboration} {CMS}),\ }\bibfield
  {title} {\bibinfo {title} {{Search for lepton flavour violating decays of the
  Higgs boson to $\mu\tau$ and e$\tau$ in proton-proton collisions at
  $\sqrt{s}=$ 13 TeV}},\ }\href {https://doi.org/10.1007/JHEP06(2018)001}
  {\bibfield  {journal} {\bibinfo  {journal} {JHEP}\ }\textbf {\bibinfo
  {volume} {06}},\ \bibinfo {pages} {001}},\ \Eprint
  {https://arxiv.org/abs/1712.07173} {arXiv:1712.07173 [hep-ex]} \BibitemShut
  {NoStop}%
\bibitem [{\citenamefont {Akers}\ \emph {et~al.}(1995)\citenamefont {Akers}
  \emph {et~al.}}]{OPAL:1995grn}%
  \BibitemOpen
  \bibfield  {author} {\bibinfo {author} {\bibfnamefont {R.}~\bibnamefont
  {Akers}} \emph {et~al.} (\bibinfo {collaboration} {OPAL}),\ }\bibfield
  {title} {\bibinfo {title} {{A Search for lepton flavor violating Z0
  decays}},\ }\href {https://doi.org/10.1007/BF01553981} {\bibfield  {journal}
  {\bibinfo  {journal} {Z. Phys. C}\ }\textbf {\bibinfo {volume} {67}},\
  \bibinfo {pages} {555} (\bibinfo {year} {1995})}\BibitemShut {NoStop}%
\bibitem [{\citenamefont {Abreu}\ \emph {et~al.}(1997)\citenamefont {Abreu}
  \emph {et~al.}}]{DELPHI:1996iox}%
  \BibitemOpen
  \bibfield  {author} {\bibinfo {author} {\bibfnamefont {P.}~\bibnamefont
  {Abreu}} \emph {et~al.} (\bibinfo {collaboration} {DELPHI}),\ }\bibfield
  {title} {\bibinfo {title} {{Search for lepton flavor number violating Z0
  decays}},\ }\href {https://doi.org/10.1007/s002880050313} {\bibfield
  {journal} {\bibinfo  {journal} {Z. Phys. C}\ }\textbf {\bibinfo {volume}
  {73}},\ \bibinfo {pages} {243} (\bibinfo {year} {1997})}\BibitemShut
  {NoStop}%
\bibitem [{\citenamefont {Aad}\ \emph {et~al.}(2014)\citenamefont {Aad} \emph
  {et~al.}}]{ATLAS:2014vur}%
  \BibitemOpen
  \bibfield  {author} {\bibinfo {author} {\bibfnamefont {G.}~\bibnamefont
  {Aad}} \emph {et~al.} (\bibinfo {collaboration} {ATLAS}),\ }\bibfield
  {title} {\bibinfo {title} {{Search for the lepton flavor violating decay
  Z\textrightarrow{}e\ensuremath{\mu} in pp collisions at $\sqrt{s}$ TeV with
  the ATLAS detector}},\ }\href {https://doi.org/10.1103/PhysRevD.90.072010}
  {\bibfield  {journal} {\bibinfo  {journal} {Phys. Rev. D}\ }\textbf {\bibinfo
  {volume} {90}},\ \bibinfo {pages} {072010} (\bibinfo {year} {2014})},\
  \Eprint {https://arxiv.org/abs/1408.5774} {arXiv:1408.5774 [hep-ex]}
  \BibitemShut {NoStop}%
\bibitem [{\citenamefont {Aad}\ \emph {et~al.}(2021)\citenamefont {Aad} \emph
  {et~al.}}]{ATLAS:2020zlz}%
  \BibitemOpen
  \bibfield  {author} {\bibinfo {author} {\bibfnamefont {G.}~\bibnamefont
  {Aad}} \emph {et~al.} (\bibinfo {collaboration} {ATLAS}),\ }\bibfield
  {title} {\bibinfo {title} {{Search for charged-lepton-flavour violation in
  $Z$-boson decays with the ATLAS detector}},\ }\href
  {https://doi.org/10.1038/s41567-021-01225-z} {\bibfield  {journal} {\bibinfo
  {journal} {Nature Phys.}\ }\textbf {\bibinfo {volume} {17}},\ \bibinfo
  {pages} {819} (\bibinfo {year} {2021})},\ \Eprint
  {https://arxiv.org/abs/2010.02566} {arXiv:2010.02566 [hep-ex]} \BibitemShut
  {NoStop}%
\bibitem [{\citenamefont {Aad}\ \emph {et~al.}(2022{\natexlab{b}})\citenamefont
  {Aad} \emph {et~al.}}]{ATLAS:2021bdj}%
  \BibitemOpen
  \bibfield  {author} {\bibinfo {author} {\bibfnamefont {G.}~\bibnamefont
  {Aad}} \emph {et~al.} (\bibinfo {collaboration} {ATLAS}),\ }\bibfield
  {title} {\bibinfo {title} {{Search for lepton-flavor-violation in $Z$-boson
  decays with $\tau$-leptons with the ATLAS detector}},\ }\href
  {https://doi.org/10.1103/PhysRevLett.127.271801} {\bibfield  {journal}
  {\bibinfo  {journal} {Phys. Rev. Lett.}\ }\textbf {\bibinfo {volume} {127}},\
  \bibinfo {pages} {271801} (\bibinfo {year} {2022}{\natexlab{b}})},\ \Eprint
  {https://arxiv.org/abs/2105.12491} {arXiv:2105.12491 [hep-ex]} \BibitemShut
  {NoStop}%
\bibitem [{\citenamefont {Abbiendi}\ \emph {et~al.}(2001)\citenamefont
  {Abbiendi} \emph {et~al.}}]{OPAL:2001qhh}%
  \BibitemOpen
  \bibfield  {author} {\bibinfo {author} {\bibfnamefont {G.}~\bibnamefont
  {Abbiendi}} \emph {et~al.} (\bibinfo {collaboration} {OPAL}),\ }\bibfield
  {title} {\bibinfo {title} {{Search for lepton flavor violation in e+ e-
  collisions at s**(1/2) = 189-GeV - 209-GeV}},\ }\href
  {https://doi.org/10.1016/S0370-2693(01)01086-3} {\bibfield  {journal}
  {\bibinfo  {journal} {Phys. Lett. B}\ }\textbf {\bibinfo {volume} {519}},\
  \bibinfo {pages} {23} (\bibinfo {year} {2001})},\ \Eprint
  {https://arxiv.org/abs/hep-ex/0109011} {arXiv:hep-ex/0109011} \BibitemShut
  {NoStop}%
\bibitem [{\citenamefont {Han}\ \emph {et~al.}(2010)\citenamefont {Han},
  \citenamefont {Lewis},\ and\ \citenamefont {Sher}}]{Han:2010sa}%
  \BibitemOpen
  \bibfield  {author} {\bibinfo {author} {\bibfnamefont {T.}~\bibnamefont
  {Han}}, \bibinfo {author} {\bibfnamefont {I.}~\bibnamefont {Lewis}},\ and\
  \bibinfo {author} {\bibfnamefont {M.}~\bibnamefont {Sher}},\ }\bibfield
  {title} {\bibinfo {title} {{Mu-Tau Production at Hadron Colliders}},\ }\href
  {https://doi.org/10.1007/JHEP03(2010)090} {\bibfield  {journal} {\bibinfo
  {journal} {JHEP}\ }\textbf {\bibinfo {volume} {03}},\ \bibinfo {pages}
  {090}},\ \Eprint {https://arxiv.org/abs/1001.0022} {arXiv:1001.0022 [hep-ph]}
  \BibitemShut {NoStop}%
\bibitem [{\citenamefont {Davidson}\ \emph {et~al.}(2012)\citenamefont
  {Davidson}, \citenamefont {Lacroix},\ and\ \citenamefont
  {Verdier}}]{Davidson:2012wn}%
  \BibitemOpen
  \bibfield  {author} {\bibinfo {author} {\bibfnamefont {S.}~\bibnamefont
  {Davidson}}, \bibinfo {author} {\bibfnamefont {S.}~\bibnamefont {Lacroix}},\
  and\ \bibinfo {author} {\bibfnamefont {P.}~\bibnamefont {Verdier}},\
  }\bibfield  {title} {\bibinfo {title} {{LHC sensitivity to lepton flavour
  violating Z boson decays}},\ }\href {https://doi.org/10.1007/JHEP09(2012)092}
  {\bibfield  {journal} {\bibinfo  {journal} {JHEP}\ }\textbf {\bibinfo
  {volume} {09}},\ \bibinfo {pages} {092}},\ \Eprint
  {https://arxiv.org/abs/1207.4894} {arXiv:1207.4894 [hep-ph]} \BibitemShut
  {NoStop}%
\bibitem [{\citenamefont {Cai}\ and\ \citenamefont
  {Schmidt}(2016)}]{Cai:2015poa}%
  \BibitemOpen
  \bibfield  {author} {\bibinfo {author} {\bibfnamefont {Y.}~\bibnamefont
  {Cai}}\ and\ \bibinfo {author} {\bibfnamefont {M.~A.}\ \bibnamefont
  {Schmidt}},\ }\bibfield  {title} {\bibinfo {title} {{A Case Study of the
  Sensitivity to LFV Operators with Precision Measurements and the LHC}},\
  }\href {https://doi.org/10.1007/JHEP02(2016)176} {\bibfield  {journal}
  {\bibinfo  {journal} {JHEP}\ }\textbf {\bibinfo {volume} {02}},\ \bibinfo
  {pages} {176}},\ \Eprint {https://arxiv.org/abs/1510.02486} {arXiv:1510.02486
  [hep-ph]} \BibitemShut {NoStop}%
\bibitem [{\citenamefont {Cai}\ \emph {et~al.}(2018)\citenamefont {Cai},
  \citenamefont {Schmidt},\ and\ \citenamefont {Valencia}}]{Cai:2018cog}%
  \BibitemOpen
  \bibfield  {author} {\bibinfo {author} {\bibfnamefont {Y.}~\bibnamefont
  {Cai}}, \bibinfo {author} {\bibfnamefont {M.~A.}\ \bibnamefont {Schmidt}},\
  and\ \bibinfo {author} {\bibfnamefont {G.}~\bibnamefont {Valencia}},\
  }\bibfield  {title} {\bibinfo {title} {{Lepton-flavour-violating gluonic
  operators: constraints from the LHC and low energy experiments}},\ }\href
  {https://doi.org/10.1007/JHEP05(2018)143} {\bibfield  {journal} {\bibinfo
  {journal} {JHEP}\ }\textbf {\bibinfo {volume} {05}},\ \bibinfo {pages}
  {143}},\ \Eprint {https://arxiv.org/abs/1802.09822} {arXiv:1802.09822
  [hep-ph]} \BibitemShut {NoStop}%
\bibitem [{\citenamefont {Angelescu}\ \emph {et~al.}(2020)\citenamefont
  {Angelescu}, \citenamefont {Faroughy},\ and\ \citenamefont
  {Sumensari}}]{Angelescu:2020uug}%
  \BibitemOpen
  \bibfield  {author} {\bibinfo {author} {\bibfnamefont {A.}~\bibnamefont
  {Angelescu}}, \bibinfo {author} {\bibfnamefont {D.~A.}\ \bibnamefont
  {Faroughy}},\ and\ \bibinfo {author} {\bibfnamefont {O.}~\bibnamefont
  {Sumensari}},\ }\bibfield  {title} {\bibinfo {title} {{Lepton Flavor
  Violation and Dilepton Tails at the LHC}},\ }\href
  {https://doi.org/10.1140/epjc/s10052-020-8210-5} {\bibfield  {journal}
  {\bibinfo  {journal} {Eur. Phys. J. C}\ }\textbf {\bibinfo {volume} {80}},\
  \bibinfo {pages} {641} (\bibinfo {year} {2020})},\ \Eprint
  {https://arxiv.org/abs/2002.05684} {arXiv:2002.05684 [hep-ph]} \BibitemShut
  {NoStop}%
\bibitem [{\citenamefont {Murakami}\ and\ \citenamefont
  {Tait}(2015)}]{Murakami:2014tna}%
  \BibitemOpen
  \bibfield  {author} {\bibinfo {author} {\bibfnamefont {B.}~\bibnamefont
  {Murakami}}\ and\ \bibinfo {author} {\bibfnamefont {T.~M.~P.}\ \bibnamefont
  {Tait}},\ }\bibfield  {title} {\bibinfo {title} {{Searching for lepton flavor
  violation at a future high energy e+e- collider}},\ }\href
  {https://doi.org/10.1103/PhysRevD.91.015002} {\bibfield  {journal} {\bibinfo
  {journal} {Phys. Rev. D}\ }\textbf {\bibinfo {volume} {91}},\ \bibinfo
  {pages} {015002} (\bibinfo {year} {2015})},\ \Eprint
  {https://arxiv.org/abs/1410.1485} {arXiv:1410.1485 [hep-ph]} \BibitemShut
  {NoStop}%
\bibitem [{\citenamefont {Cho}\ \emph {et~al.}(2019)\citenamefont {Cho},
  \citenamefont {Fukuda},\ and\ \citenamefont {Kono}}]{Cho:2018mro}%
  \BibitemOpen
  \bibfield  {author} {\bibinfo {author} {\bibfnamefont {G.-C.}\ \bibnamefont
  {Cho}}, \bibinfo {author} {\bibfnamefont {Y.}~\bibnamefont {Fukuda}},\ and\
  \bibinfo {author} {\bibfnamefont {T.}~\bibnamefont {Kono}},\ }\bibfield
  {title} {\bibinfo {title} {{Lepton flavor violation via four-Fermi contact
  interactions at the International Linear Collider}},\ }\href
  {https://doi.org/10.1016/j.physletb.2018.12.056} {\bibfield  {journal}
  {\bibinfo  {journal} {Phys. Lett. B}\ }\textbf {\bibinfo {volume} {789}},\
  \bibinfo {pages} {399} (\bibinfo {year} {2019})},\ \Eprint
  {https://arxiv.org/abs/1803.10475} {arXiv:1803.10475 [hep-ph]} \BibitemShut
  {NoStop}%
\bibitem [{\citenamefont {Etesami}\ \emph {et~al.}(2021)\citenamefont
  {Etesami}, \citenamefont {Jafari}, \citenamefont {Najafabadi},\ and\
  \citenamefont {Tizchang}}]{Etesami:2021hex}%
  \BibitemOpen
  \bibfield  {author} {\bibinfo {author} {\bibfnamefont {S.~M.}\ \bibnamefont
  {Etesami}}, \bibinfo {author} {\bibfnamefont {R.}~\bibnamefont {Jafari}},
  \bibinfo {author} {\bibfnamefont {M.~M.}\ \bibnamefont {Najafabadi}},\ and\
  \bibinfo {author} {\bibfnamefont {S.}~\bibnamefont {Tizchang}},\ }\bibfield
  {title} {\bibinfo {title} {{Searching for lepton flavor violating
  interactions at future electron-positron colliders}},\ }\href
  {https://doi.org/10.1103/PhysRevD.104.015034} {\bibfield  {journal} {\bibinfo
   {journal} {Phys. Rev. D}\ }\textbf {\bibinfo {volume} {104}},\ \bibinfo
  {pages} {015034} (\bibinfo {year} {2021})},\ \Eprint
  {https://arxiv.org/abs/2107.00545} {arXiv:2107.00545 [hep-ph]} \BibitemShut
  {NoStop}%
\bibitem [{\citenamefont {Altmannshofer}\ \emph {et~al.}(2023)\citenamefont
  {Altmannshofer}, \citenamefont {Munbodh},\ and\ \citenamefont
  {Oh}}]{Altmannshofer:2023tsa}%
  \BibitemOpen
  \bibfield  {author} {\bibinfo {author} {\bibfnamefont {W.}~\bibnamefont
  {Altmannshofer}}, \bibinfo {author} {\bibfnamefont {P.}~\bibnamefont
  {Munbodh}},\ and\ \bibinfo {author} {\bibfnamefont {T.}~\bibnamefont {Oh}},\
  }\bibfield  {title} {\bibinfo {title} {{Probing lepton flavor violation at
  Circular Electron-Positron Colliders}},\ }\href
  {https://doi.org/10.1007/JHEP08(2023)026} {\bibfield  {journal} {\bibinfo
  {journal} {JHEP}\ }\textbf {\bibinfo {volume} {08}},\ \bibinfo {pages}
  {026}},\ \Eprint {https://arxiv.org/abs/2305.03869} {arXiv:2305.03869
  [hep-ph]} \BibitemShut {NoStop}%
\bibitem [{\citenamefont {Altmannshofer}\ and\ \citenamefont
  {Munbodh}(2025)}]{Altmannshofer:2025nbp}%
  \BibitemOpen
  \bibfield  {author} {\bibinfo {author} {\bibfnamefont {W.}~\bibnamefont
  {Altmannshofer}}\ and\ \bibinfo {author} {\bibfnamefont {P.}~\bibnamefont
  {Munbodh}},\ }\bibfield  {title} {\bibinfo {title} {{Probing Lepton Flavor
  Violation at Linear Electron-Positron Colliders}},\ }\href@noop {} {\
  (\bibinfo {year} {2025})},\ \Eprint {https://arxiv.org/abs/2505.11653}
  {arXiv:2505.11653 [hep-ph]} \BibitemShut {NoStop}%
\bibitem [{\citenamefont {Moreno-S\'anchez}\ and\ \citenamefont
  {Palavri\'c}(2025)}]{Moreno-Sanchez:2025bzz}%
  \BibitemOpen
  \bibfield  {author} {\bibinfo {author} {\bibfnamefont {A.}~\bibnamefont
  {Moreno-S\'anchez}}\ and\ \bibinfo {author} {\bibfnamefont {A.}~\bibnamefont
  {Palavri\'c}},\ }\bibfield  {title} {\bibinfo {title} {{Leptonic Flavor from
  Modular $A_4$: UV Mediators and SMEFT Realizations}},\ }\href@noop {} {\
  (\bibinfo {year} {2025})},\ \Eprint {https://arxiv.org/abs/2505.01535}
  {arXiv:2505.01535 [hep-ph]} \BibitemShut {NoStop}%
\bibitem [{\citenamefont {Palavri\'c}(2024)}]{Palavric:2024gvu}%
  \BibitemOpen
  \bibfield  {author} {\bibinfo {author} {\bibfnamefont {A.}~\bibnamefont
  {Palavri\'c}},\ }\bibfield  {title} {\bibinfo {title} {{Discrete leptonic
  flavor symmetries: UV mediators and phenomenology}},\ }\href
  {https://doi.org/10.1103/PhysRevD.110.115025} {\bibfield  {journal} {\bibinfo
   {journal} {Phys. Rev. D}\ }\textbf {\bibinfo {volume} {110}},\ \bibinfo
  {pages} {115025} (\bibinfo {year} {2024})},\ \Eprint
  {https://arxiv.org/abs/2408.16044} {arXiv:2408.16044 [hep-ph]} \BibitemShut
  {NoStop}%
\bibitem [{\citenamefont {Calibbi}\ \emph {et~al.}(2025)\citenamefont
  {Calibbi}, \citenamefont {Hagedorn}, \citenamefont {Schmidt},\ and\
  \citenamefont {Vandeleur}}]{Calibbi:2025fzi}%
  \BibitemOpen
  \bibfield  {author} {\bibinfo {author} {\bibfnamefont {L.}~\bibnamefont
  {Calibbi}}, \bibinfo {author} {\bibfnamefont {C.}~\bibnamefont {Hagedorn}},
  \bibinfo {author} {\bibfnamefont {M.~A.}\ \bibnamefont {Schmidt}},\ and\
  \bibinfo {author} {\bibfnamefont {J.}~\bibnamefont {Vandeleur}},\ }\bibfield
  {title} {\bibinfo {title} {{Selection rules for charged lepton flavour
  violating processes from residual flavour groups}},\ }\href@noop {} {\
  (\bibinfo {year} {2025})},\ \Eprint {https://arxiv.org/abs/2505.24350}
  {arXiv:2505.24350 [hep-ph]} \BibitemShut {NoStop}%
\bibitem [{\citenamefont {Kriewald}\ \emph {et~al.}(2025)\citenamefont
  {Kriewald}, \citenamefont {Pinsard},\ and\ \citenamefont
  {Teixeira}}]{Kriewald:2024cnt}%
  \BibitemOpen
  \bibfield  {author} {\bibinfo {author} {\bibfnamefont {J.}~\bibnamefont
  {Kriewald}}, \bibinfo {author} {\bibfnamefont {E.}~\bibnamefont {Pinsard}},\
  and\ \bibinfo {author} {\bibfnamefont {A.~M.}\ \bibnamefont {Teixeira}},\
  }\bibfield  {title} {\bibinfo {title} {{High-energy cLFV at
  \ensuremath{\mu}TRISTAN: HNL extensions of the Standard Model}},\ }\href
  {https://doi.org/10.1007/JHEP02(2025)116} {\bibfield  {journal} {\bibinfo
  {journal} {JHEP}\ }\textbf {\bibinfo {volume} {02}},\ \bibinfo {pages}
  {116}},\ \Eprint {https://arxiv.org/abs/2412.04331} {arXiv:2412.04331
  [hep-ph]} \BibitemShut {NoStop}%
\bibitem [{\citenamefont {Batell}\ \emph {et~al.}(2024)\citenamefont {Batell},
  \citenamefont {Davoudiasl}, \citenamefont {Marcarelli}, \citenamefont
  {Neil},\ and\ \citenamefont {Trojanowski}}]{Batell:2024cdl}%
  \BibitemOpen
  \bibfield  {author} {\bibinfo {author} {\bibfnamefont {B.}~\bibnamefont
  {Batell}}, \bibinfo {author} {\bibfnamefont {H.}~\bibnamefont {Davoudiasl}},
  \bibinfo {author} {\bibfnamefont {R.}~\bibnamefont {Marcarelli}}, \bibinfo
  {author} {\bibfnamefont {E.~T.}\ \bibnamefont {Neil}},\ and\ \bibinfo
  {author} {\bibfnamefont {S.}~\bibnamefont {Trojanowski}},\ }\bibfield
  {title} {\bibinfo {title} {{Lepton-flavor-violating ALP signals with
  TeV-scale muon beams}},\ }\href {https://doi.org/10.1103/PhysRevD.110.075039}
  {\bibfield  {journal} {\bibinfo  {journal} {Phys. Rev. D}\ }\textbf {\bibinfo
  {volume} {110}},\ \bibinfo {pages} {075039} (\bibinfo {year} {2024})},\
  \Eprint {https://arxiv.org/abs/2407.15942} {arXiv:2407.15942 [hep-ph]}
  \BibitemShut {NoStop}%
\bibitem [{\citenamefont {Calibbi}\ \emph {et~al.}(2024)\citenamefont
  {Calibbi}, \citenamefont {Li}, \citenamefont {Mukherjee},\ and\ \citenamefont
  {Yang}}]{Calibbi:2024rcm}%
  \BibitemOpen
  \bibfield  {author} {\bibinfo {author} {\bibfnamefont {L.}~\bibnamefont
  {Calibbi}}, \bibinfo {author} {\bibfnamefont {T.}~\bibnamefont {Li}},
  \bibinfo {author} {\bibfnamefont {L.}~\bibnamefont {Mukherjee}},\ and\
  \bibinfo {author} {\bibfnamefont {Y.}~\bibnamefont {Yang}},\ }\bibfield
  {title} {\bibinfo {title} {{Probing ALP lepton flavor violation at
  \ensuremath{\mu}TRISTAN}},\ }\href
  {https://doi.org/10.1103/PhysRevD.110.115009} {\bibfield  {journal} {\bibinfo
   {journal} {Phys. Rev. D}\ }\textbf {\bibinfo {volume} {110}},\ \bibinfo
  {pages} {115009} (\bibinfo {year} {2024})},\ \Eprint
  {https://arxiv.org/abs/2406.13234} {arXiv:2406.13234 [hep-ph]} \BibitemShut
  {NoStop}%
\bibitem [{\citenamefont {Ding}\ \emph {et~al.}(2025)\citenamefont {Ding},
  \citenamefont {Li}, \citenamefont {Lu}, \citenamefont {You}, \citenamefont
  {Wang},\ and\ \citenamefont {Li}}]{Ding:2024zaj}%
  \BibitemOpen
  \bibfield  {author} {\bibinfo {author} {\bibfnamefont {R.}~\bibnamefont
  {Ding}}, \bibinfo {author} {\bibfnamefont {J.}~\bibnamefont {Li}}, \bibinfo
  {author} {\bibfnamefont {M.}~\bibnamefont {Lu}}, \bibinfo {author}
  {\bibfnamefont {Z.}~\bibnamefont {You}}, \bibinfo {author} {\bibfnamefont
  {Z.}~\bibnamefont {Wang}},\ and\ \bibinfo {author} {\bibfnamefont
  {Q.}~\bibnamefont {Li}},\ }\bibfield  {title} {\bibinfo {title} {{Study of
  charged Lepton Flavor Violation in electron muon interactions}},\ }\href
  {https://doi.org/10.1007/JHEP01(2025)165} {\bibfield  {journal} {\bibinfo
  {journal} {JHEP}\ }\textbf {\bibinfo {volume} {01}},\ \bibinfo {pages}
  {165}},\ \Eprint {https://arxiv.org/abs/2405.09417} {arXiv:2405.09417
  [hep-ex]} \BibitemShut {NoStop}%
\bibitem [{\citenamefont {Santiago}\ \emph {et~al.}(2024)\citenamefont
  {Santiago}, \citenamefont {Portillo-S\'anchez}, \citenamefont
  {Hern\'andez-Tom\'e},\ and\ \citenamefont {Rend\'on}}]{Santiago:2024zpc}%
  \BibitemOpen
  \bibfield  {author} {\bibinfo {author} {\bibfnamefont {J.~L.~G.}\
  \bibnamefont {Santiago}}, \bibinfo {author} {\bibfnamefont {D.}~\bibnamefont
  {Portillo-S\'anchez}}, \bibinfo {author} {\bibfnamefont {G.}~\bibnamefont
  {Hern\'andez-Tom\'e}},\ and\ \bibinfo {author} {\bibfnamefont
  {J.}~\bibnamefont {Rend\'on}},\ }\bibfield  {title} {\bibinfo {title}
  {{Authentic Majorana versus singlet Dirac neutrino contributions to
  \ensuremath{\mu}+\ensuremath{\mu}+\textrightarrow{}\ensuremath{\ell}+\ensuremath{\ell}+
  (\ensuremath{\ell}=e,\ensuremath{\tau}) transitions}},\ }\href
  {https://doi.org/10.1103/PhysRevD.110.053006} {\bibfield  {journal} {\bibinfo
   {journal} {Phys. Rev. D}\ }\textbf {\bibinfo {volume} {110}},\ \bibinfo
  {pages} {053006} (\bibinfo {year} {2024})},\ \Eprint
  {https://arxiv.org/abs/2405.02819} {arXiv:2405.02819 [hep-ph]} \BibitemShut
  {NoStop}%
\bibitem [{\citenamefont {Heeck}\ and\ \citenamefont
  {Sokhashvili}(2024)}]{Heeck:2024uiz}%
  \BibitemOpen
  \bibfield  {author} {\bibinfo {author} {\bibfnamefont {J.}~\bibnamefont
  {Heeck}}\ and\ \bibinfo {author} {\bibfnamefont {M.}~\bibnamefont
  {Sokhashvili}},\ }\bibfield  {title} {\bibinfo {title} {{Lepton flavor
  violation by two units}},\ }\href
  {https://doi.org/10.1016/j.physletb.2024.138621} {\bibfield  {journal}
  {\bibinfo  {journal} {Phys. Lett. B}\ }\textbf {\bibinfo {volume} {852}},\
  \bibinfo {pages} {138621} (\bibinfo {year} {2024})},\ \Eprint
  {https://arxiv.org/abs/2401.09580} {arXiv:2401.09580 [hep-ph]} \BibitemShut
  {NoStop}%
\bibitem [{\citenamefont {Goudelis}\ \emph {et~al.}(2024)\citenamefont
  {Goudelis}, \citenamefont {Kriewald}, \citenamefont {Pinsard},\ and\
  \citenamefont {Teixeira}}]{Goudelis:2023yni}%
  \BibitemOpen
  \bibfield  {author} {\bibinfo {author} {\bibfnamefont {A.}~\bibnamefont
  {Goudelis}}, \bibinfo {author} {\bibfnamefont {J.}~\bibnamefont {Kriewald}},
  \bibinfo {author} {\bibfnamefont {E.}~\bibnamefont {Pinsard}},\ and\ \bibinfo
  {author} {\bibfnamefont {A.~M.}\ \bibnamefont {Teixeira}},\ }\bibfield
  {title} {\bibinfo {title} {{cLFV leptophilic $Z^\prime $ as a dark matter
  portal: prospects for colliders}},\ }\href
  {https://doi.org/10.1140/epjc/s10052-024-13158-x} {\bibfield  {journal}
  {\bibinfo  {journal} {Eur. Phys. J. C}\ }\textbf {\bibinfo {volume} {84}},\
  \bibinfo {pages} {804} (\bibinfo {year} {2024})},\ \Eprint
  {https://arxiv.org/abs/2312.14103} {arXiv:2312.14103 [hep-ph]} \BibitemShut
  {NoStop}%
\bibitem [{\citenamefont {Lichtenstein}\ \emph {et~al.}(2023)\citenamefont
  {Lichtenstein}, \citenamefont {Schmidt}, \citenamefont {Valencia},\ and\
  \citenamefont {Volkas}}]{Lichtenstein:2023iut}%
  \BibitemOpen
  \bibfield  {author} {\bibinfo {author} {\bibfnamefont {G.}~\bibnamefont
  {Lichtenstein}}, \bibinfo {author} {\bibfnamefont {M.~A.}\ \bibnamefont
  {Schmidt}}, \bibinfo {author} {\bibfnamefont {G.}~\bibnamefont {Valencia}},\
  and\ \bibinfo {author} {\bibfnamefont {R.~R.}\ \bibnamefont {Volkas}},\
  }\bibfield  {title} {\bibinfo {title} {{Complementarity of $\mu$TRISTAN and
  Belle II in searches for charged-lepton flavour violation}},\ }\href
  {https://doi.org/10.1016/j.physletb.2023.138144} {\bibfield  {journal}
  {\bibinfo  {journal} {Phys. Lett. B}\ }\textbf {\bibinfo {volume} {845}},\
  \bibinfo {pages} {138144} (\bibinfo {year} {2023})},\ \Eprint
  {https://arxiv.org/abs/2307.11369} {arXiv:2307.11369 [hep-ph]} \BibitemShut
  {NoStop}%
\bibitem [{\citenamefont {Jahedi}\ and\ \citenamefont
  {Sarkar}(2024)}]{Jahedi:2024kvi}%
  \BibitemOpen
  \bibfield  {author} {\bibinfo {author} {\bibfnamefont {S.}~\bibnamefont
  {Jahedi}}\ and\ \bibinfo {author} {\bibfnamefont {A.}~\bibnamefont
  {Sarkar}},\ }\bibfield  {title} {\bibinfo {title} {{Exploring optimal
  sensitivity of lepton flavor violating effective couplings at the e+e-
  colliders}},\ }\href {https://doi.org/10.1103/PhysRevD.110.095021} {\bibfield
   {journal} {\bibinfo  {journal} {Phys. Rev. D}\ }\textbf {\bibinfo {volume}
  {110}},\ \bibinfo {pages} {095021} (\bibinfo {year} {2024})},\ \Eprint
  {https://arxiv.org/abs/2408.00190} {arXiv:2408.00190 [hep-ph]} \BibitemShut
  {NoStop}%
\bibitem [{\citenamefont {Hamada}\ \emph {et~al.}(2022)\citenamefont {Hamada},
  \citenamefont {Kitano}, \citenamefont {Matsudo}, \citenamefont {Takaura},\
  and\ \citenamefont {Yoshida}}]{Hamada:2022mua}%
  \BibitemOpen
  \bibfield  {author} {\bibinfo {author} {\bibfnamefont {Y.}~\bibnamefont
  {Hamada}}, \bibinfo {author} {\bibfnamefont {R.}~\bibnamefont {Kitano}},
  \bibinfo {author} {\bibfnamefont {R.}~\bibnamefont {Matsudo}}, \bibinfo
  {author} {\bibfnamefont {H.}~\bibnamefont {Takaura}},\ and\ \bibinfo {author}
  {\bibfnamefont {M.}~\bibnamefont {Yoshida}},\ }\bibfield  {title} {\bibinfo
  {title} {{$\mu$TRISTAN}},\ }\href {https://doi.org/10.1093/ptep/ptac059}
  {\bibfield  {journal} {\bibinfo  {journal} {PTEP}\ }\textbf {\bibinfo
  {volume} {2022}},\ \bibinfo {pages} {053B02} (\bibinfo {year} {2022})},\
  \Eprint {https://arxiv.org/abs/2201.06664} {arXiv:2201.06664 [hep-ph]}
  \BibitemShut {NoStop}%
\bibitem [{\citenamefont {Chen}\ \emph
  {et~al.}(2024{\natexlab{a}})\citenamefont {Chen}, \citenamefont {Chiang},\
  and\ \citenamefont {Su}}]{Chen:2023eof}%
  \BibitemOpen
  \bibfield  {author} {\bibinfo {author} {\bibfnamefont {C.-H.}\ \bibnamefont
  {Chen}}, \bibinfo {author} {\bibfnamefont {C.-W.}\ \bibnamefont {Chiang}},\
  and\ \bibinfo {author} {\bibfnamefont {C.-W.}\ \bibnamefont {Su}},\
  }\bibfield  {title} {\bibinfo {title} {{Top-quark FCNC decays, LFVs, lepton g
  {\ensuremath{-}} 2, and W mass anomaly with inert charged Higgses}},\ }\href
  {https://doi.org/10.1088/1361-6471/ad560e} {\bibfield  {journal} {\bibinfo
  {journal} {J. Phys. G}\ }\textbf {\bibinfo {volume} {51}},\ \bibinfo {pages}
  {085001} (\bibinfo {year} {2024}{\natexlab{a}})},\ \Eprint
  {https://arxiv.org/abs/2301.07070} {arXiv:2301.07070 [hep-ph]} \BibitemShut
  {NoStop}%
\bibitem [{\citenamefont {Crivellin}\ \emph {et~al.}(2013)\citenamefont
  {Crivellin}, \citenamefont {Kokulu},\ and\ \citenamefont
  {Greub}}]{Crivellin:2013wna}%
  \BibitemOpen
  \bibfield  {author} {\bibinfo {author} {\bibfnamefont {A.}~\bibnamefont
  {Crivellin}}, \bibinfo {author} {\bibfnamefont {A.}~\bibnamefont {Kokulu}},\
  and\ \bibinfo {author} {\bibfnamefont {C.}~\bibnamefont {Greub}},\ }\bibfield
   {title} {\bibinfo {title} {{Flavor-phenomenology of two-Higgs-doublet models
  with generic Yukawa structure}},\ }\href
  {https://doi.org/10.1103/PhysRevD.87.094031} {\bibfield  {journal} {\bibinfo
  {journal} {Phys. Rev. D}\ }\textbf {\bibinfo {volume} {87}},\ \bibinfo
  {pages} {094031} (\bibinfo {year} {2013})},\ \Eprint
  {https://arxiv.org/abs/1303.5877} {arXiv:1303.5877 [hep-ph]} \BibitemShut
  {NoStop}%
\bibitem [{\citenamefont {Gon{\c{c}}alves}\ \emph {et~al.}(2023)\citenamefont
  {Gon{\c{c}}alves}, \citenamefont {Knauss},\ and\ \citenamefont
  {Sher}}]{Goncalves:2023ydf}%
  \BibitemOpen
  \bibfield  {author} {\bibinfo {author} {\bibfnamefont {B.~L.}\ \bibnamefont
  {Gon{\c{c}}alves}}, \bibinfo {author} {\bibfnamefont {M.}~\bibnamefont
  {Knauss}},\ and\ \bibinfo {author} {\bibfnamefont {M.}~\bibnamefont {Sher}},\
  }\bibfield  {title} {\bibinfo {title} {{Lepton flavor specific extended Higgs
  model}},\ }\href {https://doi.org/10.1103/PhysRevD.107.095001} {\bibfield
  {journal} {\bibinfo  {journal} {Phys. Rev. D}\ }\textbf {\bibinfo {volume}
  {107}},\ \bibinfo {pages} {095001} (\bibinfo {year} {2023})},\ \Eprint
  {https://arxiv.org/abs/2301.08641} {arXiv:2301.08641 [hep-ph]} \BibitemShut
  {NoStop}%
\bibitem [{\citenamefont {Dor{\v{s}}ner}\ \emph {et~al.}(2016)\citenamefont
  {Dor{\v{s}}ner}, \citenamefont {Fajfer}, \citenamefont {Greljo},
  \citenamefont {Kamenik},\ and\ \citenamefont
  {Ko{\v{s}}nik}}]{Dorsner:2016wpm}%
  \BibitemOpen
  \bibfield  {author} {\bibinfo {author} {\bibfnamefont {I.}~\bibnamefont
  {Dor{\v{s}}ner}}, \bibinfo {author} {\bibfnamefont {S.}~\bibnamefont
  {Fajfer}}, \bibinfo {author} {\bibfnamefont {A.}~\bibnamefont {Greljo}},
  \bibinfo {author} {\bibfnamefont {J.~F.}\ \bibnamefont {Kamenik}},\ and\
  \bibinfo {author} {\bibfnamefont {N.}~\bibnamefont {Ko{\v{s}}nik}},\
  }\bibfield  {title} {\bibinfo {title} {{Physics of leptoquarks in precision
  experiments and at particle colliders}},\ }\href
  {https://doi.org/10.1016/j.physrep.2016.06.001} {\bibfield  {journal}
  {\bibinfo  {journal} {Phys. Rept.}\ }\textbf {\bibinfo {volume} {641}},\
  \bibinfo {pages} {1} (\bibinfo {year} {2016})},\ \Eprint
  {https://arxiv.org/abs/1603.04993} {arXiv:1603.04993 [hep-ph]} \BibitemShut
  {NoStop}%
\bibitem [{\citenamefont {Davidson}\ \emph {et~al.}(1994)\citenamefont
  {Davidson}, \citenamefont {Bailey},\ and\ \citenamefont
  {Campbell}}]{Davidson:1993qk}%
  \BibitemOpen
  \bibfield  {author} {\bibinfo {author} {\bibfnamefont {S.}~\bibnamefont
  {Davidson}}, \bibinfo {author} {\bibfnamefont {D.~C.}\ \bibnamefont
  {Bailey}},\ and\ \bibinfo {author} {\bibfnamefont {B.~A.}\ \bibnamefont
  {Campbell}},\ }\bibfield  {title} {\bibinfo {title} {{Model independent
  constraints on leptoquarks from rare processes}},\ }\href
  {https://doi.org/10.1007/BF01552629} {\bibfield  {journal} {\bibinfo
  {journal} {Z. Phys. C}\ }\textbf {\bibinfo {volume} {61}},\ \bibinfo {pages}
  {613} (\bibinfo {year} {1994})},\ \Eprint
  {https://arxiv.org/abs/hep-ph/9309310} {arXiv:hep-ph/9309310} \BibitemShut
  {NoStop}%
\bibitem [{\citenamefont {Langacker}(2009)}]{Langacker:2008yv}%
  \BibitemOpen
  \bibfield  {author} {\bibinfo {author} {\bibfnamefont {P.}~\bibnamefont
  {Langacker}},\ }\bibfield  {title} {\bibinfo {title} {{The Physics of Heavy
  $Z^\prime$ Gauge Bosons}},\ }\href
  {https://doi.org/10.1103/RevModPhys.81.1199} {\bibfield  {journal} {\bibinfo
  {journal} {Rev. Mod. Phys.}\ }\textbf {\bibinfo {volume} {81}},\ \bibinfo
  {pages} {1199} (\bibinfo {year} {2009})},\ \Eprint
  {https://arxiv.org/abs/0801.1345} {arXiv:0801.1345 [hep-ph]} \BibitemShut
  {NoStop}%
\bibitem [{\citenamefont {Langacker}\ and\ \citenamefont
  {Plumacher}(2000)}]{Langacker:2000ju}%
  \BibitemOpen
  \bibfield  {author} {\bibinfo {author} {\bibfnamefont {P.}~\bibnamefont
  {Langacker}}\ and\ \bibinfo {author} {\bibfnamefont {M.}~\bibnamefont
  {Plumacher}},\ }\bibfield  {title} {\bibinfo {title} {{Flavor changing
  effects in theories with a heavy $Z^\prime$ boson with family nonuniversal
  couplings}},\ }\href {https://doi.org/10.1103/PhysRevD.62.013006} {\bibfield
  {journal} {\bibinfo  {journal} {Phys. Rev. D}\ }\textbf {\bibinfo {volume}
  {62}},\ \bibinfo {pages} {013006} (\bibinfo {year} {2000})},\ \Eprint
  {https://arxiv.org/abs/hep-ph/0001204} {arXiv:hep-ph/0001204} \BibitemShut
  {NoStop}%
\bibitem [{\citenamefont {Agashe}\ \emph {et~al.}(2005)\citenamefont {Agashe},
  \citenamefont {Contino},\ and\ \citenamefont {Pomarol}}]{Agashe:2004rs}%
  \BibitemOpen
  \bibfield  {author} {\bibinfo {author} {\bibfnamefont {K.}~\bibnamefont
  {Agashe}}, \bibinfo {author} {\bibfnamefont {R.}~\bibnamefont {Contino}},\
  and\ \bibinfo {author} {\bibfnamefont {A.}~\bibnamefont {Pomarol}},\
  }\bibfield  {title} {\bibinfo {title} {{The Minimal composite Higgs model}},\
  }\href {https://doi.org/10.1016/j.nuclphysb.2005.04.035} {\bibfield
  {journal} {\bibinfo  {journal} {Nucl. Phys. B}\ }\textbf {\bibinfo {volume}
  {719}},\ \bibinfo {pages} {165} (\bibinfo {year} {2005})},\ \Eprint
  {https://arxiv.org/abs/hep-ph/0412089} {arXiv:hep-ph/0412089} \BibitemShut
  {NoStop}%
\bibitem [{\citenamefont {Feruglio}\ \emph {et~al.}(2015)\citenamefont
  {Feruglio}, \citenamefont {Paradisi},\ and\ \citenamefont
  {Pattori}}]{Feruglio:2015gka}%
  \BibitemOpen
  \bibfield  {author} {\bibinfo {author} {\bibfnamefont {F.}~\bibnamefont
  {Feruglio}}, \bibinfo {author} {\bibfnamefont {P.}~\bibnamefont {Paradisi}},\
  and\ \bibinfo {author} {\bibfnamefont {A.}~\bibnamefont {Pattori}},\
  }\bibfield  {title} {\bibinfo {title} {{Lepton Flavour Violation in Composite
  Higgs Models}},\ }\href {https://doi.org/10.1140/epjc/s10052-015-3807-9}
  {\bibfield  {journal} {\bibinfo  {journal} {Eur. Phys. J. C}\ }\textbf
  {\bibinfo {volume} {75}},\ \bibinfo {pages} {579} (\bibinfo {year} {2015})},\
  \Eprint {https://arxiv.org/abs/1509.03241} {arXiv:1509.03241 [hep-ph]}
  \BibitemShut {NoStop}%
\bibitem [{\citenamefont {Altmannshofer}\ \emph {et~al.}(2025)\citenamefont
  {Altmannshofer}, \citenamefont {Balme}, \citenamefont {Donohue},
  \citenamefont {Gori},\ and\ \citenamefont
  {Mukundhan}}]{Altmannshofer:2025lun}%
  \BibitemOpen
  \bibfield  {author} {\bibinfo {author} {\bibfnamefont {W.}~\bibnamefont
  {Altmannshofer}}, \bibinfo {author} {\bibfnamefont {Z.}~\bibnamefont
  {Balme}}, \bibinfo {author} {\bibfnamefont {C.~M.}\ \bibnamefont {Donohue}},
  \bibinfo {author} {\bibfnamefont {S.}~\bibnamefont {Gori}},\ and\ \bibinfo
  {author} {\bibfnamefont {S.~V.}\ \bibnamefont {Mukundhan}},\ }\bibfield
  {title} {\bibinfo {title} {{Targets for Flavor-Violating Top Decay}},\
  }\href@noop {} {\  (\bibinfo {year} {2025})},\ \Eprint
  {https://arxiv.org/abs/2504.18664} {arXiv:2504.18664 [hep-ph]} \BibitemShut
  {NoStop}%
\bibitem [{\citenamefont {Lu}\ \emph {et~al.}(2021)\citenamefont {Lu},
  \citenamefont {Levin}, \citenamefont {Li}, \citenamefont {Agapitos},
  \citenamefont {Li}, \citenamefont {Meng}, \citenamefont {Qian}, \citenamefont
  {Xiao},\ and\ \citenamefont {Yang}}]{Lu:2020dkx}%
  \BibitemOpen
  \bibfield  {author} {\bibinfo {author} {\bibfnamefont {M.}~\bibnamefont
  {Lu}}, \bibinfo {author} {\bibfnamefont {A.~M.}\ \bibnamefont {Levin}},
  \bibinfo {author} {\bibfnamefont {C.}~\bibnamefont {Li}}, \bibinfo {author}
  {\bibfnamefont {A.}~\bibnamefont {Agapitos}}, \bibinfo {author}
  {\bibfnamefont {Q.}~\bibnamefont {Li}}, \bibinfo {author} {\bibfnamefont
  {F.}~\bibnamefont {Meng}}, \bibinfo {author} {\bibfnamefont {S.}~\bibnamefont
  {Qian}}, \bibinfo {author} {\bibfnamefont {J.}~\bibnamefont {Xiao}},\ and\
  \bibinfo {author} {\bibfnamefont {T.}~\bibnamefont {Yang}},\ }\bibfield
  {title} {\bibinfo {title} {{The physics case for an electron-muon
  collider}},\ }\href {https://doi.org/10.1155/2021/6693618} {\bibfield
  {journal} {\bibinfo  {journal} {Adv. High Energy Phys.}\ }\textbf {\bibinfo
  {volume} {2021}},\ \bibinfo {pages} {6693618} (\bibinfo {year} {2021})},\
  \Eprint {https://arxiv.org/abs/2010.15144} {arXiv:2010.15144 [hep-ph]}
  \BibitemShut {NoStop}%
\bibitem [{\citenamefont {Bouzas}\ and\ \citenamefont
  {Larios}(2022)}]{Bouzas:2021sif}%
  \BibitemOpen
  \bibfield  {author} {\bibinfo {author} {\bibfnamefont {A.~O.}\ \bibnamefont
  {Bouzas}}\ and\ \bibinfo {author} {\bibfnamefont {F.}~\bibnamefont
  {Larios}},\ }\bibfield  {title} {\bibinfo {title} {{Two-to-Two Processes at
  an Electron-Muon Collider}},\ }\href {https://doi.org/10.1155/2022/3603613}
  {\bibfield  {journal} {\bibinfo  {journal} {Adv. High Energy Phys.}\ }\textbf
  {\bibinfo {volume} {2022}},\ \bibinfo {pages} {3603613} (\bibinfo {year}
  {2022})},\ \Eprint {https://arxiv.org/abs/2109.02769} {arXiv:2109.02769
  [hep-ph]} \BibitemShut {NoStop}%
\bibitem [{\citenamefont {Bouzas}\ and\ \citenamefont
  {Larios}(2023)}]{Bouzas:2023vba}%
  \BibitemOpen
  \bibfield  {author} {\bibinfo {author} {\bibfnamefont {A.~O.}\ \bibnamefont
  {Bouzas}}\ and\ \bibinfo {author} {\bibfnamefont {F.}~\bibnamefont
  {Larios}},\ }\bibfield  {title} {\bibinfo {title} {{An electron-muon
  collider: what can be probed with it?}},\ }\href
  {https://doi.org/10.31349/SuplRevMexFis.4.021128} {\bibfield  {journal}
  {\bibinfo  {journal} {Rev. Mex. Fis. Suppl.}\ }\textbf {\bibinfo {volume}
  {4}},\ \bibinfo {pages} {021128} (\bibinfo {year} {2023})}\BibitemShut
  {NoStop}%
\bibitem [{\citenamefont {Iwata}(1993)}]{Iwata:1993qk}%
  \BibitemOpen
  \bibfield  {author} {\bibinfo {author} {\bibfnamefont {S.}~\bibnamefont
  {Iwata}},\ }\bibfield  {title} {\bibinfo {title} {{TRISTAN experiment}},\
  }in\ \href@noop {} {\emph {\bibinfo {booktitle} {{Adriatico Research
  Conference on Mesoscopic Systems and Chaos: A Novel Approach}}}}\ (\bibinfo
  {year} {1993})\ pp.\ \bibinfo {pages} {255--275}\BibitemShut {NoStop}%
\bibitem [{\citenamefont {Dehghani}\ \emph {et~al.}(2025)\citenamefont
  {Dehghani}, \citenamefont {Frank},\ and\ \citenamefont
  {Fuks}}]{Dehghani:2025xkd}%
  \BibitemOpen
  \bibfield  {author} {\bibinfo {author} {\bibfnamefont {P.}~\bibnamefont
  {Dehghani}}, \bibinfo {author} {\bibfnamefont {M.}~\bibnamefont {Frank}},\
  and\ \bibinfo {author} {\bibfnamefont {B.}~\bibnamefont {Fuks}},\ }\bibfield
  {title} {\bibinfo {title} {{Vector Boson Fusion Signatures of Superheavy
  Majorana Neutrinos at Muon Colliders}},\ }\href@noop {} {\  (\bibinfo {year}
  {2025})},\ \Eprint {https://arxiv.org/abs/2506.06159} {arXiv:2506.06159
  [hep-ph]} \BibitemShut {NoStop}%
\bibitem [{\citenamefont {Bhattacharya}\ \emph
  {et~al.}(2025{\natexlab{b}})\citenamefont {Bhattacharya}, \citenamefont
  {Datta},\ and\ \citenamefont {Sarkar}}]{Bhattacharya:2025xwv}%
  \BibitemOpen
  \bibfield  {author} {\bibinfo {author} {\bibfnamefont {S.}~\bibnamefont
  {Bhattacharya}}, \bibinfo {author} {\bibfnamefont {S.}~\bibnamefont
  {Datta}},\ and\ \bibinfo {author} {\bibfnamefont {A.}~\bibnamefont
  {Sarkar}},\ }\bibfield  {title} {\bibinfo {title} {{Probing $\Delta L=2$
  lepton number violating SMEFT operators at the same-sign muon collider}},\
  }\href@noop {} {\  (\bibinfo {year} {2025}{\natexlab{b}})},\ \Eprint
  {https://arxiv.org/abs/2505.20936} {arXiv:2505.20936 [hep-ph]} \BibitemShut
  {NoStop}%
\bibitem [{\citenamefont {Bolton}\ \emph {et~al.}(2025)\citenamefont {Bolton},
  \citenamefont {Deppisch}, \citenamefont {Kulkarni}, \citenamefont
  {Majumdar},\ and\ \citenamefont {Pei}}]{Bolton:2025tqw}%
  \BibitemOpen
  \bibfield  {author} {\bibinfo {author} {\bibfnamefont {P.~D.}\ \bibnamefont
  {Bolton}}, \bibinfo {author} {\bibfnamefont {F.~F.}\ \bibnamefont
  {Deppisch}}, \bibinfo {author} {\bibfnamefont {S.}~\bibnamefont {Kulkarni}},
  \bibinfo {author} {\bibfnamefont {C.}~\bibnamefont {Majumdar}},\ and\
  \bibinfo {author} {\bibfnamefont {W.}~\bibnamefont {Pei}},\ }\bibfield
  {title} {\bibinfo {title} {{Constraining the SMEFT Extended with Sterile
  Neutrinos at FCC-ee}},\ }\href@noop {} {\  (\bibinfo {year} {2025})},\
  \Eprint {https://arxiv.org/abs/2502.06972} {arXiv:2502.06972 [hep-ph]}
  \BibitemShut {NoStop}%
\bibitem [{\citenamefont {de~Lima}\ \emph {et~al.}(2025)\citenamefont
  {de~Lima}, \citenamefont {McKeen}, \citenamefont {Ng}, \citenamefont
  {Shamma},\ and\ \citenamefont {Tuckler}}]{deLima:2024ohf}%
  \BibitemOpen
  \bibfield  {author} {\bibinfo {author} {\bibfnamefont {C.~H.}\ \bibnamefont
  {de~Lima}}, \bibinfo {author} {\bibfnamefont {D.}~\bibnamefont {McKeen}},
  \bibinfo {author} {\bibfnamefont {J.~N.}\ \bibnamefont {Ng}}, \bibinfo
  {author} {\bibfnamefont {M.}~\bibnamefont {Shamma}},\ and\ \bibinfo {author}
  {\bibfnamefont {D.}~\bibnamefont {Tuckler}},\ }\bibfield  {title} {\bibinfo
  {title} {{Probing lepton number violation at same-sign lepton colliders}},\
  }\href {https://doi.org/10.1103/PhysRevD.111.075002} {\bibfield  {journal}
  {\bibinfo  {journal} {Phys. Rev. D}\ }\textbf {\bibinfo {volume} {111}},\
  \bibinfo {pages} {075002} (\bibinfo {year} {2025})},\ \Eprint
  {https://arxiv.org/abs/2411.15303} {arXiv:2411.15303 [hep-ph]} \BibitemShut
  {NoStop}%
\bibitem [{\citenamefont {Hamada}\ \emph {et~al.}(2024)\citenamefont {Hamada},
  \citenamefont {Kitano}, \citenamefont {Matsudo}, \citenamefont {Okawa},
  \citenamefont {Takai}, \citenamefont {Takaura},\ and\ \citenamefont
  {Treuer}}]{Hamada:2024ojj}%
  \BibitemOpen
  \bibfield  {author} {\bibinfo {author} {\bibfnamefont {Y.}~\bibnamefont
  {Hamada}}, \bibinfo {author} {\bibfnamefont {R.}~\bibnamefont {Kitano}},
  \bibinfo {author} {\bibfnamefont {R.}~\bibnamefont {Matsudo}}, \bibinfo
  {author} {\bibfnamefont {S.}~\bibnamefont {Okawa}}, \bibinfo {author}
  {\bibfnamefont {R.}~\bibnamefont {Takai}}, \bibinfo {author} {\bibfnamefont
  {H.}~\bibnamefont {Takaura}},\ and\ \bibinfo {author} {\bibfnamefont
  {L.}~\bibnamefont {Treuer}},\ }\bibfield  {title} {\bibinfo {title} {{Higgs
  boson production at \ensuremath{\mu}+\ensuremath{\mu}+ colliders}},\ }\href
  {https://doi.org/10.1103/PhysRevD.110.113011} {\bibfield  {journal} {\bibinfo
   {journal} {Phys. Rev. D}\ }\textbf {\bibinfo {volume} {110}},\ \bibinfo
  {pages} {113011} (\bibinfo {year} {2024})},\ \Eprint
  {https://arxiv.org/abs/2408.01068} {arXiv:2408.01068 [hep-ph]} \BibitemShut
  {NoStop}%
\bibitem [{\citenamefont {Chen}\ \emph
  {et~al.}(2024{\natexlab{b}})\citenamefont {Chen}, \citenamefont {Iguro},\
  and\ \citenamefont {Hamada}}]{Chen:2024tqh}%
  \BibitemOpen
  \bibfield  {author} {\bibinfo {author} {\bibfnamefont {L.}~\bibnamefont
  {Chen}}, \bibinfo {author} {\bibfnamefont {S.}~\bibnamefont {Iguro}},\ and\
  \bibinfo {author} {\bibfnamefont {Y.}~\bibnamefont {Hamada}},\ }\bibfield
  {title} {\bibinfo {title} {{Determining Weak-Mixing Angle at $\mu$TRISTAN}},\
  }\href@noop {} {\  (\bibinfo {year} {2024}{\natexlab{b}})},\ \Eprint
  {https://arxiv.org/abs/2406.04500} {arXiv:2406.04500 [hep-ph]} \BibitemShut
  {NoStop}%
\bibitem [{\citenamefont {Okabe}\ and\ \citenamefont
  {Shirai}(2024)}]{Okabe:2023esr}%
  \BibitemOpen
  \bibfield  {author} {\bibinfo {author} {\bibfnamefont {R.}~\bibnamefont
  {Okabe}}\ and\ \bibinfo {author} {\bibfnamefont {S.}~\bibnamefont {Shirai}},\
  }\bibfield  {title} {\bibinfo {title} {{Indirect probe of
  electroweak-interacting particles at the \ensuremath{\mu}TRISTAN
  \ensuremath{\mu}+\ensuremath{\mu}+ collider}},\ }\href
  {https://doi.org/10.1103/PhysRevD.110.035002} {\bibfield  {journal} {\bibinfo
   {journal} {Phys. Rev. D}\ }\textbf {\bibinfo {volume} {110}},\ \bibinfo
  {pages} {035002} (\bibinfo {year} {2024})},\ \Eprint
  {https://arxiv.org/abs/2310.08434} {arXiv:2310.08434 [hep-ph]} \BibitemShut
  {NoStop}%
\bibitem [{\citenamefont {Fukuda}\ \emph {et~al.}(2024)\citenamefont {Fukuda},
  \citenamefont {Moroi}, \citenamefont {Niki},\ and\ \citenamefont
  {Wei}}]{Fukuda:2023yui}%
  \BibitemOpen
  \bibfield  {author} {\bibinfo {author} {\bibfnamefont {H.}~\bibnamefont
  {Fukuda}}, \bibinfo {author} {\bibfnamefont {T.}~\bibnamefont {Moroi}},
  \bibinfo {author} {\bibfnamefont {A.}~\bibnamefont {Niki}},\ and\ \bibinfo
  {author} {\bibfnamefont {S.-F.}\ \bibnamefont {Wei}},\ }\bibfield  {title}
  {\bibinfo {title} {{Search for WIMPs at future
  \ensuremath{\mu}$^{+}$\ensuremath{\mu}$^{+}$ colliders}},\ }\href
  {https://doi.org/10.1007/JHEP02(2024)214} {\bibfield  {journal} {\bibinfo
  {journal} {JHEP}\ }\textbf {\bibinfo {volume} {02}},\ \bibinfo {pages}
  {214}},\ \Eprint {https://arxiv.org/abs/2310.07162} {arXiv:2310.07162
  [hep-ph]} \BibitemShut {NoStop}%
\bibitem [{\citenamefont {Dev}\ \emph {et~al.}(2024)\citenamefont {Dev},
  \citenamefont {Heeck},\ and\ \citenamefont {Thapa}}]{Dev:2023nha}%
  \BibitemOpen
  \bibfield  {author} {\bibinfo {author} {\bibfnamefont {P.~S.~B.}\
  \bibnamefont {Dev}}, \bibinfo {author} {\bibfnamefont {J.}~\bibnamefont
  {Heeck}},\ and\ \bibinfo {author} {\bibfnamefont {A.}~\bibnamefont {Thapa}},\
  }\bibfield  {title} {\bibinfo {title} {{Neutrino mass models at $\mu
  $TRISTAN}},\ }\href {https://doi.org/10.1140/epjc/s10052-024-12496-0}
  {\bibfield  {journal} {\bibinfo  {journal} {Eur. Phys. J. C}\ }\textbf
  {\bibinfo {volume} {84}},\ \bibinfo {pages} {148} (\bibinfo {year} {2024})},\
  \Eprint {https://arxiv.org/abs/2309.06463} {arXiv:2309.06463 [hep-ph]}
  \BibitemShut {NoStop}%
\bibitem [{\citenamefont {Buchmuller}\ and\ \citenamefont
  {Wyler}(1986)}]{Buchmuller:1985jz}%
  \BibitemOpen
  \bibfield  {author} {\bibinfo {author} {\bibfnamefont {W.}~\bibnamefont
  {Buchmuller}}\ and\ \bibinfo {author} {\bibfnamefont {D.}~\bibnamefont
  {Wyler}},\ }\bibfield  {title} {\bibinfo {title} {{Effective Lagrangian
  Analysis of New Interactions and Flavor Conservation}},\ }\href
  {https://doi.org/10.1016/0550-3213(86)90262-2} {\bibfield  {journal}
  {\bibinfo  {journal} {Nucl. Phys. B}\ }\textbf {\bibinfo {volume} {268}},\
  \bibinfo {pages} {621} (\bibinfo {year} {1986})}\BibitemShut {NoStop}%
\bibitem [{\citenamefont {Grzadkowski}\ \emph {et~al.}(2010)\citenamefont
  {Grzadkowski}, \citenamefont {Iskrzynski}, \citenamefont {Misiak},\ and\
  \citenamefont {Rosiek}}]{Grzadkowski:2010es}%
  \BibitemOpen
  \bibfield  {author} {\bibinfo {author} {\bibfnamefont {B.}~\bibnamefont
  {Grzadkowski}}, \bibinfo {author} {\bibfnamefont {M.}~\bibnamefont
  {Iskrzynski}}, \bibinfo {author} {\bibfnamefont {M.}~\bibnamefont {Misiak}},\
  and\ \bibinfo {author} {\bibfnamefont {J.}~\bibnamefont {Rosiek}},\
  }\bibfield  {title} {\bibinfo {title} {{Dimension-Six Terms in the Standard
  Model Lagrangian}},\ }\href {https://doi.org/10.1007/JHEP10(2010)085}
  {\bibfield  {journal} {\bibinfo  {journal} {JHEP}\ }\textbf {\bibinfo
  {volume} {10}},\ \bibinfo {pages} {085}},\ \Eprint
  {https://arxiv.org/abs/1008.4884} {arXiv:1008.4884 [hep-ph]} \BibitemShut
  {NoStop}%
\bibitem [{\citenamefont {Weinberg}(1979)}]{Weinberg:1979sa}%
  \BibitemOpen
  \bibfield  {author} {\bibinfo {author} {\bibfnamefont {S.}~\bibnamefont
  {Weinberg}},\ }\bibfield  {title} {\bibinfo {title} {{Baryon and Lepton
  Nonconserving Processes}},\ }\href
  {https://doi.org/10.1103/PhysRevLett.43.1566} {\bibfield  {journal} {\bibinfo
   {journal} {Phys. Rev. Lett.}\ }\textbf {\bibinfo {volume} {43}},\ \bibinfo
  {pages} {1566} (\bibinfo {year} {1979})}\BibitemShut {NoStop}%
\bibitem [{\citenamefont {Durieux}\ \emph {et~al.}(2015)\citenamefont
  {Durieux}, \citenamefont {Maltoni},\ and\ \citenamefont
  {Zhang}}]{Durieux:2014xla}%
  \BibitemOpen
  \bibfield  {author} {\bibinfo {author} {\bibfnamefont {G.}~\bibnamefont
  {Durieux}}, \bibinfo {author} {\bibfnamefont {F.}~\bibnamefont {Maltoni}},\
  and\ \bibinfo {author} {\bibfnamefont {C.}~\bibnamefont {Zhang}},\ }\bibfield
   {title} {\bibinfo {title} {{Global approach to top-quark flavor-changing
  interactions}},\ }\href {https://doi.org/10.1103/PhysRevD.91.074017}
  {\bibfield  {journal} {\bibinfo  {journal} {Phys. Rev. D}\ }\textbf {\bibinfo
  {volume} {91}},\ \bibinfo {pages} {074017} (\bibinfo {year} {2015})},\
  \Eprint {https://arxiv.org/abs/1412.7166} {arXiv:1412.7166 [hep-ph]}
  \BibitemShut {NoStop}%
\bibitem [{\citenamefont {Aebischer}\ \emph {et~al.}(2025)\citenamefont
  {Aebischer}, \citenamefont {Buras},\ and\ \citenamefont
  {Kumar}}]{Aebischer:2025qhh}%
  \BibitemOpen
  \bibfield  {author} {\bibinfo {author} {\bibfnamefont {J.}~\bibnamefont
  {Aebischer}}, \bibinfo {author} {\bibfnamefont {A.~J.}\ \bibnamefont
  {Buras}},\ and\ \bibinfo {author} {\bibfnamefont {J.}~\bibnamefont {Kumar}},\
  }\bibfield  {title} {\bibinfo {title} {{SMEFT ATLAS: The Landscape Beyond the
  Standard Model}},\ }\href@noop {} {\  (\bibinfo {year} {2025})},\ \Eprint
  {https://arxiv.org/abs/2507.05926} {arXiv:2507.05926 [hep-ph]} \BibitemShut
  {NoStop}%
\bibitem [{\citenamefont {Delzanno}\ \emph {et~al.}(2025)\citenamefont
  {Delzanno}, \citenamefont {Fuyuto}, \citenamefont {Gonz{\`a}lez-Sol{\'\i}s},\
  and\ \citenamefont {Mereghetti}}]{Delzanno:2024ooj}%
  \BibitemOpen
  \bibfield  {author} {\bibinfo {author} {\bibfnamefont {F.}~\bibnamefont
  {Delzanno}}, \bibinfo {author} {\bibfnamefont {K.}~\bibnamefont {Fuyuto}},
  \bibinfo {author} {\bibfnamefont {S.}~\bibnamefont
  {Gonz{\`a}lez-Sol{\'\i}s}},\ and\ \bibinfo {author} {\bibfnamefont
  {E.}~\bibnamefont {Mereghetti}},\ }\bibfield  {title} {\bibinfo {title}
  {{Global analysis of {\ensuremath{\mu}} {\textrightarrow} e interactions in
  the SMEFT}},\ }\href {https://doi.org/10.1007/JHEP07(2025)283} {\bibfield
  {journal} {\bibinfo  {journal} {JHEP}\ }\textbf {\bibinfo {volume} {07}},\
  \bibinfo {pages} {283}},\ \Eprint {https://arxiv.org/abs/2411.13497}
  {arXiv:2411.13497 [hep-ph]} \BibitemShut {NoStop}%
\bibitem [{\citenamefont {Chala}\ \emph {et~al.}(2019)\citenamefont {Chala},
  \citenamefont {Santiago},\ and\ \citenamefont {Spannowsky}}]{Chala:2018agk}%
  \BibitemOpen
  \bibfield  {author} {\bibinfo {author} {\bibfnamefont {M.}~\bibnamefont
  {Chala}}, \bibinfo {author} {\bibfnamefont {J.}~\bibnamefont {Santiago}},\
  and\ \bibinfo {author} {\bibfnamefont {M.}~\bibnamefont {Spannowsky}},\
  }\bibfield  {title} {\bibinfo {title} {{Constraining four-fermion operators
  using rare top decays}},\ }\href {https://doi.org/10.1007/JHEP04(2019)014}
  {\bibfield  {journal} {\bibinfo  {journal} {JHEP}\ }\textbf {\bibinfo
  {volume} {04}},\ \bibinfo {pages} {014}},\ \Eprint
  {https://arxiv.org/abs/1809.09624} {arXiv:1809.09624 [hep-ph]} \BibitemShut
  {NoStop}%
\bibitem [{\citenamefont {Roney}(2021)}]{Roney:2021pwz}%
  \BibitemOpen
  \bibfield  {author} {\bibinfo {author} {\bibfnamefont {J.~M.}\ \bibnamefont
  {Roney}} (\bibinfo {collaboration} {Belle II SuperKEKB e- Polarization
  Upgrade Working Group}),\ }\bibfield  {title} {\bibinfo {title} {{Upgrading
  SuperKEKB with polarized $e^-$ beams}},\ }\href
  {https://doi.org/10.22323/1.390.0699} {\bibfield  {journal} {\bibinfo
  {journal} {PoS}\ }\textbf {\bibinfo {volume} {ICHEP2020}},\ \bibinfo {pages}
  {699} (\bibinfo {year} {2021})}\BibitemShut {NoStop}%
\bibitem [{\citenamefont {Alloul}\ \emph {et~al.}(2014)\citenamefont {Alloul},
  \citenamefont {Christensen}, \citenamefont {Degrande}, \citenamefont {Duhr},\
  and\ \citenamefont {Fuks}}]{Alloul:2013bka}%
  \BibitemOpen
  \bibfield  {author} {\bibinfo {author} {\bibfnamefont {A.}~\bibnamefont
  {Alloul}}, \bibinfo {author} {\bibfnamefont {N.~D.}\ \bibnamefont
  {Christensen}}, \bibinfo {author} {\bibfnamefont {C.}~\bibnamefont
  {Degrande}}, \bibinfo {author} {\bibfnamefont {C.}~\bibnamefont {Duhr}},\
  and\ \bibinfo {author} {\bibfnamefont {B.}~\bibnamefont {Fuks}},\ }\bibfield
  {title} {\bibinfo {title} {{FeynRules 2.0 - A complete toolbox for tree-level
  phenomenology}},\ }\href {https://doi.org/10.1016/j.cpc.2014.04.012}
  {\bibfield  {journal} {\bibinfo  {journal} {Comput. Phys. Commun.}\ }\textbf
  {\bibinfo {volume} {185}},\ \bibinfo {pages} {2250} (\bibinfo {year}
  {2014})},\ \Eprint {https://arxiv.org/abs/1310.1921} {arXiv:1310.1921
  [hep-ph]} \BibitemShut {NoStop}%
\bibitem [{\citenamefont {Darm\'e}\ \emph {et~al.}(2023)\citenamefont {Darm\'e}
  \emph {et~al.}}]{Darme:2023jdn}%
  \BibitemOpen
  \bibfield  {author} {\bibinfo {author} {\bibfnamefont {L.}~\bibnamefont
  {Darm\'e}} \emph {et~al.},\ }\bibfield  {title} {\bibinfo {title} {{UFO 2.0:
  the \textquoteleft{}Universal Feynman Output\textquoteright{} format}},\
  }\href {https://doi.org/10.1140/epjc/s10052-023-11780-9} {\bibfield
  {journal} {\bibinfo  {journal} {Eur. Phys. J. C}\ }\textbf {\bibinfo {volume}
  {83}},\ \bibinfo {pages} {631} (\bibinfo {year} {2023})},\ \Eprint
  {https://arxiv.org/abs/2304.09883} {arXiv:2304.09883 [hep-ph]} \BibitemShut
  {NoStop}%
\bibitem [{\citenamefont {Alwall}\ \emph {et~al.}(2011)\citenamefont {Alwall},
  \citenamefont {Herquet}, \citenamefont {Maltoni}, \citenamefont {Mattelaer},\
  and\ \citenamefont {Stelzer}}]{Alwall:2011uj}%
  \BibitemOpen
  \bibfield  {author} {\bibinfo {author} {\bibfnamefont {J.}~\bibnamefont
  {Alwall}}, \bibinfo {author} {\bibfnamefont {M.}~\bibnamefont {Herquet}},
  \bibinfo {author} {\bibfnamefont {F.}~\bibnamefont {Maltoni}}, \bibinfo
  {author} {\bibfnamefont {O.}~\bibnamefont {Mattelaer}},\ and\ \bibinfo
  {author} {\bibfnamefont {T.}~\bibnamefont {Stelzer}},\ }\bibfield  {title}
  {\bibinfo {title} {{MadGraph 5 : Going Beyond}},\ }\href
  {https://doi.org/10.1007/JHEP06(2011)128} {\bibfield  {journal} {\bibinfo
  {journal} {JHEP}\ }\textbf {\bibinfo {volume} {06}},\ \bibinfo {pages}
  {128}},\ \Eprint {https://arxiv.org/abs/1106.0522} {arXiv:1106.0522 [hep-ph]}
  \BibitemShut {NoStop}%
\bibitem [{\citenamefont {Bierlich}\ \emph {et~al.}(2022)\citenamefont
  {Bierlich} \emph {et~al.}}]{Bierlich:2022pfr}%
  \BibitemOpen
  \bibfield  {author} {\bibinfo {author} {\bibfnamefont {C.}~\bibnamefont
  {Bierlich}} \emph {et~al.},\ }\bibfield  {title} {\bibinfo {title} {{A
  comprehensive guide to the physics and usage of PYTHIA 8.3}},\ }\href
  {https://doi.org/10.21468/SciPostPhysCodeb.8} {\bibfield  {journal} {\bibinfo
   {journal} {SciPost Phys. Codeb.}\ }\textbf {\bibinfo {volume} {2022}},\
  \bibinfo {pages} {8} (\bibinfo {year} {2022})},\ \Eprint
  {https://arxiv.org/abs/2203.11601} {arXiv:2203.11601 [hep-ph]} \BibitemShut
  {NoStop}%
\bibitem [{\citenamefont {de~Favereau}\ \emph {et~al.}(2014)\citenamefont
  {de~Favereau}, \citenamefont {Delaere}, \citenamefont {Demin}, \citenamefont
  {Giammanco}, \citenamefont {Lema\^\i{}tre}, \citenamefont {Mertens},\ and\
  \citenamefont {Selvaggi}}]{deFavereau:2013fsa}%
  \BibitemOpen
  \bibfield  {author} {\bibinfo {author} {\bibfnamefont {J.}~\bibnamefont
  {de~Favereau}}, \bibinfo {author} {\bibfnamefont {C.}~\bibnamefont
  {Delaere}}, \bibinfo {author} {\bibfnamefont {P.}~\bibnamefont {Demin}},
  \bibinfo {author} {\bibfnamefont {A.}~\bibnamefont {Giammanco}}, \bibinfo
  {author} {\bibfnamefont {V.}~\bibnamefont {Lema\^\i{}tre}}, \bibinfo {author}
  {\bibfnamefont {A.}~\bibnamefont {Mertens}},\ and\ \bibinfo {author}
  {\bibfnamefont {M.}~\bibnamefont {Selvaggi}} (\bibinfo {collaboration}
  {DELPHES 3}),\ }\bibfield  {title} {\bibinfo {title} {{DELPHES 3, A modular
  framework for fast simulation of a generic collider experiment}},\ }\href
  {https://doi.org/10.1007/JHEP02(2014)057} {\bibfield  {journal} {\bibinfo
  {journal} {JHEP}\ }\textbf {\bibinfo {volume} {02}},\ \bibinfo {pages}
  {057}},\ \Eprint {https://arxiv.org/abs/1307.6346} {arXiv:1307.6346 [hep-ex]}
  \BibitemShut {NoStop}%
\bibitem [{\citenamefont {Abramowicz}\ \emph {et~al.}(2013)\citenamefont
  {Abramowicz} \emph {et~al.}}]{Behnke:2013lya}%
  \BibitemOpen
  \bibfield  {author} {\bibinfo {author} {\bibfnamefont {H.}~\bibnamefont
  {Abramowicz}} \emph {et~al.},\ }\bibfield  {title} {\bibinfo {title} {{The
  International Linear Collider Technical Design Report - Volume 4:
  Detectors}},\ }\href@noop {} {\  (\bibinfo {year} {2013})},\ \Eprint
  {https://arxiv.org/abs/1306.6329} {arXiv:1306.6329 [physics.ins-det]}
  \BibitemShut {NoStop}%
\bibitem [{\citenamefont {Navas}\ \emph {et~al.}(2024)\citenamefont {Navas}
  \emph {et~al.}}]{ParticleDataGroup:2024cfk}%
  \BibitemOpen
  \bibfield  {author} {\bibinfo {author} {\bibfnamefont {S.}~\bibnamefont
  {Navas}} \emph {et~al.} (\bibinfo {collaboration} {Particle Data Group}),\
  }\bibfield  {title} {\bibinfo {title} {{Review of particle physics}},\ }\href
  {https://doi.org/10.1103/PhysRevD.110.030001} {\bibfield  {journal} {\bibinfo
   {journal} {Phys. Rev. D}\ }\textbf {\bibinfo {volume} {110}},\ \bibinfo
  {pages} {030001} (\bibinfo {year} {2024})}\BibitemShut {NoStop}%
\bibitem [{\citenamefont {Wilks}(1938)}]{Wilks:1938dza}%
  \BibitemOpen
  \bibfield  {author} {\bibinfo {author} {\bibfnamefont {S.~S.}\ \bibnamefont
  {Wilks}},\ }\bibfield  {title} {\bibinfo {title} {{The Large-Sample
  Distribution of the Likelihood Ratio for Testing Composite Hypotheses}},\
  }\href {https://doi.org/10.1214/aoms/1177732360} {\bibfield  {journal}
  {\bibinfo  {journal} {Annals Math. Statist.}\ }\textbf {\bibinfo {volume}
  {9}},\ \bibinfo {pages} {60} (\bibinfo {year} {1938})}\BibitemShut {NoStop}%
\bibitem [{\citenamefont {Greljo}\ \emph {et~al.}(2023)\citenamefont {Greljo},
  \citenamefont {Salko}, \citenamefont {Smolkovi{\v{c}}},\ and\ \citenamefont
  {Stangl}}]{Greljo:2022jac}%
  \BibitemOpen
  \bibfield  {author} {\bibinfo {author} {\bibfnamefont {A.}~\bibnamefont
  {Greljo}}, \bibinfo {author} {\bibfnamefont {J.}~\bibnamefont {Salko}},
  \bibinfo {author} {\bibfnamefont {A.}~\bibnamefont {Smolkovi{\v{c}}}},\ and\
  \bibinfo {author} {\bibfnamefont {P.}~\bibnamefont {Stangl}},\ }\bibfield
  {title} {\bibinfo {title} {{Rare b decays meet high-mass Drell-Yan}},\ }\href
  {https://doi.org/10.1007/JHEP05(2023)087} {\bibfield  {journal} {\bibinfo
  {journal} {JHEP}\ }\textbf {\bibinfo {volume} {05}},\ \bibinfo {pages}
  {087}},\ \Eprint {https://arxiv.org/abs/2212.10497} {arXiv:2212.10497
  [hep-ph]} \BibitemShut {NoStop}%
\bibitem [{\citenamefont {Pruna}\ and\ \citenamefont
  {Signer}(2014)}]{Pruna:2014asa}%
  \BibitemOpen
  \bibfield  {author} {\bibinfo {author} {\bibfnamefont {G.~M.}\ \bibnamefont
  {Pruna}}\ and\ \bibinfo {author} {\bibfnamefont {A.}~\bibnamefont {Signer}},\
  }\bibfield  {title} {\bibinfo {title} {{The $\mu\to e\gamma$ decay in a
  systematic effective field theory approach with dimension 6 operators}},\
  }\href {https://doi.org/10.1007/JHEP10(2014)014} {\bibfield  {journal}
  {\bibinfo  {journal} {JHEP}\ }\textbf {\bibinfo {volume} {10}},\ \bibinfo
  {pages} {014}},\ \Eprint {https://arxiv.org/abs/1408.3565} {arXiv:1408.3565
  [hep-ph]} \BibitemShut {NoStop}%
\bibitem [{\citenamefont {Kumar}(2022)}]{Kumar:2021yod}%
  \BibitemOpen
  \bibfield  {author} {\bibinfo {author} {\bibfnamefont {J.}~\bibnamefont
  {Kumar}},\ }\bibfield  {title} {\bibinfo {title} {{Renormalization group
  improved implications of semileptonic operators in SMEFT}},\ }\href
  {https://doi.org/10.1007/JHEP01(2022)107} {\bibfield  {journal} {\bibinfo
  {journal} {JHEP}\ }\textbf {\bibinfo {volume} {01}},\ \bibinfo {pages}
  {107}},\ \Eprint {https://arxiv.org/abs/2107.13005} {arXiv:2107.13005
  [hep-ph]} \BibitemShut {NoStop}%
\end{thebibliography}%
\end{document}